\documentclass[journal=nalefd,manuscript=letter]{achemso}

\usepackage[version=3]{mhchem} 
\usepackage{amsmath}
\usepackage{graphicx}
\usepackage{dcolumn}
\usepackage{bm}
\usepackage{xcolor}
\usepackage{amssymb}
\usepackage{dcolumn}
\usepackage{bm}
\SectionNumbersOn

\author{Daniel M. Markiewitz}
\affiliation{Department of Chemical Engineering, Massachusetts Institute of Technology, Cambridge, Massachusetts  02139, USA}

\author{Zachary A. H. Goodwin}
\affiliation{John A. Paulson School of Engineering and Applied Sciences, Harvard University, Cambridge, Massachusetts 02138, United States}
\alsoaffiliation{Department of Materials, University of Oxford, Parks Road, Oxford OX1 3PH, United Kingdom}

\author{Qianlu Zheng}
\affiliation{Department of Civil and Environmental Engineering, University of Illinois Urbana–Champaign, Urbana, IL, 61801 USA}

\author{Michael McEldrew}
\affiliation{Department of Chemical Engineering, Massachusetts Institute of Technology, Cambridge, Massachusetts  02139, USA}

\author{Rosa M. Espinosa-Marzal}
\affiliation{Department of Civil and Environmental Engineering, University of Illinois Urbana–Champaign, Urbana, IL, 61801 USA}
\alsoaffiliation{Department of Materials Science and Engineering, University of Illinois Urbana–Champaign, Urbana, IL, 61801 USA}

\author{Martin Z. Bazant}
\email{bazant@mit.edu}
\affiliation{Department of Chemical Engineering, Massachusetts Institute of Technology, Cambridge, Massachusetts  02139, USA}
\alsoaffiliation{Department of Mathematics, Massachusetts Institute of Technology, Cambridge, Massachusetts 02139, USA}

\title[Ionic Associations and Hydration in the Electrical Double Layer of Water-in-Salt Electrolytes]{Ionic Associations and Hydration in the Electrical Double Layer of Water-in-Salt Electrolytes}

\title{Ionic Associations and Hydration in the Electrical Double Layer of Water-in-Salt Electrolytes}


\begin{document}

\begin{tocentry} 
     \includegraphics[]{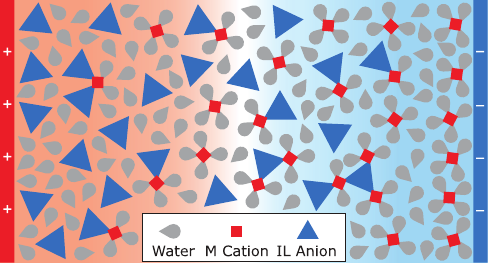}
\end{tocentry}

\begin{abstract}
Water-in-Salt-Electrolytes (WiSEs) are an exciting class of concentrated electrolytes finding applications in energy storage devices because of their expanded electrochemical stability window, good conductivity and cation transference number, and fire-extinguishing properties. These distinct properties are thought to originate from the presence of an anion-dominated ionic network and interpenetrating water channels for cation transport, which indicates that associations in WiSEs are crucial to understanding their properties. Currently, associations have mainly been investigated in the bulk, while little attention has been given to the electrolyte structure near electrified interfaces. Here, we develop a theory for the electrical double layer (EDL) of WiSEs, where we consistently account for the thermoreversible associations of species into Cayley tree aggregates. The theory predicts an asymmetric structure of the EDL. At negative voltages, hydrated Li$^+$ dominate and cluster aggregation is initially slightly enhanced before disintegration at larger voltages. At positive voltages when compared to the bulk, clusters are strictly diminished. Performing atomistic molecular dynamics (MD) simulations of the EDL of WiSE provides EDL data for validation and bulk data for parameterization of our theory. Validating the predictions of our theory against MD showed good qualitative agreement. Furthermore, we performed electrochemical impendence measurements to determine the differential capacitance of the studied LiTFSI WiSE and also found reasonable agreement with our theory. Overall, the developed approach can be used to investigate ionic aggregation and solvation effects in the EDL, which amongst other properties, can be used to understand the pre-cursers for solid-electrolyte interphase formation.
\end{abstract}


\section{Introduction}
Water-in-Salt Electrolytes (WiSEs) have emerged as a promising class of electrolytes for applications in batteries and supercapacitors~\cite{Suo2013,Suo2015,Smith2015,Wang2016,Wang2018,Wang2018a,Sun2017,Sodeyama2014,Yamada2016,kuhnel2017,suo2017water,leonard2018,thareja2021water,park2022redox}. In contrast to conventional organic Li-ion battery electrolytes, WiSEs have dramatically improved safety and stability, owing to the use of water as a solvent instead of flammable carbonate solvents~\cite{Suo2013,Suo2015,haregewoin2016electrolyte,suo2017water,dou2018safe}. In classical dilute aqueous electrolytes, water is known to electrolyze around 1.23~V, giving their small electrochemical stability windows (ESW), but the super-concentrated WiSE regime displays enhanced ESWs up to 4~V~\cite{Yamada2016,zhang2018water,yang2019aqueous}. The reductive stability of WiSEs is attributed to the formation of a passivating solid-electrolyte interphase (SEI) at the anode, similar to conventional Li-ion battery electrolytes~\cite{Suo2015,yang2017,borodin2020uncharted,sayah2022super}. At the same time, the oxidative stability originates from the thermodynamic activity of water reducing in the super-concentrated regime~\cite{vatamanu2017,mceldrew2018,borodin2020uncharted,mceldrew2021ion,sayah2022super}.

As WiSEs are often used in the super-concentrated salt regime, such as 21m LiTFSI, it is perhaps not surprising that the aggregation of ions has been discovered to be important in numerous simulation and experimental studies~\cite{borodin2017liquid,zheng2018understanding,lim2018,choi2018graph,han2020origin,andersson2020ion,yu2020asymmetric,lewis2020signatures,gonzalez2020nanoscale,zhang2020potential,mceldrew2021ion,Han2021WiSE,liu2021microscopic,groves2021surface}. At the optimal concentration of 21m LiTFSI, the electrolyte obtained an operating voltage of $\sim$2.3 V while maintaining reasonable conductivity~\cite{Suo2015}. They found that a predominantly Li-anion ionic network exists~\cite{borodin2017liquid} that is interpenetrated by nano-channels of water-rich domains containing Li cations~\cite{borodin2017liquid,lim2018,mceldrew2021ion}. The existence of these nano-channels enables the facile transport of Li cations, which is crucial for their operating performance. In addition, the existence and equilibrium of aggregates at interfaces have been revealed by surface force apparatus (SFA) and atomic force microscopy (AFM) measurements~\cite{Han2021WiSE,groves2021surface,zhang2020potential,ichii2020solvation} and molecular dynamics simulations~\cite{vatamanu2017,li2022unconventional,mceldrew2018}, where it was found that hydrated Li$^+$ exists near the interface.

To understand these simulations and experiments in the electrical double layer (EDL) of WiSEs, it is useful to have a theory to rationalize the observations. One of the first theories for the EDL of WiSEs came from McEldrew \textit{et al.}~\cite{mceldrew2018}, where the Bazant-Storey-Kornyshev theory~\cite{Bazant2011} was incorporated with the Langevin fluctuating dipole model for ``free'' water molecules, with most of the water assumed to be rigidly bound to Li cations. This theory was able to rationalize the overall changes in the composition of the EDL; but, it contains no explicit information of the ionic associations, and assumes that the water bound to each of Li$^+$ had infinitely strong associations. Later, McEldrew \textit{et al.}~\cite{mceldrew2020theory} developed a theory for thermo-reversible aggregation and gelation in WiSEs~\cite{mceldrew2021ion}, amongst other electrolytes~\cite{mceldrew2020corr,McEldrewsalt2021,Goodwin2023}, which allowed the cluster distributions and percolating ionic networks to be understood within the framework of Flory, Stockmayers and Tanaka's famous polymer work~\cite{flory1942thermodynamics,flory1953principles,stockmayer1943theory,stockmayer1944theory,stockmayer1952molecular,tanaka1989,tanaka1990thermodynamic,tanaka1994,tanaka1995,ishida1997,tanaka1998,tanaka1999,tanaka2002}. More recently, Goodwin \textit{et al.}~\cite{Goodwin2022EDL,Goodwin2022Kornyshev} and Markiewitz \textit{et al.}~\cite{Markiewitz2024} extended this formalism to tackle the EDL, but the theory has yet to be rigorously tested against MD simulations and experiments for real electrolytes of interest in the battery and supercapacitor community.

\begin{figure}[h!]
 \centering
 \includegraphics[width= 0.7\textwidth]{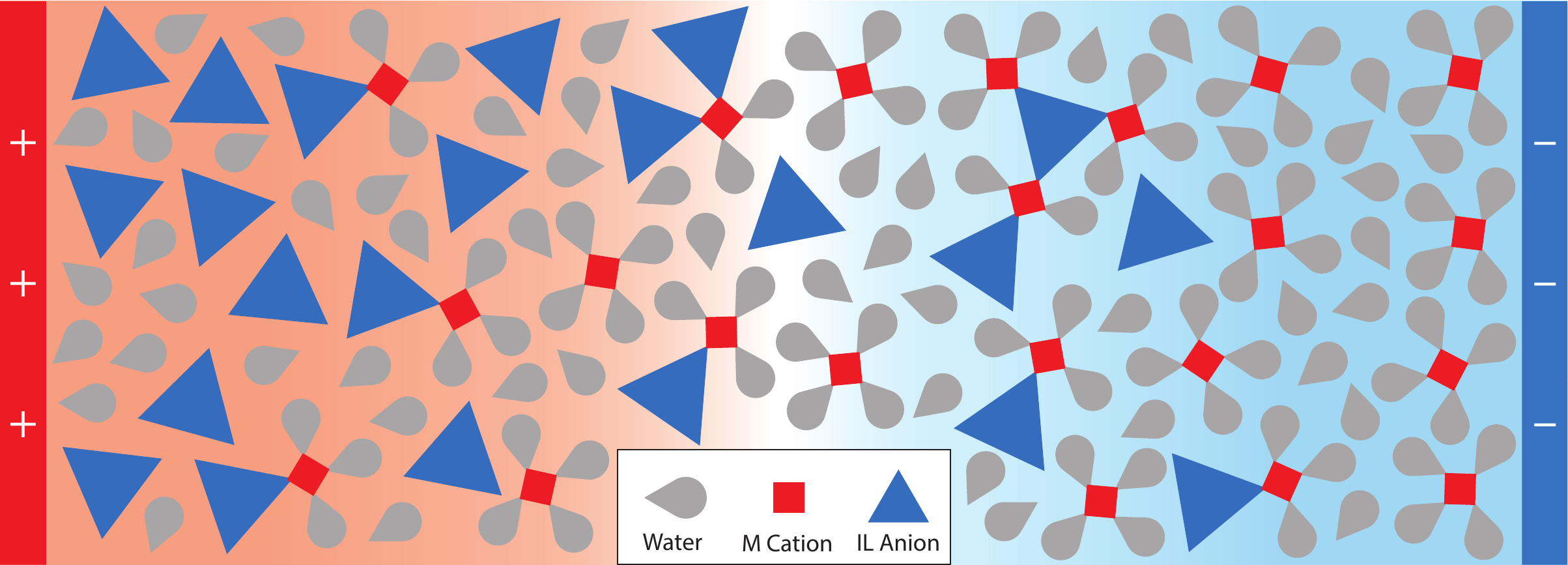}
 \caption{Schematic of the modulation of aggregation occurring in the EDL of WiSE near positively (left) and negatively (right) charged electrodes. Here, the alkali metal cations can form up to 4 associations, the IL anions can form up to 3 associations, and water can form up to 1 association. Ion associations are shown by touching vertices.}
 \label{fig:vis_edl}
\end{figure}

In this paper, we develop a theory for the EDL of WiSEs based on previous work~\cite{mceldrew2020theory,mceldrew2021ion,Goodwin2022EDL,Markiewitz2024}, where thermoreversible associations are treated consistently, and test it against MD simulations and experiments. This theory, MD simulations, and experiments allow us to analyze the role these associations have in the structure and properties of WiSEs in the EDL. We found three main conclusions from this investigation. First, we found that at negative voltages, in both theory and MD simulations, electric field-induced associations are present at small negative electrostatic potentials, i.e. over a range of potentials the WiSE \textit{becomes more associated than in bulk}. Moreover, we found the hydrated Li$^+$ becomes increasingly dominant with larger negative potentials. At positive voltages, it appears the clusters are strictly diminished when compared to the bulk. The distinct difference in the EDL structure and behavior at negative and positive voltages can be seen through our theory and is presented schematically in Fig.~\ref{fig:vis_edl}. Second, we found the theory was able to reproduce trends observed in MD simulations as well as aggregation length scales inferred from AFM measurements. Third, we found that the trends in the theory's predicted differential capacitance were promising when compared against experimental measurements. Overall, we found that the simple theory presented here can capture how associations change \textit{within} the EDL, which has never been quantified with any theory. 

\section{Theory}
Here, we consider our system to be an incompressible lattice gas model~\cite{mceldrew2020theory} composed of alkali metal cations (+), ionic liquid anions (-), and water (0), where we define the size of a lattice site by the volume of a water molecule ($v_0$), with the volume ratios of all species, $\xi_j$ = $v_j$/$v_0$. From these ratios, we can define the dimensionless concentration of a species as $c_j$ = $\phi_j$/$\xi_j$, where ($\phi_j$) is the volume fraction of each species and \textit{j} is +,-,0.

Similar to previous works~\cite{mceldrew2021ion,Goodwin2023}, we consider the formation of associations between the cations and anions as well as between cations and water. The cations can form a maximum of $f_+$ associations, anions can form a maximum of $f_-$ associations, and water can form a maximum of 1 association; these maximum associations are defined as the functionality of each species. The functionality of a species can be obtained from the maximum coordination numbers in the first solvation shell from MD simulations; and therefore, this is not a free parameter of our theory. When the functionality of the associating species is greater than 1 they can form a set of polydisperse clusters, which can be classified by the rank $lms$ of the cluster. This rank specifies the number of cations $l$, anions $m$, and water $s$ that comprise the cluster, with the dimensionless concentration of the rank $lms$ cluster being $c_{lms}$. We assume the clusters only form Cayley-tree-like structures, i.e., no loops are present in a cluster~\cite{mceldrew2020theory}. This Cayley tree assumption for the clusters is necessary to keep this theory analytically tractable and physically intuitive~\cite{mceldrew2020theory}. This approximation is known to breakdown for some electrolytes~\cite{mceldrew2021ion}, but it was shown to work well for WiSEs~\cite{McEldrewsalt2021}.

When the functionalities of cations and anions are equal to or greater than 2, then a percolating ionic network can form~\cite{mceldrew2020theory}. This transition is referred to as gelation and is a second-order phase transition. In this gel regime, we employ Flory's post-gel convention to determine the volume fraction of each species in the sol ($\phi_{j}^{sol}$ this phase contains both the clusters and the free species) and in the gel phase ($\phi_{j}^{gel}$ this phase contains only the percolating network), where $\phi_{j} = \phi_{j}^{sol} + \phi_{j}^{gel}$ and \textit{j} is +,-,0. The total dimensionless concentration of each species is given by summing over all possible clusters, for example $c_+ = \sum_{lms} l c_{lms} + c_+^{gel}$.

The free energy functional ($\mathcal{F}$) is proposed to take the following form
\begin{align}
    \label{HEnergy} 
    \beta\mathcal{F} =& \int_V  \,d\textbf{r} \left\{-\beta \frac{\epsilon_0\epsilon_r}{2}\big(\nabla\Phi\big)^2 + \beta \rho_e \Phi  - \frac{c_{001}}{v_0} \ln\left(\frac{\sinh(\beta P |\nabla \Phi|)}{\beta P |\nabla \Phi|}\right)\right\}  \nonumber \\
    &+ \frac{1}{v_0}\int_V  \,d\textbf{r} \left\{\sum_{lms} \left(c_{lms}\ln\phi_{lms} + \beta c_{lms}\Delta_{lms}\right) + \beta\Delta^{gel}_+ c^{gel}_+ + \beta\Delta^{gel}_- c^{gel}_- + \beta\Delta^{gel}_0 c^{gel}_0\right\} \nonumber \\
    &+ \int_V  \,d\textbf{r} \left\{\Lambda \left(1-\sum_{lms}(\xi_+l+\xi_-m+s)c_{lms} - \xi_+c^{gel}_+ - \xi_-c^{gel}_- - c^{gel}_0 \right)\right\} 
\end{align}

Here the following variables, electrostatic potential, $\Phi$(\textbf{r}), charge density, $\rho_e$(\textbf{r}), volume fractions/dimensionless concentrations, $\phi$(\textbf{r})/$c$(\textbf{r}), and the Lagrange multiplier, $\Lambda$(\textbf{r}), all vary in space away from the interface and are integrated over the entire electrolyte domain. The first three terms represent the electrostatic contribution to the free energy: the first subtracts the self-energy of the electrostatic field, the second is the self-interaction energy of the charge density interacting with the mean-field electrostatic potential, and the third is the energy from the fluctuating Langevin dipoles (free water) interacting with the electric field ($-\nabla\Phi$). The first two terms come from a Legendre transform to enforce Poisson's equation while taking the variation with $\Phi$~\cite{grimley1947general,borukhov1997steric,Kornyshev2007,Bazant2009a,Bazant2011}. The third term comes from the classical theory of random walks with drift, applied to the dipole alignment to the electric field initially by Langevin~\cite{hughes1996random}. This mean-field refinement has been implemented in prior double layer theories~\cite{abrashkin2007,gongadze2013spatial,pedro2022polar} and in modeling WiSEs~\cite{mceldrew2018}. The bound water molecules are assumed to not act as fluctuating dipoles. Here $\epsilon_0$ and $\epsilon_r$, respectively, represent the permittivity of free space and the relative dielectric constant, $\Phi$ is the electrostatic potential, $\rho_e$ is the charge density, given by $\rho_e = \frac{e}{v_0}(c_+-c_-)$, with $e$ denoting the elemental charge, and $P$ is the dipole moment of the free water. The fourth term is the ideal entropy of mixing from the clusters of rank $lms$. The fifth term is the free energy for forming clusters, where $\Delta_{lms}$ is the free energy of forming the clusters of rank $lms$; this variable is discussed in detail later. The sixth, seventh, and eighth terms represent the free energy of species associating with the gel, $\Delta_j^{gel}$, which is a function of $\phi_{\pm}$ for thermodynamic consistency. The final term is the Lagrange multiplier, which is used to enforce the incompressibility; similar to previous works, this requires one to solve for $\Lambda$(\textbf{r})~\cite{gongadze2013spatial,Markiewitz2024}.

We can consider the free energy of formation of a $lms$ ranked cluster to consist of three contributions,
    \begin{equation}
        \label{dfrom}
        \Delta_{lms} = \Delta^{comb}_{lms} + \Delta^{bind}_{lms},
    \end{equation}

\noindent where $\Delta^{comb}_{lms}$ is the combinatorial entropy and $\Delta^{bind}_{lms}$ is the binding free energy.

The combinatorial entropy comes from the number of ways the ions and water molecules can be arranged in each cluster; in the context of polymers, this was first derived by Stockmayer~\cite{stockmayer1943theory,stockmayer1952molecular} for the combinatorial entropy for Cayley tree associations and can be extended to the case of WiSEs~\cite{mceldrew2020theory,Goodwin2023},
    \begin{equation}
        \label{dcomb}
        \Delta^{comb}_{lms} = k_BT\ln\{f_+^lf_-^m W_{lms}\},
    \end{equation}

\noindent where
    \begin{equation}
        \label{Wlms}
        W_{lms}=\frac{(f_{+}l-l)!(f_{-}m-m)!}{l!m!s!(f_{+}l-l-m-s+1)!(f_{-}m-m-l+1)!}.
    \end{equation}

The binding free energy for an $lms$ cluster with $l>0$ is simply given by,
    \begin{equation}
        \label{dbind}
        \Delta^{bind}_{lms} = (l+m-1)\Delta f_{+-} + s \Delta f_{+0},
    \end{equation}

\noindent where $\Delta f_{+i}=\Delta f_{i+}$ is the free energy of an association between a cation and the $i^{th}$ species (anions or water). In the case where $l=0$, the binding free energy is zero. Note previously, the binding free energy was split up into two terms, the binding energy and the conformational entropy
~\cite{mceldrew2020theory,mceldrew2020corr,Goodwin2023}.
    
We can calculate the chemical potential of the clusters in the bulk and the EDL, where an overbar will indicate that the variable is the EDL version and $\Phi$ is non-zero,
    \begin{align}
    \label{ChemPot}
        \beta \bar{\mu}_{lms} =& (l-m) \beta e \Phi - \ln\left( \frac{\text{Sinh}(\beta P |\nabla \Phi|)}{\beta P |\nabla \Phi|} \right)\delta_{l,0}\delta_{m,0}\delta_{s,1} + 1 + \ln(\bar{\phi}_{lms}) + \beta\Delta_{lms} \nonumber \\
        & - (\xi_+l+\xi_-m+s)\Lambda + (\xi_+l+\xi_-m+s)\beta\bar{d}'
    \end{align}

\noindent where $\bar{d}'=\bar{c}_+^{gel}\partial\bar{\Delta}_+^{gel}+\bar{c}_-^{gel}\partial\bar{\Delta}_-^{gel}+\bar{c}_0^{gel}\partial\bar{\Delta}_0^{gel}$, with the derivative being with respect to $\bar{\phi}_{lms}$.
  
In the bulk with zero electrostatic potential and field, by asserting the clusters are in equilibrium with the bare species, it follows that,
    \begin{equation}
    \label{BulkBCEq}
        l\mu_{100} + m\mu_{010} + s\mu_{001} = \mu_{lms}.
    \end{equation}

From this equilibrium, we can predict the cluster distribution in the bulk with the bare species,
    \begin{equation}
    \label{BulkClusterD}
        c_{lms}=\frac{W_{lms}}{\lambda_{+-}} \left(\frac{f_+\phi_{100}\lambda_{+-}}{\xi_+}\right)^l \left(\frac{f_-\phi_{010}\lambda_{+-}}{\xi_-}\right)^m \left(\phi_{001}\lambda_{+0}\right)^{s},
    \end{equation}

\noindent where $\lambda_{+-}$ is the cation-anion association constant and $\lambda_{+0}$ is the cation-water association constant are given respectively by,
    \begin{equation}
    \label{AssoBpm}
        \lambda_{+-} = \exp\{-\beta\Delta f_{+-}\}
    \end{equation}
    \begin{equation}
    \label{AssoBp0}
        \lambda_{+0} = \exp\{-\beta\Delta f_{+0}\}.
    \end{equation}

\subsection{EDL Equilibrium}

By establishing the equilibrium between the free species and the clusters \textit{within} the EDL it follows, 
    \begin{equation}
    \label{EDLBCEq}
        l\bar{\mu}_{100} + m\bar{\mu}_{010} + s\bar{\mu}_{001} = \bar{\mu}_{lms}.
    \end{equation}

We obtain an analogous solution to the bulk's for the EDL cluster distribution given the volume fractions of the bare species in the EDL,
    \begin{equation}
    \label{EDLClusterD}
        \bar{c}_{lms}=\frac{W_{lms}}{\lambda_{+-}} \left(\frac{f_+\bar{\phi}_{100}\lambda_{+-}}{\xi_+}\right)^l \left(\frac{f_-\bar{\phi}_{010}\lambda_{+-}}{\xi_-}\right)^m \left(\bar{\phi}_{001}\bar{\lambda}_{+0}\right)^{s},
    \end{equation}

\noindent where
    \begin{equation}
    \label{AssoEDLp0}
        \bar{\lambda}_{+0}= \lambda_{+0}\frac{\beta P|\nabla \Phi|}{\text{Sinh}(\beta P|\nabla \Phi|)}
    \end{equation}

Note that establishing the equilibrium \textit{within} the EDL allows the aggregation to be consistently treated at the interface, in contrast to previous approaches which only considered the equilibrium in the bulk~\cite{Chen2017,goodwin2017underscreening,Yufan2020}. In the Supplemental Material (SM), we verify that $\bar{\lambda}_{+0}$ varies in the EDL, supporting the assumptions of our theory. The problem then boils down to consistently linking the equilibriums in the bulk and in the EDL. These approximations are expected to still hold under dynamics. For example, when the bulk goes out of equilibrium, but the EDL and the bulk ``just outside" the EDL remains in equilibrium from the effective boundary conditions. This result comes from the asymptotic analysis for thin EDLs~\cite{Bazant2009a}.

Following the work of Markiewitz and Goodwin \textit{et al.}~\cite{Goodwin2022EDL,Markiewitz2024}, we can connect the bulk and EDL cluster distributions to the Poisson-Boltzmann equation through closure relations. Here, the closure relations are based on the pre-gel regime; hence we limit the current analysis to this regime, as other terms should be accounted for in the post-gel regime~\cite{Goodwin2022EDL}. This is achieved by equating the bare species in the bulk to those in the EDL. For the bare cations
    \begin{equation}
        \label{bcatp}
        \bar{\phi}_{100} = \phi_{100}\text{exp}(-\beta e \Phi + \xi_+ \Lambda),
    \end{equation}

\noindent there are only contributions from the electrostatic potential and excluded volume effects. For the bare anions
    \begin{equation}
        \label{banp}
        \bar{\phi}_{010} = \phi_{010}\text{exp}(\beta e \Phi + \xi_- \Lambda),
    \end{equation}

\noindent there are only contributions from the electrostatic potential and excluded volume effects. For the free water
    \begin{equation}
        \label{fwatp}
        \bar{\phi}_{001} = \phi_{001}\frac{\text{Sinh}(\beta P |\nabla \Phi|)}{\beta P |\nabla \Phi|}\text{exp}(\Lambda),
    \end{equation}

\noindent there are only contributions from the fluctuating Langevin dipoles and excluded volume effects. It is important to note here that in previous versions~\cite{Goodwin2022EDL,Goodwin2022Kornyshev}, a parameter $\alpha$~\cite{goodwin2017mean} was introduced to account for the short-ranged correlations between ions, beyond the mean-field that is accounted for here. One can simply introduce the $\alpha$-parameter by replacing $\Phi$ with $\alpha\Phi$. For simplicity, $\alpha$ is excluded here for all but the differential capacitance predictions, where we set $\alpha$ to be 0.1, which has proven to be a reasonable value~\cite{Jitvisate2018}. For more details on $\alpha$'s implementation see the SM~\cite{goodwin2017mean,Goodwin2022EDL,Goodwin2022Kornyshev}.

Lastly, to solve for the volume fraction of bare species, we need to introduce the idea of association probabilities, conservation of associations, and the law of mass action on said associations~\cite{mceldrew2020theory}. Knowing the volume fraction of bare species $\phi_{100}$, $\phi_{010}$, and $\phi_{001}$ \textit{a priori} is uncommon, requiring MD simulations~\cite{mceldrew2021ion}. This motivates the theory to be designed around the volume fraction of each species that comprise the solution, $\phi_i$. One can obtain the desired volume fraction of bare species by introducing the association probabilities $p_{ij}$, where $i$ is the species of interest and $j$ is the species that it can form associations with, for this paper they are ($p_{+-}$,$p_{+0}$,$p_{-+}$,$p_{0+}$) as it has been shown that associations between water and anions are negligible~\cite{mceldrew2018,mceldrew2021ion}. Similar to previous works~\cite{mceldrew2020theory,mceldrew2021ion,McEldrewsalt2021,Goodwin2022EDL,Goodwin2023,Markiewitz2024}, we can use these probabilities and the functionality of the species to determine the bare species volume fractions, $\phi_{100} = \phi_+(1-p_{+-}-p_{+0})^{f_+}$, $\phi_{010} = \phi_-(1-p_{-+})^{f_-}$, and $\phi_{001} = \phi_0(1-p_{0+})$.

In order to solve for the bare species, we need four additional equations to determine the association probabilities. These equations are obtained via the conservation of associations and using the law of mass action on the open and occupied association sites~\cite{mceldrew2020theory,mceldrew2021ion}. The conservation of cation-anion association produces,
    \begin{equation}
        \label{cac}
        \psi_+p_{+-} = \psi_-p_{-+} = \zeta
    \end{equation}

\noindent where $\psi_+ = f_+\phi_+/\xi_+$ and $\psi_- = f_-\phi_-/\xi_-$ correspond to the number of cation and anion association sites per lattice site and $\zeta$ represents the number of cation-anion associations per lattice site~\cite{mceldrew2020theory,mceldrew2021ion}. The conservation of cation-water association provides,
    \begin{equation}
        \label{csc}
        \psi_+p_{+0} = \phi_0p_{0+} = \Gamma
    \end{equation}

\noindent where $\Gamma$ represents the number of cation-water associations per lattice site~\cite{mceldrew2020theory,mceldrew2021ion}. Using the law of mass action on the open and occupied association sites for cation-anion associations produces,
    \begin{equation}
        \label{cal}
        \lambda_{+-}\zeta = \frac{p_{+-}p_{-+}}{(1-p_{+-}-p_{+0})(1-p_{-+})}.
    \end{equation}

For cation-anion associations it produces,
    \begin{equation}
        \label{csl}
        \lambda_{+0}\Gamma = \frac{p_{+0}p_{0+}}{(1-p_{+-}-p_{+0})(1-p_{0+})}.
    \end{equation}

Where $\lambda_{+-}$ and $\lambda_{+0}$ are the association constants for cation-anion associations and cation-water associations, respectively, as they are determined by the equilibrium between the open and occupied association sites. These bulk parameters can be extracted from bulk MD simulations, and thus they are not free parameters. Their EDL counterparts are determined by their bulk value and state variables of the EDL, i.e. they are also not free parameters. An analogous version of these association probability equations uses their EDL quantities and are assumed to hold and smoothly vary in space across the EDL.

\subsection{Sticky-Cation Approximation}

The complexity of this model can be further reduced by utilizing the sticky-cation approximation, first introduced in Ref.~\citenum{mceldrew2021ion} where it is asserted and shown that in lithium-based WiSEs the cation associations are sufficiently strong as to fully populate its first solvation shell, i.e. on the timescales of interest the lithium ions always have their max amount of associations. Physically, one can motivate this assumption for Li$^+$ as it is the smallest cation, other than a proton, so it will have the largest local electric field. Thus Li$^+$ is expected to have the strongest and longest lasting associations with species in its solvation shell. Therefore, it follows that its first solvation shell will be completely filled. This treatment relaxes the previous treatment of the first solvation shell of Li$^+$ for WiSE in the EDL seen in McEldrew \textit{et al}.~\cite{mceldrew2018}, where four waters and Li$^+$ were assumed to form a single effective cation. Our current treatment allows one to have different associations with the sticky cation. This assumption fundamentally reduces down to the constraint $p_{+-}+p_{+0} = 1$. Additionally, this leads to singularities in the law of mass action equations, Eq.~\eqref{cal} \& Eq.~\eqref{csl}; this can be overcome by taking the ratio of these equations,
    \begin{equation}
    \label{lambdaeq}
        \lambda = \frac{\lambda_{+-}}{\lambda_{+0}} = \frac{p_{-+}(1-p_{0+})}{p_{0+}(1-p_{-+})},
    \end{equation}

\noindent where $\lambda$ is the cation association constant ratio. Furthermore as this assumption is incompatible with having bare cations, Eq.~\eqref{bcatp} must be replaced; this is achieved by considering the equilibrium between the fully hydrated cation in the bulk and the EDL,
    \begin{equation}
    \label{sbarec}
        \bar{\phi}_{10f_+} = \phi_{10f_+}\text{exp}(-\beta e \Phi + (\xi_+ + f_+) \Lambda)
    \end{equation}

\noindent where $\phi_{10f_+} = \left(1+f_+/\xi_+\right)\phi_+(1-p_{+-})^{f_+}$. Additionally, as the sticky cation approximation requires that $s=f_+l-l-m+1$, the sticky-cation cluster distribution simplifies to
    \begin{equation}
    \label{sBulkClusterD}
        c_{lm}=\frac{\phi_0\alpha_0 W_{lm}}{{\lambda}}\left({\lambda} \frac{\psi_+\alpha_{+-}}{\phi_0\alpha_0}\right)^{l}\left({\lambda} \frac{\psi_-\alpha_-}{\phi_0\alpha_0}\right)^{m}
    \end{equation}

    \noindent where
    \begin{align}
        \label{sWlm}
        W_{lm}=\frac{(f_+ l -l)!(f_- m -m)!}{l!m!(f_+l-l-m+1)!(f_-m-l-m+1)!}.
    \end{align}

Here, $\alpha_0=1-p_{0+}$, $\alpha_{+-}=(1-p_{+-})^{f_+}$ and $\alpha_-=(1-p_{-+})^{f_-}$ are the fraction of free water molecules, fully hydrated cations, and free anions, respectively. Additionally, an analogous set of equations exists for the EDL. Lastly by using $p_{+-}+p_{+0} = 1$ with Eq.~\eqref{cac}, Eq.~\eqref{csc}, and Eq.~\eqref{lambdaeq}, one can explicitly solve for the association probabilities in terms of $\lambda$, $\psi_+$, $\psi_-$, and $\phi_0$, the equations for which are shown in the SM.

\subsection{Modified Poisson-Boltzmann}

To predict WiSEs behavior in the EDL, we can derive our modified Poisson-Boltzmann equation by taking the functional derivative of the free energy with respect to the electrostatic potential,
    \begin{equation}
        \label{mpb}
        \nabla\cdot(\epsilon\nabla \Phi) = -\rho_e = -\frac{e}{v_0}(\bar{c}_+-\bar{c}_-) ,
    \end{equation}

    \noindent where
    \begin{equation}
        \label{epsilon}
        \epsilon = \epsilon_0\epsilon_r + P\frac{\bar{c}_{001}}{v_0}\dfrac{L(\beta P|\nabla\Phi|)}{|\nabla\Phi|},
    \end{equation}
    
\noindent and $L(x) = \coth(x) - 1/x$ is the Langevin function. Here, we can define the inverse Debye length, $\kappa = \sqrt{e^2\beta(c_++c_-)/v_0\epsilon_0\epsilon_r}$, which will be used to express the distances from the electrode as dimensionless values. A drawback of this simple modified Poisson-Boltzmann equation is that it still makes the same approximations utilized in the typical Poisson-Boltzmann (PB) equation, which follows from the mean-field local approximation of Coulomb correlations. This approach models the correlations of point-like charges interacting through a uniform dielectric media, which is only technically valid in the dilute electrolyte limit~\cite{guldbrand1984electrical,kjellander1986interaction,grochowski2008continuum}. While fluctuations from the dipole moment of free water molecules are accounted for here, non-local effects still exist that are not captured in this simple modified PB equation. Other modified PB equations have been developed to correct for the effects of finite ion sizes, Coulomb correlations, and non-local dielectric responses~\cite{netz2001electrostatistics,Bazant2009a,Bazant2011,Pedro2020,pedro2022polar,avni2020charge,adar2019screening,gongadze2013spatial}. Similar to the theory of SiILs in the EDL~\cite{Markiewitz2024}, our simple model may indirectly capture these corrections through the short-range associations that promote the formation of ionic clusters with solvent decorations. These clusters have a ``spin-glass" ordering, i.e. ions favor oppositely charged neighbors~\cite{levy2019spin}. The procedure implemented to solve our system of equations coupled to the modified PB equations is discussed in detail in the SM.

\subsection{EDL Prediction}

Beyond the spatial profiles, cluster distributions, and how aggregated the WiSE is in the EDL, one can extract other informative quantities such as the screening length and the length scale of the aggregates. The screening length can be extracted by fitting an exponential function to the electrostatic potential at small potentials for various molalities. Obtaining the screening length allows one to test the consistency of the model's electrostatic predictions. The length scale of the aggregates ($\ell_A$) can be obtained in the pre-gel regime by solving,
    \begin{equation}
        \label{LSAgg}
        \ell_A^3 = v_0\sum_{lms}(\xi_+l+\xi_-m+s)^2 \bar{c}_{lms}.
    \end{equation}

The $\ell_A$ may be used to understand more thoroughly the structuring near charged interfaces, allowing for qualitative comparison against experimental results~\cite{Han2021WiSE}.

A valuable aspect of mean-field models is their tendency to provide reasonable predictions for integrated quantities such as the excess surface concentrations~\cite{hiemenz2016principles,chu2007surface,mceldrew2018}, the interfacial concentration of water~\cite{mceldrew2018}, and the differential capacitance. The excess surface concentrations provide an integrated perspective on how the composition of the electrolyte is affected by being in the presence of a charged interface,
    \begin{equation}
        \label{exsurfc}
        \Gamma_i(q_s) = \frac{1}{v_0}\int_0^{\infty} \left(\bar{c}_i (x,q_s)-c_i^{\text{bulk}}\right)\,dx,
    \end{equation}

\noindent where q$_s$ is the surface charge density at the interface, $c_i^{\text{bulk}}$ is the dimensionless concentration of the bulk electrolyte solution, and where $x$ is the dimensional distance from the interface. We obtain these predictions by directly integrating the numerical solutions from the modified PB equations and the MD simulations. 

The interfacial concentration of water provides deeper insight into the average composition of the EDL and the amount of free interfacial water present. This measurement can provide insight into how accessible water is to undergo reactions at the interface~\cite{mceldrew2018}. This quantity can be obtained from both theory and MD simulations by integrating a distance $\ell_w$ from the charged interface and normalizing their bulk value over this distance,
    \begin{equation}
        \label{wsurf}
        \tilde{\rho}_{w,n}^{ads}(q_s) = \frac{\int_0^{\ell_w} \bar{c}_n(x,q_s)\,dx}{\ell_wc_0^{\text{bulk}}}.
    \end{equation}

\noindent Here $\ell_w$ was chosen to be 5 \AA, and n indicates the form of water, i.e. total is $\bar{c}_{0}$, free is $\bar{c}_{001}$, and bound is $\bar{c}_0-\bar{c}_{001}$. 

The differential capacitance, C, also known as the double layer capacitance, can be calculated as,
    \begin{equation}
        \label{DC}
        C = \frac{d q_s}{d \Phi_s}.
    \end{equation}

Here, $\Phi_s$ is the electrostatic potential at the charged interface, equivalent to the potential drop across the EDL.

\section{Results}
Here, we will mainly discuss the EDL properties of 15m water-in-LiTFSI (12m  water-in-LiTFSI is shown in the SM), describing and comparing the theory's predictions under the sticky cation approximation (non-stick case shown in SM) against the predictions from MD simulations and later experimental data. The theory and its predictions discussed here build strongly on previous studies of WiSEs in bulk, such as Ref.~\citenum{mceldrew2021ion} and concentrated electrolytes in the EDL, see Ref.~\citenum{Goodwin2022EDL} and Ref.~\citenum{Markiewitz2024}. In the SM, we have included the MD simulation methodology, experimental protocols, sticky cation approximation, numerical maps of how the WiSEs properties change as a function of the electric potential and the magnitude of the electric field, and the results under different conditions.

\subsection{Bulk WiSE}

Before investigating the EDL properties of WiSE, it is first prudent to overview the aggregation behavior in the bulk. In order to model the bulk, one must first find the salt's functionalities, which can be extracted from the coordination number distributions. Here we found that in our simulations of pre-gel water-in-LiTFSI, Li$^+$ has an average coordination number greater than 4 (approximately 4.5). This is in agreement with experiments, where they find the first hydration shell of $Li^+$ contains 4-5 water molecules as determined by neutron diffraction experiments~\cite{mason2015neutron}. This finding is also similar to previous work~\cite{mceldrew2021ion}, in which they set the functionality to be 4 and created the sticky cation approximation, i.e. the cation is always fully associated, as it better modeled the simulation data. As our finding suggested a Li$^+$ functionality of either 4 or 5, we tested and verified the adequacy of using the sticky-cation approximation in the EDL with functionality of 4, and non-sticky case with functionality 5. This was also done with 12m water-in-LiTFSI and is discussed in detail in the SM. Based on the behaviors of TFSI$^-$ associations, its maximum cation coordination is 3~\cite{mceldrew2021ion}. Therefore, we concluded Li$^+$ functionality to be 4 under the sticky-cation approximation and the functionality of TFSI$^-$ to be 3, both values agree with the previous study~\cite{mceldrew2021ion}. 

The association probabilities can also be extracted from the simulations, which are discussed in detail in the SM. From the bulk association probabilities and mass action laws, we were able to extract the cation association constant ratio, $\lambda$, using Eq.~\eqref{lambdaeq} to find it is 0.231 for 15m water-in-LiTFSI. From the molality, we can obtain the volume fraction of each species in the WiSE. Using the volume fractions, functionalities, and $\lambda$ one can predict the bulk association probabilities from our theory. Lastly, we can also predict the gel-point, which is given by $1-(f_+-1)(f_--1)p_{+-}^*p_{-+}^* = 0$~\cite{mceldrew2020theory}. This criterion comes from calculating the critical probabilities for which the ionic backbone of these clusters can become infinitely large. The proximity of $p_{+-}p_{-+}$ to $p_{+-}^*p_{-+}^*$ provides insight into how large the aggregates are and how close the solution is to gelation. For 15m LiTFSI, we find both from the sticky-cation theory and simulations that the WiSE is just under the gelation point.

\begin{figure}[h!]
     \centering
     \includegraphics[width= 1\textwidth]{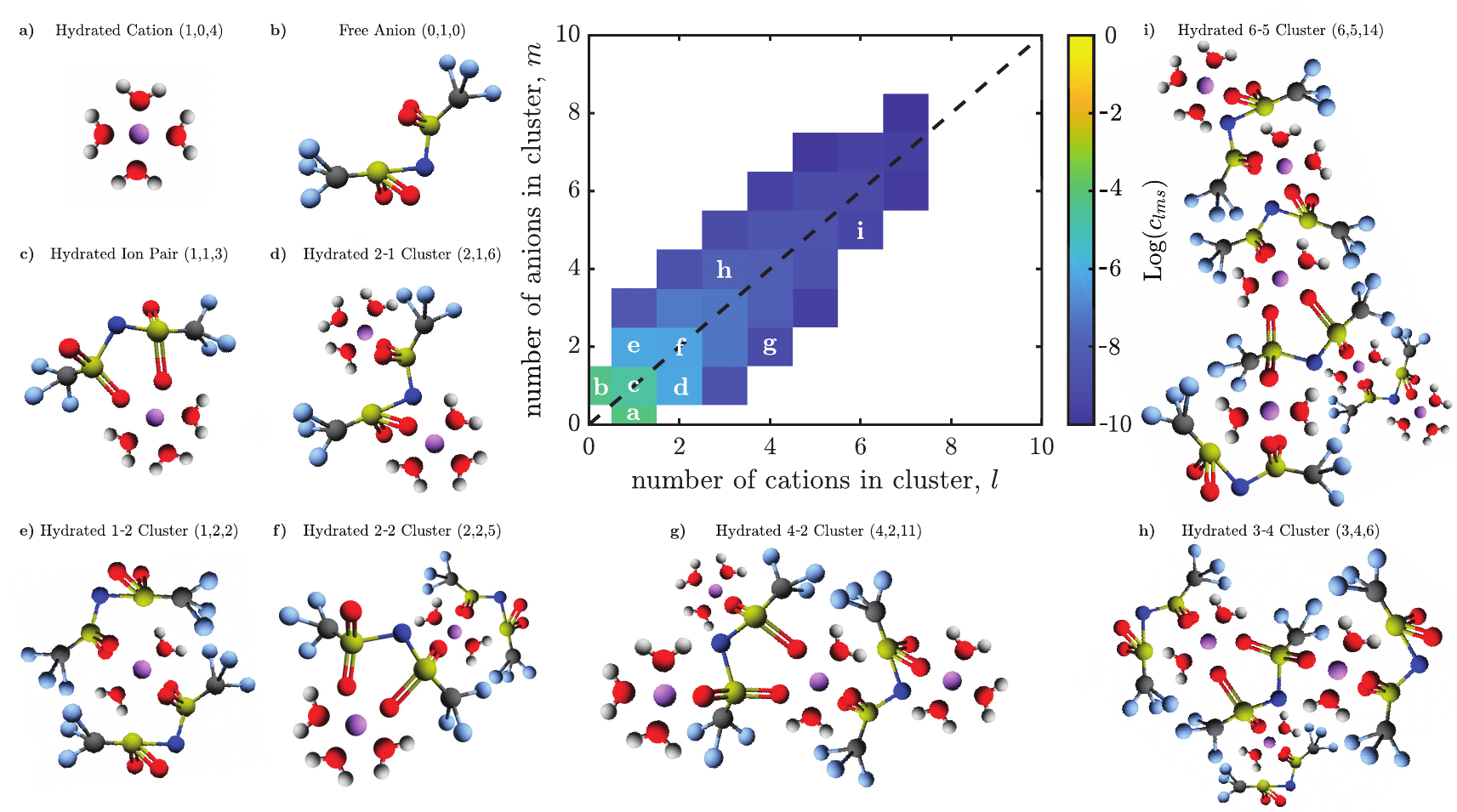}
     \caption{Cluster distribution of bulk 15m water-in-LiTFSI. Here we use $f_+=4$, $f_-=3$, $\xi_{0}=1$, $\xi_+=0.4$, $\xi_-=10.8$, $\epsilon_r = 10.1$, $\lambda = 0.231$, $P$ = 4.995 Debye, and $v_0 = 22.5$ \AA$^3$. The cluster distribution is surrounded by a sample of schematics of the common clusters in bulk 15m water-in-LiTFSI visualized by the software Avogadro 1.2.0~\cite{hanwell2012avogadro}.}
     \label{fig:bcdist}
\end{figure}

Since the theory is fully parameterized for bulk 15m water-in-LiTFSI, we can investigate the bulk cluster distribution using Eq.~\eqref{sBulkClusterD}. In Fig.~\ref{fig:bcdist}, we show schematics of some of the most common clusters in the LiTFSI WiSE: cations hydrated with 4 water molecules, free anions, hydrated ion pairs, and sample multi-ion clusters. In Fig.~\ref{fig:bcdist}, the bulk cluster distribution from the sticky-cation theory for 15m water-in-LiTFSI is shown. An informative quantity is $c_{10f_+}$/$c_{010}$, which tells us how positively or negatively biased the ``free" species are. Additionally from this quantity, one can deduce the sign of the net charge bias of the clusters. Moreover, it can be shown here to depend only on the ratio of anion to cation functionality ($f_-$/$f_+$), for the derivation see SM. Here one can note that the distribution beyond the hydrated Lithium and the free TFSI$^-$ is marginally biased towards net negative clusters, which occurs because the functionality of cations is larger than anions. This preference for slightly negative ionic aggregates is balanced by the excess amount of hydrated cations compared to free anions, giving overall electroneutrality in the bulk. 

\subsection{Anode EDL}

\begin{figure}[h!]
     \centering
     \includegraphics[width= 1\textwidth]{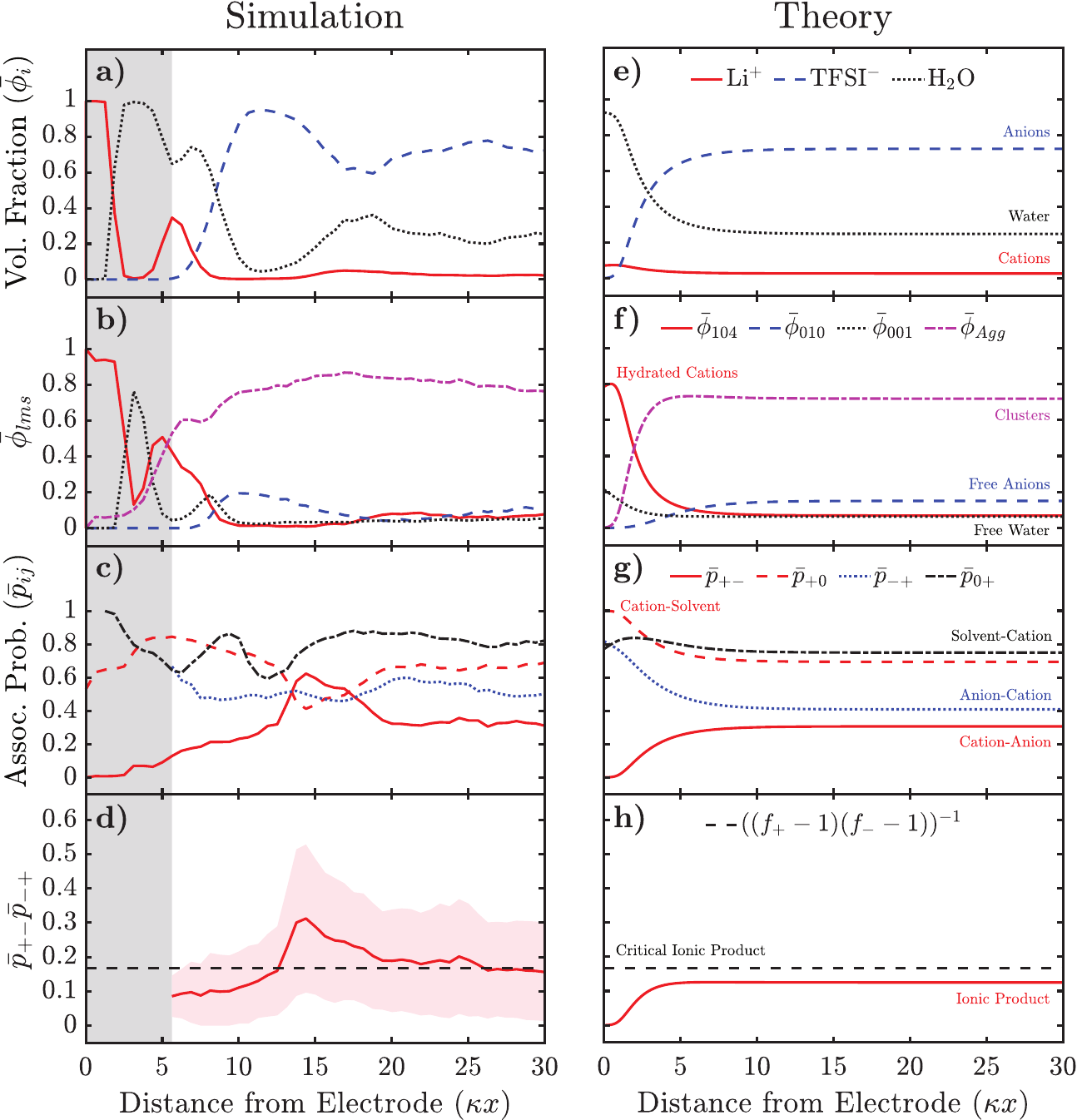}
     \caption{Distributions of properties of 15m WiSEs in the EDL as a function from the interface, in dimensionless units, where $\kappa$ is the inverse Debye length. a-d) are the results from MD simulations, and e-h) are the corresponding predictions from theory. The gray region indicates the minimum distance from the electrode at which a species was never found. a,e) Total volume fraction of each species. b,f) Volume fractions of hydrated cations, free anions, free water, and aggregates. c,g) Association probabilities. d,h) Product of the ionic association probabilities, $\bar{p}_{+-}\bar{p}_{-+}$, where the dashed line indicates the critical line for gelation. Here we use $f_+=4$, $f_-=3$, $\xi_{0}=1$, $\xi_+=0.4$, $\xi_-=10.8$, $\epsilon_r = 10.1$, $\lambda = 0.231$, $P$ = 4.995 Debye, $v_0 = 22.5$ \AA$^3$, and q$_s$ = -0.2 C/m$^2$.}
     \label{fig:neg}
\end{figure}

In Fig.~\ref{fig:neg}, we show the predicted properties of the EDL for negatively charged interfaces from both MD and theory. In Fig.~\ref{fig:pos}, we present the analogous figures for positively charged interfaces. All plots are shown as a function from the charged interface, in dimensionless units of distance, which is normalized by the inverse Debye length, $\kappa$, here 1/$\kappa = \lambda_D \sim 0.5$ \AA\ for the EDL plots shown here. Here, we let $\epsilon_r = 10.1$ and $P$ = 4.995 Debye, which respectively comes from the relative dielectric constant and the dipole moment of water-in-LiTFSI obtained in previous MD simulations of water-in-LiTFSI solutions~\cite{mceldrew2018}. The gray regions in the simulation plots below indicate the area closest to the interface where only one species did not exist in the simulations, i.e. the distance where one species center-of-mass is excluded. In this region, accurately extracting the probabilities can be challenging; hence, the theory is expected to break down near this condensed layer.

In Fig.~\ref{fig:neg}.a) and Fig.~\ref{fig:neg}.e) the volume fraction of the Li$^+$ ($\bar{\phi}_+$), TFSI$^-$ ($\bar{\phi}_-$), and H$_2$O ($\bar{\phi}_0$) from, respectively, the simulation and theory are displayed. In Fig.~\ref{fig:neg}.a), one can identify three distinct regions for the cations, with minimal population in between. The first layer is found at the interface where the cations have saturated, i.e. its volume fraction reaches 1, followed by a depleted region where water dominates. Following this hydration layer, another cation peak is found with water at 6$\lambda_D$, followed by a large volume fraction of anions. Lastly, there is a small third layer of cations at 17$\lambda_D$, after which the volume fraction of cations fluctuates around the bulk value. In Fig.~\ref{fig:neg}.a), the water forms two distinct layers, the first being the hydration layer around 3-7$\lambda_D$, which follows the saturated layer of cations and is smeared out into the second cation peak before being depleted by the large anion layer. The second water peak is after the anion layer around 19$\lambda_D$ and decays into the bulk oscillations of the system. Lastly, the anions are depleted just before the condensed layer; before this, they peak around 11$\lambda_D$ and subsequently fluctuate around their bulk value.

These MD results can be compared against the theory in Fig.~\ref{fig:neg}.e). The theory predicts a monotonic increase in the cation volume fraction approaching the interface before negligibly decreasing close to the interface, which agrees with the simulation trends, albeit without the oscillations and surface structuring observed in the simulations. This deviation occurs from the local approximations of our theory. Similarly, the water volume fraction slowly increases till close to the interface where it rapidly increases. This predicted trend by the theory agrees roughly with the simulations, as the oscillations and surface structuring, respectively, leads to fluctuations throughout the EDL and more refined structuring in the condensed layer. Lastly, the anion volume fraction slowly decreases until it rapidly goes to zero closer to the interface. In this case, the theory more accurately captures the trends seen in the simulations; this result follows from the anions not being present in the condensed layer making the diffuse nature of our theory even more apt. However, we still see deviation from the simulations through oscillations, but this is expected from the local approximations used in our theory not capturing the overscreening behavior of the concentrated electrolyte (which will be described in more detail in the Discussion Section). The value of the presented theory is not in predicting the exact distribution of each species in the EDL, but in investigating how the associations change within the EDL, which will now be described.

Considering the volume fraction of specific clusters in the EDL provides deeper insight into the structure of the volume fraction of each species. This quantity is a prediction that no prior EDL theory for WiSEs has been capable of producing. This observable could provide key insight into interfacial reactions, composition of solid-electrolyte interfaces, and more. Observe first the free anion volume fraction for the simulation in Fig.~\ref{fig:neg}.b), $\bar{\phi}_{010}$, is found to be significantly smaller than $\bar{\phi}_-$ as much of the anions are in clusters. In Fig.~\ref{fig:neg}.f), the decaying trend in $\bar{\phi}_{010}$ towards the electrode's surface is replicated qualitatively by the theory. 

Second, let us consider the free water volume fraction, $\bar{\phi}_{001}$. From the simulation in Fig.~\ref{fig:neg}.b), a sharp peak in $\bar{\phi}_{001}$ is observed in the middle of the condensed layer at 3$\lambda_D$. A little outside of the condense layer, $\bar{\phi}_{001}$ peaks again at 8$\lambda_D$ to around a fourth of its first peak's amplitude. Following this, $\bar{\phi}_{001}$ decays and fluctuates around its bulk value. In Fig.~\ref{fig:neg}.f), the theory predicts $\bar{\phi}_{001}$ monotonically increases as it approaches the charged interface, albeit the absolute values are quite different. This enrichment of $\bar{\phi}_{001}$ near the negatively charged interface qualitatively captures the behavior of $\bar{\phi}_{001}$ observed in the simulation. 

Third, one can consider the volume fraction of aggregates containing more than one ion, $\bar{\phi}_{Agg}$. From the simulation in Fig.~\ref{fig:neg}.b), one can observe that this profile slightly increases entering the EDL followed by a gradual decay before a rapid decay to a near zero value through the condensed layer. In Fig.~\ref{fig:neg}.f), this qualitative trend is found in the theory with the local maximum in aggregation being notable in this curve around 6$\lambda_D$. This local maximum emerges from both the cations and anions existing in similar and appreciable amounts, instead of there being one dominant ion present, which maximizes the aggregation emerging. This kind of phenomenon has been predicted to occur in other concentrated electrolytes such as Salt-in-Ionic-Liquids~\cite{Markiewitz2024}.

Lastly, the volume fraction of various degrees of hydration ($x$) cations, $\bar{\phi}_{10x}$, from the simulation, as seen in Fig.~\ref{fig:neg}.b). At the surface, $\bar{\phi}_{10x}$ saturates excluding free water from the interface. Following this layer, $\bar{\phi}_{10x}$ decays slightly before rapidly increasing, producing a second $\bar{\phi}_{10x}$ peak in the condensed layer. This second $\bar{\phi}_{10x}$ peak occurs at 5$\lambda_D$, which is around the same location as the second $\bar{\phi}_{+}$ peak. After this peak, $\bar{\phi}_{10x}$ is depleted until further away from the interface where $\bar{\phi}_{10x}$ obtains its third and final peak in the EDL at 21$\lambda_D$ after which it fluctuates around its bulk value. All the peaks in $\bar{\phi}_{10x}$ correspond to the peaks in $\bar{\phi}_{+}$ in Fig.~\ref{fig:neg}.a), but the peaks in $\bar{\phi}_{10x}$ are broader and dissipate slower. In Fig.~\ref{fig:neg}.f), the theory predicts that the volume fraction of hydrated cations increases when approaching the negatively charged interface, with the fastest increase within 5$\lambda_D$ of the interface. Close to the interface, a local maximum in $\bar{\phi}_{104}$ is predicted. The theory is able to capture qualitatively the increase of $\bar{\phi}_{104}$ in the diffused EDL and dominate presence of $\bar{\phi}_{104}$ close to the interface.

At this time, let us consider how the negatively charged electrode impacts the association probabilities, which no prior theory of WiSEs in EDL has been able to predict. From the simulation in Fig.~\ref{fig:neg}.c), one can note that the association probability of cations being bound to an anion, $\bar{p}_{+-}$, increases initially after entering the EDL before gradually decaying to zero as one approaches the interface. The association probability of cations being bound to a water, $\bar{p}_{+0}$ initially decreases before increasing around 15$\lambda_D$ from the interface, which is the same turning point as seen for $\bar{p}_{+-}$. Upon entering the condensed layer, $\bar{p}_{+0}$ decays; this behavior appears to result from the excess amount of Li$^+$'s in the condensed layer without enough water to fully fill their first solvation shell. Note in this analysis, associations of species with the wall are not considered. From the theory in Fig.~\ref{fig:neg}.g), $\bar{p}_{+-}$ gradually decays before quickly decaying near the interface, and $\bar{p}_{+0}$ gradually increases before quickly increasing near the interface. The general trends of $\bar{p}_{+-}$ and $\bar{p}_{+0}$ in the EDL seen in the simulation can be captured by the theory. 

Next, one can consider in Fig.~\ref{fig:neg}.c) that the simulation of the anion-cation association probability $\bar{p}_{-+}$, appears to slightly increase while undergoing broad fluctuations when approaching the negatively charged interface and is ill-defined in the condensed layer. While the direction of $\bar{p}_{-+}$ in the simulation is captured in the theory seen in Fig.~\ref{fig:neg}.g), the magnitude of the change in $\bar{p}_{-+}$ is larger in the theory.

Lastly, in Fig.~\ref{fig:neg}.c), the water-cation association probability from the simulation, $\bar{p}_{0+}$ is shown. Here, $\bar{p}_{0+}$ slightly increases until $\sim$17$\lambda_D$. After which $\bar{p}_{0+}$ strongly oscillates with a negative trend till the condensed layer. In the condensed layer, $\bar{p}_{0+}$ strongly increases towards 1. The theory can be seen in Fig.~\ref{fig:neg}.g) where $\bar{p}_{0+}$ slowly increases till close to the interface, where it then decreases slightly. Overall the theory can capture the rough behaviour of $\bar{p}_{0+}$ seen in the simulation.

Finally, one can consider how ionic associations are impacted by the EDL through the product of the cation-anion and anion-cation association probabilities, $\bar{p}_{+-}\bar{p}_{-+}$, another observable which prior theories have not be capable of providing insight on for WiSE in the EDL. From the simulation in Fig.~\ref{fig:neg}.d), one can observe an initial increase in $\bar{p}_{+-}\bar{p}_{-+}$ where it even crosses this critical value before the electrostatic potential and electric field strength achieve sufficiently strong values to melt the induced gel. This effect has been predicted to occur in concentrated electrolyte systems such as in salt-in-ionic-liquids~\cite{Markiewitz2024}. Comparing this result against the theory in Fig.~\ref{fig:neg}.h), $\bar{p}_{+-}\bar{p}_{-+}$ increases very slightly, peaking around 7$\lambda_D$, before rapidly decaying to zero, shown in greater resolution in the SM. Once again, the theory appears to predict qualitative trends in the WiSE solution in the EDL, but the overall changes are smoother. Note that the association probabilities can also be combined to compute how the association constants vary in the EDL, as we show in the SM, where good agreement with the theory is found.

Overall, by considering the simulation results and theory's predictions near a negatively charged interface in Fig.~\ref{fig:neg}, one can consider the EDL to be structured by the following regions: (1) a condensed cation layer with some bound water molecules, (2) a hydration layer filled with free water, and some hydrated cations, (3) a cation rich layer with fully hydrated cations, aggregates, and trace amounts of free water, (4) a small peak in aggregation, and (5) bulk region. Additionally, while the exact predictions differ slightly between the simulations and theory, the qualitative trends in the diffuse sections of the EDL produced by the theory appear to agree with the MD simulation results sufficiently to provide previously inaccessible insights into the association environments in the EDL, both in terms of ionic associations and solvent associations. The importance of the cluster distribution, local associations, and structuring in the EDL, and ways to garner insight into them experimentally, are discussed later.

\subsection{Cathode EDL}

\begin{figure}[h!]
     \centering
     \includegraphics[width= 1\textwidth]{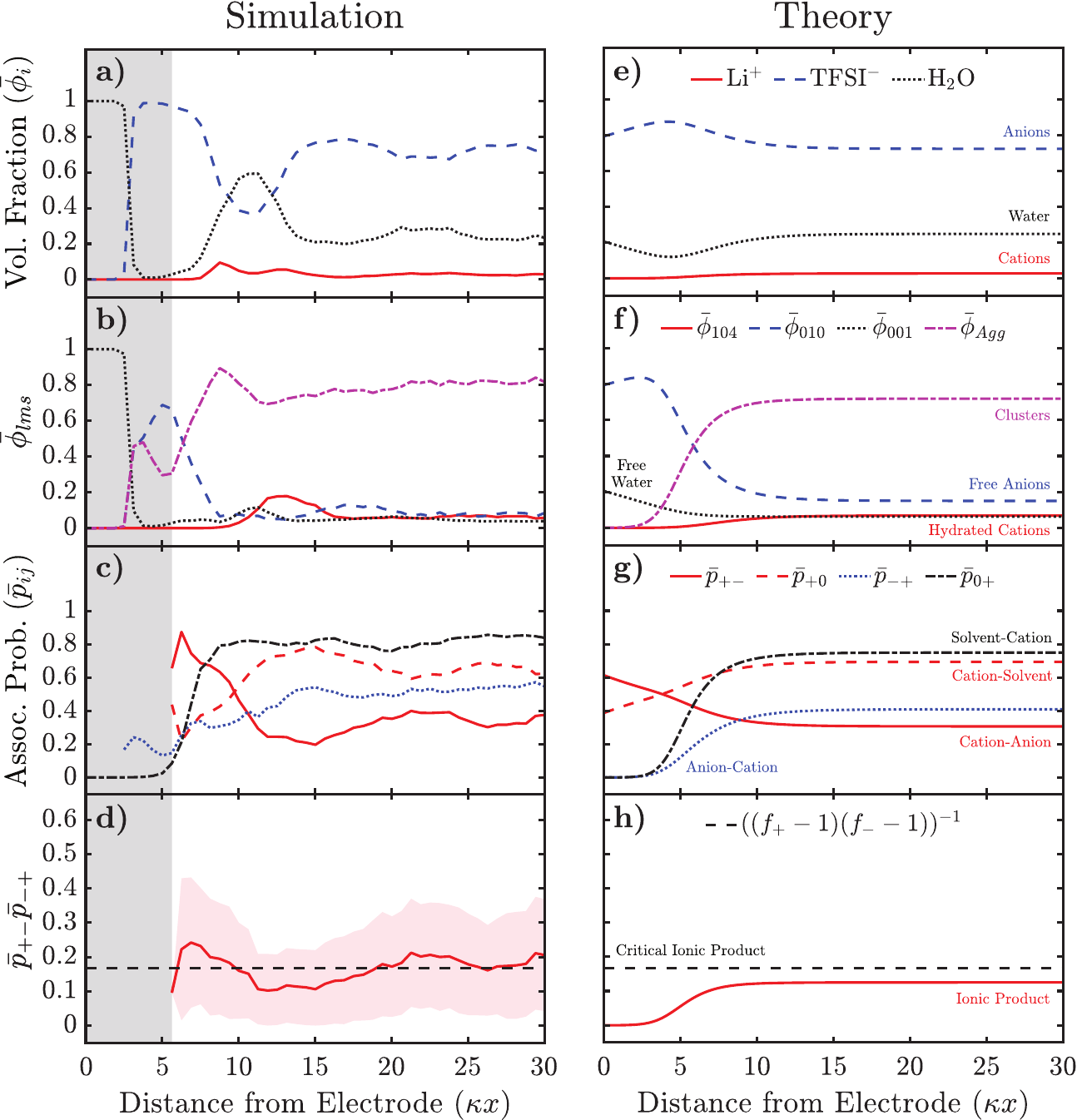}
     \caption{Distributions of properties of 15m WiSEs in the EDL as a function from the interface, in dimensionless units, where $\kappa$ is the inverse Debye length. a-d) are the results from MD simulations, and e-h) are the corresponding predictions from theory. The gray region indicates the minimum distance from the electrode at which a species was never found. a,e) Total volume fraction of each species. b,f) Volume fractions of hydrated cations, free anions, free water, and aggregates. c,g) Association probabilities. d,h) Product of the ionic association probabilities, $\bar{p}_{+-}\bar{p}_{-+}$, where the dashed line indicates the critical line for gelation. Here we use $f_+=4$, $f_-=3$, $\xi_{0}=1$, $\xi_+=0.4$, $\xi_-=10.8$, $\epsilon_r = 10.1$, $\lambda = 0.231$, $P$ = 4.995 Debye, $v_0 = 22.5$ \AA$^3$, and q$_s$ = 0.2 C/m$^2$.}
     \label{fig:pos}
\end{figure}

A similar analysis can be conducted for a positively charged electrode in Fig.~\ref{fig:pos}. The total volume fractions obtained with MD simulation can be seen in Fig.~\ref{fig:pos}.a). In Fig.~\ref{fig:pos}.a) as one approaches the interface, the volume fraction of Li$^+$, $\bar{\phi}_+$, fluctuates around its bulk value before increasing slightly for a short region and then dropping to zero, which is consistent with overscreening. While the overarching trend in $\bar{\phi}_+$ depleting near the interface is captured by the theory, as seen in Fig.~\ref{fig:pos}.e), the higher order correlation effects such as overscreening are not captured, which will be discussed in greater detail later. 

Considering the simulation prediction for the volume fraction of H$_2$O, $\bar{\phi}_0$, in Fig.~\ref{fig:pos}.a) one notes the two peaks. The first being the hydration layer at the interface followed by a depletion region where the first anion layer can be found. This layer is followed by the second peak at 11$\lambda_D$ in $\bar{\phi}_0$ before it drops to its bulk value. The theory predicts an initial decay in $\bar{\phi}_0$ entering the EDL before quickly increasing close to the interface. This behavior qualitatively captures the key features of the $\bar{\phi}_0$ curve from the MD simulation. 

Now, one can consider the behavior of the volume fraction of TFSI$^-$, $\bar{\phi}_-$, in the EDL displayed in Fig.~\ref{fig:pos}.a). Upon entering the EDL, $\bar{\phi}_-$ retains its bulk value before decaying to about half its bulk value. Following this region, there is an anion layer with little water present around 5$\lambda_D$. Entering the condensed layer close to the interface, $\bar{\phi}_-$ goes to zero as the hydration layer excludes the anions from the interface. Considering the theory's prediction for $\bar{\phi}_-$ in Fig.~\ref{fig:pos}.e), the anions initially increase their presence in the EDL before gently decaying near the interface.

Next, one can consider the volume fraction of a specific cluster near a positively charged electrode. First, from simulation in Fig.~\ref{fig:pos}.b), the volume 
fraction of free anions, $\bar{\phi}_{010}$, stays near the bulk value in the EDL until reaching the anion layer around 5$\lambda_D$ where it then increases significantly, but seemingly a little less than half of these anions remain in a cluster. Following this, $\bar{\phi}_{010}$ decays quickly to zero in the hydration layer. The theory's prediction is seen in Fig.~\ref{fig:pos}.f), with $\bar{\phi}_{010}$ increasing and reaching a local maximum close to the interface before slowly decaying. Here, the theory appears to capture the general behavior but lacks the ability to account for the finite nature of these clusters, which is expected given the current theory's point-like treatment of species. These effects, along with specific surface interactions, become increasingly important at closer distances to an interface.

Second by simulation in Fig.~\ref{fig:pos}.b), one can see that the volume fraction of free water's, $\bar{\phi}_{001}$, enhancements in the diffuse part of the EDL are in line with $\bar{\phi}_0$'s but scaled down significantly as it happens that most of the water is bound in the bulk solution. However, close to the interface, a ``hydration" layer is filled with free water molecules. Similar to the theory's prediction for the $\bar{\phi}_{010}$, its prediction for $\bar{\phi}_{001}$ is consistent with a qualitative trend from the MD simulation, with the condensed layer producing more complicated effects, such as the ``hydration" layer at the interface.

Third, one can consider how EDL influences multi-ion aggregates obtained through simulation in Fig.~\ref{fig:pos}.b). In Fig.~\ref{fig:pos}.b) one can observe that the volume fraction of aggregates with more than one ion, $\bar{\phi}_{Agg}$, is zero through the ``hydration" layer after which it takes on half of its bulk value in the anion layer, before returning to around its bulk value. This result demonstrates how the TFSI$^-$ extended nature allows it to form clusters across the EDL. This occurs as the anion can form associations with cations further away from its center-of-mass, which hinders local theories' ability to capture oscillations as short-ranged ordering is averaged out. As expected in Fig.~\ref{fig:pos}.f), the theory predicts $\bar{\phi}_{Agg}$ to decay towards the interface monotonically. This is loosely agrees with the trend seen in the MD simulation, without the strong oscillations.

Lastly, one can consider the simulation's prediction for the volume fraction of hydrated cations, $\bar{\phi}_{10x}$, in Fig.~\ref{fig:pos}.b). Approaching the interface, $\bar{\phi}_{10x}$ fluctuates around its bulk value, until it increases and obtains a peak value around 13$\lambda_D$ from the interface before rapidly decaying to zero. This peak at 13$\lambda_D$ is at a similar location to the second local maximum of $\bar{\phi}_+$ in Fig.~\ref{fig:pos}.a) but has a larger peak amplitude. Considering $\bar{\phi}_+$ in Fig.~\ref{fig:pos}.a) with higher resolution, there are two local maximum's at 8$\lambda_D$ and 13$\lambda_D$. The first maximum at 8$\lambda_D$ appears to correspond to cations in multi-ion clusters as the peak is missing from the $\bar{\phi}_{10x}$'s profile, but is present in the $\bar{\phi}_{Agg}$ profile. From this information, we can infer that the majority of the cations present in the second peak in $\bar{\phi}_+$ around 13$\lambda_D$ exist as hydrated cations. The theory in Fig.~\ref{fig:pos}.f) predicts that the hydrated cations monotonically decay, which qualitatively captures the diffuse EDL's global behavior, but misses this local maximum.

Now, let us consider how the association probabilities are influenced near a positively charged electrode. This is another key prediction that could provide insight into the various properties of WiSEs and one which prior theories of WiSEs in the EDL were incapable of predicting. In Fig.~\ref{fig:pos}.c) from the simulations, one can note that the association probability of cations being bound to an anion, $\bar{p}_{+-}$, fluctuates around its bulk value until it quickly increases close to the condensed layer. In Fig.~\ref{fig:pos}.g), the theory predicts that $\bar{p}_{+-}$ increases into the EDL in a similar fashion. Comparing $\bar{p}_{+-}$ profile from the simulation and the theory displays an adequate match in the qualitative trends. In Fig.~\ref{fig:pos}.c), the value of the association probability of cations being bound to water, $\bar{p}_{+0}$, fluctuates around its bulk value before quickly decreasing near the condensed layer, in a roughly reversed response to the EDL than $\bar{p}_{+-}$. In Fig.~\ref{fig:pos}.g), the theory predicts an equivalent behavior for $\bar{p}_{+0}$. Hence, the theory is able to qualitatively capture the trends in $\bar{p}_{+0}$ seen in the simulation.

Next, as seen in the MD simulation in Fig.~\ref{fig:pos}.c), the association probability of anions being bound to an cation, $\bar{p}_{-+}$, slowly decreases through the EDL. The decay of $\bar{p}_{-+}$ near the interface is more gentle in the simulation compared to the theory's prediction. In Fig.~\ref{fig:pos}.g.), the theory predicts $\bar{p}_{-+}$ to behave similarly to the simulation trends with minor decreases in the bulk of the EDL. Lastly through simulation in Fig.~\ref{fig:pos}.c), the association probability of waters being bound to a cation, $\bar{p}_{0+}$, fluctuates around its bulk value for much of EDL before rapidly decaying to zero before and through the condensed layer. From the theory in Fig.~\ref{fig:pos}.g), $\bar{p}_{0+}$ is observed to have a strikingly similar behavior to the simulation prediction with negligible decay in much of the EDL before rapidly decaying close to the interface. Overall, the theory appears to adequately agree with the qualitative trends presented in the MD predictions for the association probabilities in the EDL.

Finally, let us consider how ionic associations are impacted by the EDL through the product of the ionic association probabilities, $\bar{p}_{+-}\bar{p}_{-+}$, is impacted by the positively charged interface, once again an interesting prediction which previous theories of WiSEs in the EDL are unable to predict. In Fig.~\ref{fig:pos}.d), $\bar{p}_{+-}\bar{p}_{-+}$ can be seen fluctuating, in a seemingly decreasing fashion. Here, it appears to generally exist close to the critical threshold. In Fig.~\ref{fig:pos}.h), $\bar{p}_{+-}\bar{p}_{-+}$ is predicted to gently decay before rapidly decaying close to the interface. The differences between these curves further support the important role the condensed layer may have on the structure of the EDL itself and how these effects propagate into the bulk.

Both the simulation and theory can give a detailed picture of the EDL structure near a positively charged electrode. Here, we have found it has four distinct regions: (1) a hydration layer of free water molecules, (2) an anion-rich layer filled with free anions and anions associated with cations from further away from the interface, (3) an enriched cluster and hydrated cation layer, and (4) the bulk. Overall, the theory appears to adequately capture the qualitative trends from the MD simulations, with its spatial profiles tending to be compressed at points, but this is a known consequence of having a short screening length, which will be discussed later.

\subsection{Size and Structure of EDL Aggregates}

\begin{figure}[h!]
     \centering
     \includegraphics[width= .89\textwidth]{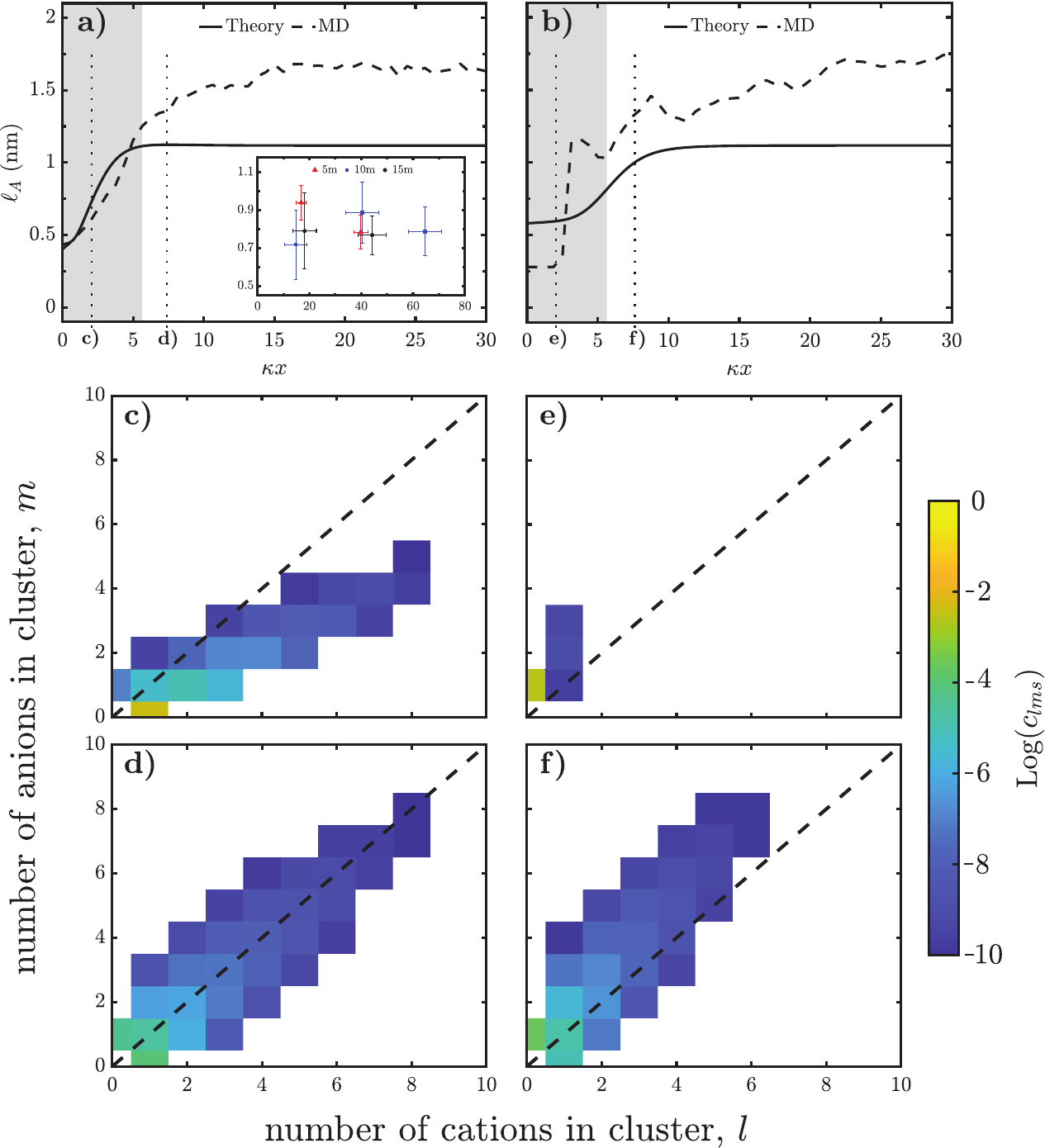}
     \caption{Aggregation length scale and cluster distributions through the EDL in WiSEs. a) Aggregate length scale of 15m water-in-LiTFSI at q$_s$ = -0.2 C/m$^2$ as a function of distance from the interface in dimensionless units, where $\kappa$ is the inverse Debye length. Inset of a) Experimental aggregate length scale for water-in-LiTFSI at mica surface as a function of distance from the interface in dimensionless units~\cite{Han2021WiSE}. b) Aggregate length scale of 15m WiSE at q$_s$ = 0.2  C/m$^2$ as a function of distance from the interface in dimensionless units. c) Cluster distribution $\sim$2$\lambda_D$ from the q$_s$ = -0.2 C/m$^2$ interface. d) Cluster distribution $\sim$7.5$\lambda_D$ from the q$_s$ = -0.2 C/m$^2$ interface. e) Cluster distribution $\sim$2$\lambda_D$ from the q$_s$ = 0.2 C/m$^2$ interface. f) Cluster distribution $\sim$7.5$\lambda_D$ from the q$_s$ = 0.2 C/m$^2$ interface. Here we use $f_+=4$, $f_-=3$, $\xi_{0}=1$, $\xi_+=0.4$, $\xi_-=10.8$, $\epsilon_r = 10.1$, $\lambda = 0.231$, $P$ = 4.995 Debye, and $v_0 = 22.5$ \AA$^3$.}
     \label{fig:comp}
\end{figure}

One direct way associations impact the EDL is through the length scale of aggregates, $\ell_{A}$, which provides insight into the weighted average cluster size throughout the EDL. The length scales of aggregates can be measured through SFA and AFM measurements, which means this quantity is physically observable. In Fig.~\ref{fig:comp}.a), $\ell_{A}$ predicted from theory near a negatively charged electrode is displayed along with its predicted value from MD simulations and in the inset inferred from force-distance measurements by AFM on mica~\cite{Han2021WiSE}. These predictions provide insight into the structure of layers near the interface and can be qualitatively compared against experimental measurements. In Fig.~\ref{fig:comp}.a), the MD simulation predicts that for a negatively charged surface $\ell_{A}$ fluctuates around its bulk value far from the interface, before gradually decaying as it approaches the interface. Then $\ell_{A}$ rapidly decays in the condensed layer, where it takes on a final value around the length scale of fully hydrated Li cation. The theory captures the general trend, but with the bulk value deviating from the MD results.

Turning to the inset of Fig.~\ref{fig:comp}.a), the experimental data~\cite{Han2021WiSE} predicts a lower length scale compared to the MD simulation and the theory. Moreover, the experiments show that clusters are located further away from the surface. Experimentally this may be the result of AFM as the position of the surface has an uncertainty, meaning it is possible that the tip cannot displace the last layer of strongly bound cations. Additionally, the deviations might be due to a difference in the surface charge of mica compared to the surface charge in simulation and theory. One could expect the magnitude of the surface charge for mica to be up to 0.33 C/m$^2$. The experimental data measures the distances in the force peak heights in the AFM/SFA measurements, which can be interpreted as representing the minimum size of the ions/clusters in that region that are squeezed out together. However, the experimental measurements in the diffuse portion of the EDL appear to suggest the presence of a local maximum in $\ell_{A}$, indicating enhanced aggregation compared to the bulk before decaying towards the surface. The local maximum seen in 10m measurements~\cite{Han2021WiSE} is in line with this hypothesis. This qualitative comparison suggests an experimental agreement with the theory's and MD simulation's prediction of the local enhancement in associations near a negatively charged interface.

Moving to the bulk, one can compare the MD and the theory predictions for $\ell_{A}$ against previous experimental measurements from scattering~\cite{borodin2017liquid} and SFA~\cite{Han2021WiSE}. Here, we find that for 15m water-in-LiTFSI, the characteristic length scale of the clusters should be around 1 to 1.7 nm. This result is consistent with findings from scattering, which found the length scale of nano-heterogeneity from 9 to 14m water-in-LiTFSI being around 1 to 2 nm~\cite{borodin2017liquid}, and this was later found to be consistent against SFA results~\cite{Han2021WiSE}. To this end, additional experimental investigations into the local structure at the interface could provide additional grounding for refining the current understanding of interfacial effects and theories.

Next, one can consider the aggregation length scale near a positively charged interface from the theory and MD simulation, as shown in Fig.~\ref{fig:comp}.b). The MD simulation predicts that $\ell_{A}$ fluctuates with a decaying trend before rapidly decaying in the condensed layer to its final value around the length scale of water molecules. The theory captures the general trend with both the final bulk and interfacial value deviating from the MD results. The deviation at the interface could be resulting from the distinct interactions expected at the interface with different interactions and forces dominating the physics. This reasoning is in line with the ``hydration" layer seen in the MD, but not in the theory. 

One of the powerful aspects of our theory is that we are able to investigate in more detail the predicted cluster distributions in different regions of the EDL. In our previous descriptions of the EDL we only described what happens to the aggregates and free species, as is often done, but we can resolve all possible clusters. In Fig.~\ref{fig:comp}.c) \& Fig.~\ref{fig:comp}.d) we show the cluster distribution for the positions in the EDL of the negative electrode that are indicated in Fig.~\ref{fig:comp}.a). In Fig.~\ref{fig:comp}.c), the cluster distribution is $\sim$2$\lambda_D$ from the interface is considered. The theory predicts that the cluster distribution here to be strongly skewed towards positively charged clusters, with the majority of the clusters being hydrated cations. Moving further away from the interface, at $\sim$7.5$\lambda_D$ as shown in Fig.~\ref{fig:comp}.d), the cluster distribution initially appears similar to the bulk; however, comparing the two distributions closely proves otherwise. This comparison highlights that the distribution of clusters in Fig.~\ref{fig:comp}.d) is elongated indicating that larger clusters occur more here compared to the bulk shown in Fig.~\ref{fig:bcdist}. This suggests that a local maximum in the aggregation length scale would be around this location. Moreover, this suggests an increase in associations occurring around $\sim$7.5$\lambda_D$. This finding is consistent with the previous result where the maximum in $\bar{p}_{+-}\bar{p}_{-+}$ was found around this location for q$_s$ = -0.2 C/m$^2$ interface, which is shown with higher resolution in the SM. 

Similarly one can consider the changes in the cluster distribution as one approaches a positively charged electrode in Fig.~\ref{fig:comp}.e) \& Fig.~\ref{fig:comp}.f), as indicated in Fig.~\ref{fig:comp}.b). In Fig.~\ref{fig:comp}.e), the distribution $\sim$2$\lambda_D$ from the interface is shown. Here the theory predicts that the clusters will mainly be free anions, with minor amounts of small negatively charged clusters present. Further away from the interface at $\sim$7.5$\lambda_D$ is depicted in Fig.~\ref{fig:comp}.f), here the cluster distribution is slightly shifted in favor of negatively charged clusters, with the most common cluster being the free anion. Overall, the novel ability of the theory to predict the cluster distribution provides meaningful insights into the local solvation environment that are consistent in their implications drawn from the other novel properties of our theory. Moreover, our theory's ability to provide a deep understanding of how these cluster distributions are modulated in the EDL, as well as by the composition and other experimentally tunable variables, makes it valuable for better understanding the role of electrolytes' local solvation structure in energy storage.

\subsection{Integrated Quantities}
\subsubsection{Differential Capacitance}
\begin{figure}[h!]
     \centering
     \includegraphics[width= 1\textwidth]{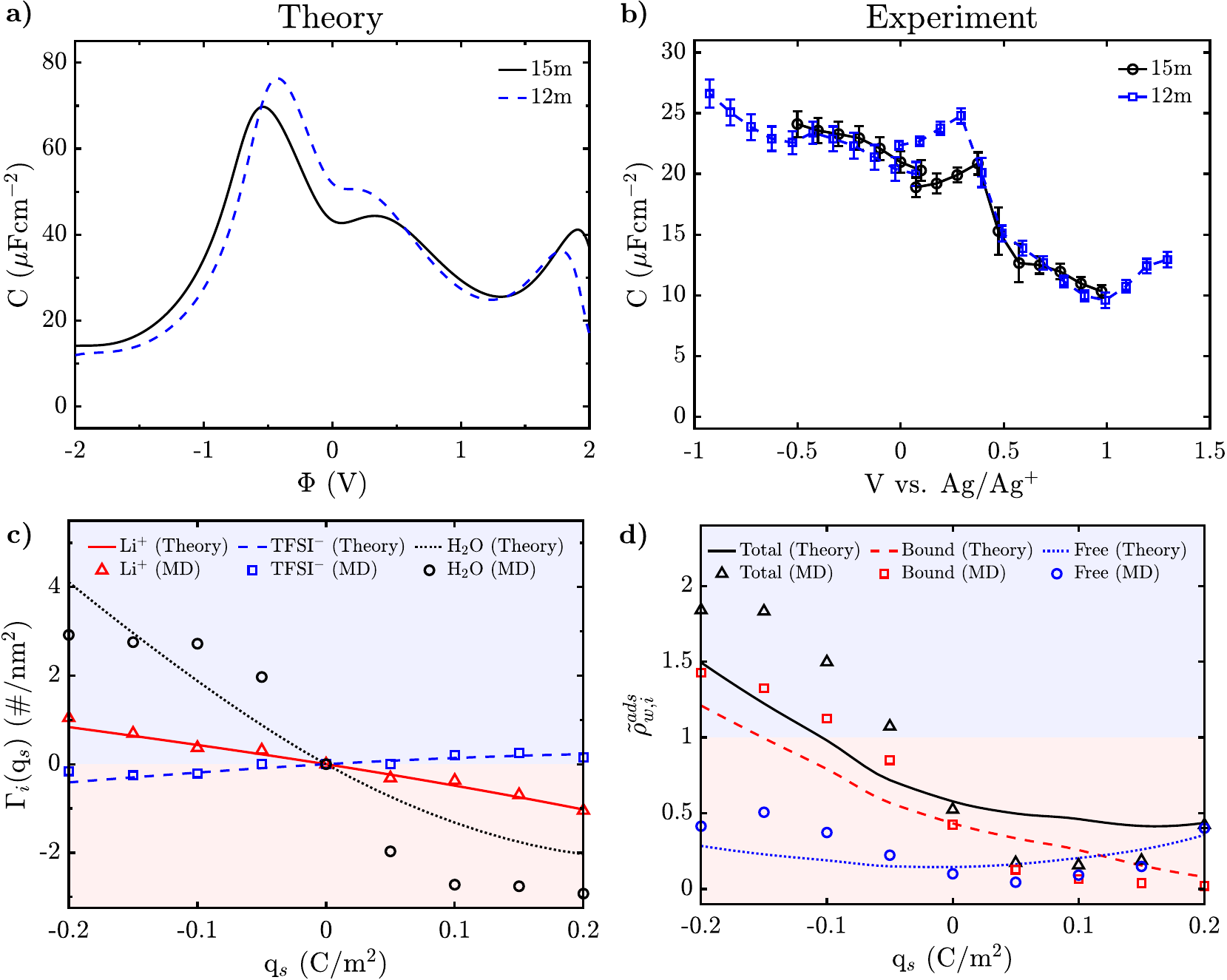}
     \caption{EDL predictions of WiSEs. a) Theory prediction for the differential capacitance of water-in-LiTFSI as a function of the electrostatic potential, $\alpha$ = 0.1. b) Experimental measurement of the differential capacitance of water-in-LiTFSI as a function of the applied voltage. c) Excess surface concentrations for 15m water-in-LiTFSI as a function of surface charge. d) Interfacial concentration of water for 15m water-in-LiTFSI as a function of surface charge. Here we use $f_+=4$, $f_-=3$, $\xi_{0}=1$, $\xi_+=0.4$, $\xi_-=10.8$, $\epsilon_r = 10.1$, and $P$ = 4.995 Debye. For 15m, $\lambda = 0.231$ and $v_0 = 22.5$ \AA$^3$. For 12m, $\lambda = 0.226$ and $v_0 = 22.9$ \AA$^3$.}
     \label{fig:comp2}
\end{figure}

Next, we can consider how the differential capacitance of water-in-LiTFSI varies as a function of electrostatic potential in the theory, as shown in Fig.~\ref{fig:comp2}.a). Here, we have introduced the $\alpha$-parameter~\cite{goodwin2017mean}, which accounts for short-ranged correlations between ions and stretches the theory's voltage range to that from MD simulations and experiments. We set $\alpha$ to be 0.1, which has proven to be a reasonable value~\cite{Jitvisate2018}. At low potentials in Fig.~\ref{fig:comp2}.a), we see that the theory predicts the differential capacitance for pre-gel water-in-LiTFSI at 12m and 15m takes on a strongly asymmetric camel shape~\cite{Kornyshev2007,kilic2007a,Bazant2009a}, with the larger peak occurring in the negative potential and a satellite peak occurring at large positive potential. One can associate each of these peaks with distinct circumstances, visualized and discussed in greater detail in the SM. The large negative peak is associated with the cation enrichment and is further amplified by the enhanced dielectric function. The moderate positive peak is associated with anion enrichment, but lacks the same dielectric enhancement as seen in the negative peak. The satellite peak at large positive potentials is associated with water enrichment and dielectric enhancement. This water-induced asymmetric satellite is analogous to having hydrophilic anions or hydrophobic cations, which is the case discussed in previous work~\cite{BBKGK}. In the previous work, Budkov \textit{et al}.~\cite{BBKGK} showed that the interactions from water with hydrophobic cations or hydrophilic anions lead to the asymmetric peak observed at low positive potentials~\cite{Zheng2023water}. In our work, this peak emerges at a large positive potential as a result of initial water diminution at a low positive potential, followed by its enhancement at a large positive potential.

Additionally, in Fig.~\ref{fig:comp2}.a), one can consider the effects of concentration on the predicted differential capacitance of the theory. Here, we see that increasing the molality stretches out the profile horizontally, as seen from the peaks of the camel shape separating in voltage and the water-satellite peak moving to a higher voltage. Moreover, the amplitude of the camel valley in the differential capacitance increases with increasing concentration, causing a larger camel shape. These features are consistent with the differential capacitance curve becoming more camel-like, which has been correlated with a decrease in the number of free charge carriers in the electrolyte~\cite{Goodwin2017,Chen2017}. Additionally, we see the water-satellite peak's amplitude increase with molality. Molalities effect on the screening length, and therefore the Debye capacitance, is discussed in the SM in more detail.

Here we also experimentally investigated the differential capacitance of water-in-LiTFSI, at different molalities, as a function of the applied voltage, as shown in Fig.~\ref{fig:comp2}.b). The differential capacitance was extracted from electrochemical impedance spectroscopy (EIS) measurements performed in a three-electrode cell, where we started from the open circuit potential and went to positive and negative applied voltages. By conducting cyclic voltammograms for LiTFSI at various concentrations, we were able to ensure our EIS measurements were collected within the electrochemical stability window. The differential capacitance was extracted by analyzing EIS data using three different methods, two Nyquist plot-based fitting methods and a Cole-Cole plot-based fitting method, all of which produced similar trends. Shown in Fig.~\ref{fig:comp2}.b) is the extracted differential capacitance by fitting to the equivalent circuit, as first introduced by Brug \textit{et al.}~\cite{brug1984analysis}, in the limit of dominant double layer resistance. For the differential capacitance predictions using the other methods, see the SM. The experimental protocol and analysis, along with predictions for 1m and 21m LiTFSI, are discussed in-depth in the SM.

In Fig.~\ref{fig:comp2}.b), one can note that the experimental differential capacitance for both 12m and 15m water-in-LiTFSI appears to have a camel-shaped curve. At both concentrations, 12m and 15m, we see a peak at moderately positive applied potentials at 0.29V vs. Ag/Ag$^+$ and peaking around 24.78 $\mu\text{F cm}^{-2}$ and at 0.37V vs. Ag/Ag$^+$ and peaking around 20.85 $\mu\text{F cm}^{-2}$ respectively~\cite{Kornyshev2007}. As the concentration increases, the magnitude of the differential capacitance curves decreases slightly everywhere, but the positive peak at 15m is smaller than the 12m. Additionally, both the 12m and 15m differential capacitance curves appear to decay monotonically after their positive peak, with some increase being measured at large positive potentials, and the differential capacitance profile appears to increase with increasingly negative applied potentials. As these profiles are collected within the electrochemical stability window, these increases in differential capacitance are believed to be part of a larger negative peak at a larger negative potential. This result suggests the differential capacitance has a camel-shaped curve centered close to 0~V vs. Ag/Ag$^+$ or at slightly negative applied potential. 

Now, one can compare the theory's predictions against experimental data for a more robust evaluation of the theory. Here, we find that in both the theory in Fig.~\ref{fig:comp2}.a) and the experimental data in Fig.~\ref{fig:comp2}.b), as the molality of the solution increases, the magnitude of the differential capacitance decreases. This effect is clearest in the positive potential peaks. The general shape of the differential capacitance between the theory and the experiments suggests a camel shape with a water-satellite peak at a large positive potential. Moreover, comparing the 1~m case results for theory and experiments in the SM shows the absence of the positive differential capacitance peak, further supporting that this experimental peak corresponds to the positive peak in the theory's prediction around 0.5~V. Lastly, the magnitude of the differential capacitance differs between the theory's predictions and the experimental data. These deviations are expected given the sophisticated nature of obtaining differential capacitance profiles experimentally and a well-known weakness of simple mean-field models, both of which are discussed in greater detail in the following section. Overall, we observe similar trends in the differential capacitance profiles in the theory and the experimental measurements.
\subsubsection{Excess Surface Concentration}
For reactive interfaces, understanding the excess surface concentrations, defined by Eq.~\eqref{exsurfc}, can provide insight into the local reaction environment~\cite{andersson2005activity,chu2007}. The excess surface concentrations from MD simulations and the theory's predictions for 15m water-in-LiTFSI at a variety of surface charges are shown in Fig.~\ref{fig:comp2}.c). For Li$^+$, the MD simulations predict that its concentration will be enhanced at negative surface charges and diminished at positive surface charges. This prediction for Li$^+$ is accurately captured by the theory. For TFSI$^-$, the MD simulations predict that the excess surface concentration will be diminished at negative surface charges and enriched at positive surface charges. Additionally, the excess TFSI$^-$ surface concentration will achieve its maximum value of reduction around -0.15 C/m$^2$ and the maximum enhancement around 0.15 C/m$^2$. The theory can accurately capture the general trend for the modulation of excess surface TFSI$^-$ although it does miss the local maximum and minimum observed from the simulations. Lastly, water is predicted to accumulate strongly at negative surface charges and be depleted at positive surface charges. Once again, the theory captures the general trend with strong enrichment and reduction at the respective surface charges. Overall, the theory appears to be capable of capturing the general trends in the accumulation and depletion of each species near the surface.
\subsubsection{Interfacial Concentration of Water}
Lastly using the results from our MD simulations, we tested the theory's prediction for the interfacial concentration of water, defined by Eq.~\eqref{wsurf} and shown in Fig.~\ref{fig:comp2}.d). From the simulations, one observes the asymmetric response in the system, which is expected from the excess surface measurements, and the existence of a depletion region near each charged interface. At increasingly negative surface charges, the total amount of water found close to the interface is enriched compared to the bulk concentration in the simulations. At increasingly positive surface charges in the simulations, the average total water concentration gradually diminishes before increasing moderately around 0.1 C/m$^2$. Note this later moderate enhancement in total water and, more importantly, free water at larger positive potentials is consistent with the water-satellite peak found at large positive potentials in the theory and experiments. In Fig.~\ref{fig:comp2}.d), the theory predicts the total amount of water increases with increasing negative surface charge, becoming enriched at a later point than seen in the simulations. Similarly the theory predicts the average total concentration of water is diminished at increasing positive surface charge. Hence, the theory appears to predict the qualitative trends seen in the MD simulations of the interfacial concentration of total water while deviating from the quantitative values and local minimum at moderate positive surface charges. Strengthening this finding, one can turn to prior experimental studies of WiSE using Surface-enhanced infrared absorption spectroscopy (SEIRAS), which can probe species enrichment near charged interfaces~\cite{zhang2020potential,hoane2024impact}. From these experiments, they found that in WiSE less water at positive than negative potentials~\cite{zhang2020potential}, and a similar result was found for WiSE with divalent cations~\cite{hoane2024impact}. Therefore, the overall enrichment (depletion) at a negatively (positively) charged interface from the MD and theory is in qualitative agreement with experimental results obtained by SEIRAS~\cite{zhang2020potential,hoane2024impact}.

We can further decompose the total water into that bound to Li cations and free water. In Fig.~\ref{fig:comp2}.d) from the MD results, the amount of bound water monotonically increases at increasingly negative surface charges and obtains a larger average concentration than bulk water around -0.1 C/m$^2$. For increasing positive surface charges the amount of bound water slowly decreases monotonically. Similarly, the theory predicts the monotonic increase in the amount of bound water near the increasingly negatively charged interface and its decrease near the increasingly positively charged interface. These qualitative trends agree with the MD simulations. The modulation of interfacial free water from MD and theory can also be seen in Fig.~\ref{fig:comp2}.d), wherein the MD simulations it increases with increasing negatively charged interfaces with a maximum near -0.15 C/m$^2$ after which it slightly decreases. Considering now increasing positive surface charges, the amount of free water initially decreases before rapidly increasing and becoming the majority of the total water near the interface. The theory predicts as the surface becomes increasingly negatively charged, the interfacial concentration of free water shows a slight decrease with a minimum at -0.01 C/m$^2$ followed by a monotonic increase. For increasing positive surface charge, the theory predicts a monotonic increase in the amount of free water. The theory adequately predicts the qualitative trends for the amount of interfacial free water. Overall as shown, the theory appears to capture the qualitative trends from the MD simulations for the interfacial concentration of water of 15m water-in-LiTFSI. Note that the amount of free water becomes dominant to the bound water at 0.1 C/m$^2$ in the MD results and a little after 0.1 C/m$^2$ in theory, which further demonstrates the utility of the theory in capturing the qualitative trends and behavior of WiSEs.

\section{Discussion}

Through testing our theory against MD simulations, we have observed the limitations of the theory in capturing overscreening and non-local effects as well as surface effects. These are expected technical limitations from the theory presented here and are seen through the theory's predictions deviating quantitatively and from the finer structure of the MD profiles. Additional conceptual and technical limitations of this style of theory are outlined in Ref.~\citenum{Goodwin2022EDL}.

The first category of limitations are non-local effects and overscreening. Overscreening is the phenomenon where an excess amount of counter-ions are pulled into the EDL, leading to a layer of co-ions being dragged into the EDL to compensate for their excess charge. Conceptually, overscreening can be represented in the thermo-reversible associations in the WiSE, as the alternating structure of cations and anions is similar to the layered ions. However, the internal structure of the clusters is not explicitly modeled in our theory, leading to layering of ions being averaged. Therefore, because of the simple construction of the theory, we do not explicitly obtain decaying oscillations in the charge density, but note that the overscreening effect is expected to be captured through associations, as seen through sophisticated but limited cluster size theory~\cite{avni2020charge}. Generally, this limitation is reflected in the theory being unable to capture the oscillations, local maximums, and specific layering associated with overscreening given the simple local point-like formulation of the modified PB equations, which can be accounted for with more intricate and non-local approaches~\cite{netz2001electrostatistics,Bazant2009a,Bazant2011,Pedro2020,pedro2022polar,avni2020charge,adar2019screening,gongadze2013spatial}. The importance of the interplay between associations, overscreening, and steric effects was highlighted in Ref.~\citenum{avni2020charge} and discussed in Ref.~\citenum{Goodwin2022EDL}. Developing more sophisticated theories that are able to succinctly capture and balance the short-ranged associations, non-local correlations, steric effects, and surface-specific interactions will be crucial for further resolving concentrated electrolytes such as WiSEs near charged interfaces. 

The non-local effects may also be critical to capturing the consequences of the finite nature of these clusters. For example as shown in Fig.~\ref{fig:neg}.h), the theory prediction for the product of the ionic association probabilities, $\bar{p}_{+-}\bar{p}_{-+}$, is smaller than from the simulation in Fig.~\ref{fig:neg}.d). This deviation could be an artifact of neglecting non-local effects by treating clusters as points. Additionally, these effects can lead to non-monotonic electrostatic potential. Now, considering that WiSEs are expected to display some induced associations at a slightly negative potential, these oscillations in potential could lead to electric field induced associations occurring at both charged electrodes. This effect could, in fact, be seen in the MD simulations in Fig.~\ref{fig:pos}.d), where the fluctuations from non-local effects and overscreening are captured. This may explain the regions of induced associations seen in Fig.~\ref{fig:pos}.d), which depicts $\bar{p}_{+-}\bar{p}_{-+}$ in the EDL near a positively charged electrode.

The other category of limitations are introduced by surface effects which may dominate the physics in the condensed layer. Even though the theory can reasonably describe the diffuse EDL, it cannot capture the condensed layer where there are interfacial layers of ions and water. This breakdown is expected as the cluster distribution should deviate from the diffuse and bulk as a result of the specific interactions with the electrode creating a significant change in the coordination and cluster distribution. In short, in the condensed region, the local solvation environment is expected to be disrupted by the interaction with the interface. These changes can be directly seen by the gray regions representing the condensed layer from the MD simulations in the left-hand column of Fig.~\ref{fig:neg} and Fig.~\ref{fig:pos}. However, the intricacies of the condensed layer seen in our MD simulations could be further compounded by the lack of sufficient statistics for co-ions in the EDL. This potential statistical shortfall could be investigated and overcome by biased sampling methods, such as metadynamics simulations~\cite{barducci2011metadynamics,valsson2016enhancing,henin2022enhanced}.

The most significant limitation in our theory comes from the overly short screening length, $\lambda_s$, being predicted as $\sim 1.1$ \AA\ for 15m water-in-LiTFSI. This $\lambda_s$ is smaller than an individual Li$^+$ ($\sim 1.6$ \AA), and this is much smaller than the length scale of aggregates in the system in the theory for 15m water-in-LiTFSI being $\sim 11$ \AA\ ($\sim 16$ \AA\ from the simulations). This result suggests that the theory is not acting self-consistently in regard to its electrostatic predictions; this is a well-known challenge of mean-field lattice gas models and is discussed in detail in Ref.~\citenum{Bazant2009a}. This common failing of mean-field theories can be corrected through more sophisticated modified PB equations that can capture higher-order correlations and non-local effects, or partially corrected through introducing $\alpha$~\cite{Goodwin2017}. This inconsistency can also be addressed to some degree by modifying local mean-field models, such as by including higher order local terms, as done in BSK theory~\cite{Bazant2011}, or by modifying the Coulomb interactions~\cite{adar2019screening}. However to overcome this inconsistency, one typically needs to employ a non-local model that can capture the entropic effects of excluded volume in a holistic fashion. This inconsistency suggests using caution when using the theory's predicted spatial profiles of its species and cluster near the charged interface~\cite{Bazant2009a}. However as discussed earlier from a qualitative perspective, the theory can capture the trends and overarching behavior seen in the spatial profiles of species and clusters near the charged interface, despite lacking the sophisticated non-local effects and overscreening seen in MD simulations. Moreover, a valuable power of mean-field models is their tendency to adequately predict integrated quantities even with mean-field models' inconsistencies. For this reason, one could expect the theory's predictions of the double-layer capacitance, excess surface concentrations, and interfacial concentration of water to be reasonable, as we demonstrated. Furthermore one could expect, as demonstrated, the general trends in WiSE properties in the EDL to be qualitatively captured along with the length scale of aggregates and, most importantly, how the association probabilities change \textit{within} the EDL.

Experimentally benchmarking any theory or simulation is desirable and timely as different approaches will capture different properties of the system. In this current work, we have been able to test the theory's prediction of the aggregate length scale and differential capacitance against experimental data. The aggregation length scale was obtained via the extraction of an estimate of the average length scale of the nano-heterogeneities in bulk solution through AFM~\cite{Han2021WiSE}. Besides the experimental data highlighted here, AFM measurements~\cite{zhang2020potential} on gold positively biased (+0.3V) and 21m LiTFSI show the presence of clusters with a max size of 0.8 nm close to the surface, then with sizes ranging from 0.3-0.6 nm closer to the surface. This finding indicates our results are in qualitative agreement with additional experimental studies~\cite{zhang2020potential}, although data at lower concentrations have yet to be obtained. Scattering experiments provide an alternative way to measure cluster sizes and the average length scale of nano-heterogeneities in the bulk~\cite{hettige2014bicontinuity,borodin2017liquid,lim2018,Zhang2024}. This previous experimental data~\cite{Han2021WiSE}, supported the theory's prediction that negatively charged interfaces can lead to an enhancement in associations along with the qualitative trends in the aggregation length scale itself compared against MD simulations and the theory. Additional experimental investigations for a more diverse set of electrolytes could provide novel insights into the role of associations and, in turn, the role of the local solvation structure on interfacial electrochemical reactions and operation of energy storage devices, using traditional to concentrated electrolytes.

Comparing the theory's predicted differential capacitance against the experimental data gave a promising conclusion. Here, we found that some of the trends seen in the theory's predictions of the differential capacitance held such as displaying camel shaped profiles with a water-satellite and the decreasing magnitude of the profiles from 12m to 15m water-in-LiTFSI. The main challenge is that a rigorous extraction of the differential capacitance profiles requires a sophisticated approach incorporating the physical details and structuring of the EDL into its differential capacitance fitting. While the current analysis here has focused on extracting trends and qualitative profiles, a more quantitative and detailed investigation into these EIS measurements could be deeply illuminating. In general the difficulty in obtaining robust and exact differential capacitance profiles is well established~\cite{pajkossy2011interfacial,druschler2011commentary,pajkossy2011response,wang2012intrinsic,roling2012comments,wang2012reply,Fedorov2014}. Beyond this challenge, the choice of electrode material and surface roughness is expected to affect the differential capacitance measurement~\cite{pajkossy1994impedance,lockett2010differential,jansch2015influence,oll2017influence,torabi2017rough,aslyamov2021electrolyte}. In the SM, these effects are highlighted. Given the simple nature of the model, it is expected to fail in terms of the absolute value as seen for the overall differential capacitance curves; however as discussed, the model’s ability to predict the general trends is desirable. Additionally, the structuring of the condensed layer is expected to play an important role in the overall capacitance measurements. Hence the condensed layer acts as one limiting factor to the current theory. Even with these deviations, the ability of the simple theory to capture some of the key trends and structure of differential capacitance remains promising.

The potential utility of the perspective our theory provides may extend beyond the limited experimental analysis shown here. Recently various investigations into the behavior of concentrated electrolytes near interfaces have worked to resolve experimentally and computationally the mystery beyond the unique interfacial properties~\cite{zhang2023evolution,finney2024properties,aggarwal2024revealing,yu2024flipping}. A common theme throughout these investigations arrive at is the importance of the local structuring, orientation, and interactions near the electrode. This idea is a core element of our theory. While the model is expected to fail in some ways the value in it is the first principles intuition it provides, which appears to be shared by top-down investigations. This perspective may aid in understanding experimental findings and improving electrolyte design.

\section{Implications for energy applications}
Understanding the structure of the EDL is essential for capturing the equilibrium properties of WiSEs and developing deeper insight into interfacial reactions occurring at electrode surfaces as well as the stored charge~\cite{Suo2015,Yamada2016,yang2017,zhang2018water,mceldrew2018,yang2019aqueous,borodin2020uncharted,mceldrew2021ion,sayah2022super}. Regarding the former, there is a strongly asymmetric response of the water, where it mainly depletes at the cathode side giving rise to the extended cathodic stability, but accumulates on the anode side. However, the increase in water on the anode side mainly corresponds to bound water, and McEldrew \textit{et al.}~\cite{mceldrew2021ion} found that solvated water has a lower activity, which means it is less likely to react. However at very large potentials (both positive and negative), there is eventually an increase in the interfacial free water that may be able to react. These findings are in line with the current understanding behind the expanded ESW in WiSEs~\cite{vatamanu2017,mceldrew2018}.

Additionally, the theory captures the association probability of the species in the EDL providing insight into the local solvation environment.  The species at the interface are precursors for  interfacial reactions. This information on the local solvation environment and species activities could provide deeper insight into which species are likely to undergo decomposition into the SEI~\cite{wu2023effect}. For example, we have found, from both theory and simulation, that at the anode there is a slight increase in aggregation at moderate potentials. Previously, McEldrew \textit{et al.}~\cite{mceldrew2021ion} found that the activity of the salt increases with concentration/aggregation, which suggests that these additional aggregates could assist in the formation of the passivating SEI layers. Understanding which species are contributing to the formation of the SEI could support rational electrolyte design~\cite{xu2007solvation,von2012correlating,pinson2012theory,mceldrew2021ion,wu2023effect}. By improving electrolyte formulation, the passivation layers produced could be more efficient and effective. Furthermore, aggregation occurring close to the electrode/electrolyte interface may also have implications for the metal cation mobility. Additional aggregation of the electrolyte and the exclusion of the IL cation from the anode can help hinder dendritic growth and form a more compact, homogeneous and stable SEI~\cite{rakov2020engineering}.

The existing studies, however, do not address how the unique microenvironment experienced by the interfacial ions and water molecules affects their reactivity and charge transfer with the electrode and other molecules.  Recent experiments have revealed that the effect of the WiSE EDL on the interfacial reactivity of redox species is significant~\cite{zhang2020potential,hoane2024impact}. These works used an ultramicroelectrode to carry out CVs, enabling faster diffusion and a higher sensitivity to the faradaic reaction. In the WiSE LiTFSI, the CVs showed a peak on the anodic scan attributed to the oxidation of Fe(CN)${_6}^{4-}$. This peak was not present in 1m LiTFSI, which indicates that the interphase concentration of Fe(CN)${_6}^{4-}$ is greatly increased in the WiSE, and is attributed to the confinement effect provided by the WiSE at the interface with the electrode. They also found that the addition of Zn$^{2+}$ led to a decrease of the surface-confined peak~\cite{zhang2020potential}. The findings suggested that the confinement effect is reduced by Zn$^{2+}$ and is enhanced by Ca$^{2+}$. The results suggest that the WiSEs EDL and structure could be a tool to enable selectivity and tunability of interfacial reactions. 

The screening length determines the stored charge in the EDL and it is a topic of controversy in the context of highly concentrated electrolytes. Experimental surface force measurements have found that concentrated electrolytes, such as in ionic liquids and water-in-Salt electrolytes, have extremely long force decay lengths~\cite{Gebbie2013,Marzal2014,Gebbie2015,Smith2016,Han2020,Han2021WiSE,groves2021surface,fung2023structure,Zhang2024}. Gebbie \textit{et al}.~\cite{Gebbie2013} asserted that these forces were electrostatic in origin, and arose from the large renormalization of the concentration of free charge carriers. This statement would imply that the screening length is about 1-2 orders of magnitude larger than the Debye length. This phenomenon was originally named underscreening, but, as the topic is controversial and unresolved~\cite{feng2019free,jones2021bayesian,krucker2021underscreening,Espinosa2023rev,jager2023screening}, it has been further classified as anomalous underscreening~\cite{hartel2023anomalous,kumar2022absence,fung2023structure}. This refinement was implemented as other experiments~\cite{Han2021WiSE,Zhang2024}, simulations~\cite{zeman2020bulk,krucker2021underscreening}, and theories~\cite{avni2020charge,goodwin2017underscreening,Goodwin2022EDL} have been able to capture an uptick in the screening length. However, the scaling seen in these works is less than originally reported~\cite{Gebbie2013,perez2017underscreening}. Moreover, as we show, we only find a modest increase in the screening length, still remaining smaller than 1 nm, which does not suggest these long force decay lengths solely arise from electrostatics. Even with the lack of consensus through studying the aggregation of ions and decoration by solvent, this approach has been widely successful in capturing the bulk and transport properties in WiSEs~\cite{mceldrew2021ion} and concentrated electrolytes~\cite{molinari2019general,molinari2019transport,McEldrewsalt2021}. Recent studies~\cite{attard1993asymptotic,leote1994decay,jager2023screening,Zhang2024,Markiewitz2024} have been converging towards an alternative hypothesis that it is steric interactions, also known as hard-core interactions, contributing towards the long-ranged interactions seen in various concentrated electrolytes and not a purely electrostatic phenomenon~\cite{Espinosa2023rev,Goodwin2021Rev}. We believe the theory presented here could aid in further resolving these kinds of measurements~\cite{Gebbie2013,Marzal2014,Gebbie2015,Smith2016,Han2020,Han2021WiSE,groves2021surface,fung2023structure,Zhang2024}, as discussed in Ref.~\citenum{Markiewitz2024}.

While our analysis here, has focused on the equilibrium properties and structuring of WiSEs, one can expand these results and theory to capture dynamics. The theory can be extended to investigate WiSE's transport properties by adapting the methodology outlined in Ref.~\citenum{mceldrew2020corr}. Moreover, as WiSEs have an active cation for intercalation, expanding the theory of coupled ion electron transfer reactions to incorporate the local solvation environment could be noteworthy~\cite{Fraggedakis2021,bazant2023unified}.

Finally, the mathematical analysis required for this theory extends beyond even the most advanced theories for patchy particle systems~\cite{Teixeira2019}, while also taking into account electrostatic interactions. These extensions could be used to account for how other interactions or drivers of concentration localization influence the cluster of polymers. For example, in 3D-printing under electric fields~\cite{zhang2020microscale} or in synthesis requiring sol-gel equilibrium's~\cite{cohen1982equilibrium,danks2016evolution}. While our theory borrowed core elements of polymer physics initially, the extensions seen here could find broader applications in polymer physics and other statistical physics.

\section{Conclusion}

Here, we have developed a theory for EDL of WiSEs that accounts for thermoreversible associations, based on McEldrew \textit{et al}.'s model for bulk WiSEs. We thoroughly tested this theory against MD simulations and found good qualitative agreement for many cases, such as: the distributions of total species, the distribution of specific clusters (free species, hydrated cations, and multi-ion aggregates), and association probabilities. Additionally, our theory’s prediction for integrated quantities such as the excess surface concentrations and the interfacial concentration of water were found to be in reasonable agreement with MD simulations. This simple theory's value is its ability to capture how the associations \textit{within} the EDL change, not previously quantified with any theory, allowing more detailed predictions of cluster distributions and ionic network formation. We found that the way cluster size changes in the EDL is similar to the changes seen in AFM measurements and matches the qualitative trends in MD simulations. Moreover, our theory’s prediction of the differential capacitance was found to be in reasonable agreement with our experimental results. The mathematical analysis here goes beyond even the most advanced theories for patchy particle systems, in addition to taking into account electrostatic interactions. These ideas might find applications in other fields of statistical physics as well as in understanding experimental results.

As WiSEs are an exciting class of electrolytes for energy storage, with lots of simulations and experiments investigating these systems for a myriad of applications, having a theory to build intuition is critical. This work can be used for the following: provide insight into the local structure through the ionic aggregation and solvation of species near electrodes, aid in predicting the formation of the solid electrolyte interphase (SEI), and shed light on surface force measurements near electrified interfaces. Overall, the applications of this theory are expansive and we hope it will inspire additional studies into the interfacial behavior of electrolytes as well as support rational electrolyte design. Looking forward, we believe developing the approach to understand the kinetics of solvation/de-solvation, charging dynamics, and coupled-ion-electron transfer reactions at interfaces could be interesting areas for exploration.

\begin{suppinfo}

See the Supplemental Materials for a detailed discussion of the theory and its evaluation, the molecular dynamics simulation methodology, the data analysis for the simulations, an additional comparison of theory as well as approximations, the experimental methods, and additional predictions and discussions of results from the theory and comparison of theory and experimental data.

\end{suppinfo}

\begin{acknowledgement}

We are grateful to J. Pedro de Souza for the helpful discussions and feedback. D.M.M. \& M.Z.B. acknowledge support from the Center for Enhanced Nanofluidic Transport 2 (CENT$^2$), an Energy Frontier Research Center funded by the U.S. Department of Energy (DOE), Office of Science, Basic Energy Sciences (BES), under award \# DE-SC0019112. D.M.M. also acknowledges support from the National Science Foundation Graduate Research Fellowship under Grant No. 2141064. M.M. and M.Z.B. acknowledge support from an Amar G. Bose Research Grant. Z.A.H.G acknowledges support through the Glasstone Research Fellowship in Materials and The Queen's College, University of Oxford. RMEM thanks the National Science for partially funding this research under National Science Foundation grants DMR 1904681 and CBET 1916609.
\end{acknowledgement}

\setcounter{equation}{0}
\setcounter{figure}{0}
\renewcommand{\theequation}{S\arabic{equation}} 
\renewcommand{\thefigure}{S\arabic{figure}}  
\renewcommand{\labelenumii}{\arabic{enumi}.\arabic{enumii}}
\renewcommand{\labelenumiii}{\arabic{enumi}.\arabic{enumii}.\arabic{enumiii}}
\renewcommand{\labelenumiv}{\arabic{enumi}.\arabic{enumii}.\arabic{enumiii}.\arabic{enumiv}}

\section*{Supplemental Material}

\section{Extended Theory Section}
In this section, the theory is discussed in greater detail, along with its implementation, and additional results.

\subsection{Sticky Cation Approximation Bulk Bias Derivation}
Here we prove in the bulk how the functionality of cations being larger than anions leads to the distribution beyond the hydrated lithium and the free TFSI$^-$ being marginally biased towards net negative clusters. To understand this result, in the sticky-cation formalism, one can note that any cation that is not in its hydrated state ($c_{10f_+}$) must belong to a multi-ion cluster. This likewise holds for the free anions ($c_{010}$). This implication means that the ratio of positive to negative charges stored in the clusters will be inversely proportional to the ratio between the hydrated cations and free anions ($c_{10f_+}$/$c_{010}$). Hence if $c_{10f_+}$/$c_{010}$ is equal to 1 the clusters have a net neutral bias, greater than 1 the clusters have a net negative bias, and less than 1 the clusters have a net positive bias. This result can be intuitively tested through limiting cases. First, consider the case where $c_{10f_+}$=$c_{010}$. Here, there are an equal number of cations and anions found in clusters as in the bulk $c_{+}$=$c_{-}$. Hence, one can conclude that the multi-ion clusters must contain equal amounts of cations and anions, i.e. the clusters would have no net charge bias. Second, consider the case where $c_{10f_+}$ is nonzero and $c_{010}$=0. Here, there are more anions in the clusters than cations as in the bulk $c_{+}$=$c_{-}$. From this implication, one can conclude that the multi-ion clusters will have a negative bias. Lastly, consider the case where $c_{10f_+}$=0 and $c_{010}$ is nonzero. In this case, there will be more cations in the clusters than anions as in the bulk $c_{+}$=$c_{-}$. This implication allows one to conclude that the clusters will have a positive bias. Note that this ratio is analogous to the Boltzmann closure relationships, but here we can see it can provide additional insightful information on the bias of the multi-ion clusters. In general, one should distinguish between the gel and the multi-ion clusters as the gel is its own phase. For this reason, we will restrict the analysis here to the pre-gel regime for simplicity, although this form of analysis could be extended for additional information. 

At this point, we can begin our proof that for water-in-salt electrolytes (WiSEs) in the pre-gel bulk, under the sticky-cation formalism, the net charge bias of the multi-ion cluster depends solely on the ratio of the anion to cation functionalities. First, recall what we just proved for the values of $c_{10f_+}$/$c_{010}$:
    \begin{equation}
    \label{pref}
        \frac{c_{10f_+}}{c_{010}}=\begin{cases} 
          \text{multi-ion clusters have net positive bias,} & <1 \\
         \text{multi-ion clusters have no net charge bias,} & =1 \\
         \text{multi-ion clusters have net negative bias,} & >1 
       \end{cases}
    \end{equation}

To make any additional discernment one must expand $c_{10f_+}$ and $c_{010}$ where we use their volume fraction equations and convert them into concentration,
   \begin{equation}
    \label{bias1}
        \frac{c_{10f_+}}{c_{010}}=\frac{c_+(1-p_{+-})^{f_+}}{c_-(1-p_{-+})^{f_-}}=\frac{(1-p_{+-})^{f_+}}{(1-p_{-+})^{f_-}}.
    \end{equation}
    
To resolve this question, convert Eq.~\eqref{bias1} into an equation that depends only on one unknown. Here we choose $p_{-+}$, where $p_{+-}$ is eliminated through recalling and using the conservation of cation-anion associations in the bulk will be $f_+p_{+-}=f_-p_{-+}$,
   \begin{equation}
    \label{bias2}
        \frac{c_{10f_+}}{c_{010}}=\frac{(1-\frac{f_-}{f_+}p_{-+})^{f_+}}{(1-p_{-+})^{f_-}}.
    \end{equation}

At this point, it is not directly obvious how any useful implications can be achieved. To proceed, one must recall that taking the logarithm of $c_{10f_+}$/$c_{010}$ and multiplication by a strictly positive constant will preserve order, 
   \begin{equation}
   \label{pref2}
        \frac{1}{f_{+}}\log\left(\frac{c_{10f_+}}{c_{010}}\right)=\begin{cases} 
          \text{multi-ion clusters have net positive bias,} & <0 \\
          \text{multi-ion clusters have no net charge bias,} & =0 \\
          \text{multi-ion clusters have net negative bias,} & >0 
       \end{cases}
   \end{equation}

This allows us to rewrite Eq.~\eqref{bias2} to be only a function of $p_{-+}$ and $f_-$/$f_+$,
    \begin{equation}
    \label{bias3}
        \frac{1}{f_+}\log\left(\frac{c_{10f_+}}{c_{010}}\right)=\log\left(1-\frac{f_-}{f_+}p_{-+}\right)-\frac{f_-}{f_+}\log\left(1-p_{-+}\right).
    \end{equation}

Now for the last key step, one must observe that the derivative of Eq.~\eqref{bias3} with respect to $p_{-+}$, is monotonic where it's defined on our regime, i.e. $p_{-+}\in[0,\text{min}(f_+/f_-,1)]$,
    \begin{equation}
    \label{bias4}
        \frac{\partial}{\partial p_{-+}}\left(\log\left(1-\frac{f_-}{f_+}p_{-+}\right)-\frac{f_-}{f_+}\log\left(1-p_{-+}\right)\right) = \frac{f_-}{f_+}\left(\frac{1}{1-p_{-+}}-\frac{1}{1-\frac{f_-}{f_+}p_{-+}}\right)
    \end{equation}

The monotonicity of Eq.~\eqref{bias4} will be critical. To see this first, it is helpful to note that not that when Eq.~\eqref{bias3} is zero when $p_{-+}$=0. Hence as Eq.~\eqref{bias4} is monotonic in the domain of interest, this implication requires that the sign of Eq.~\eqref{bias3} is determined by the sign of Eq.~\eqref{bias4}. The sign of Eq.~\eqref{bias4} is clearest at its upper bound, i.e. $p_{-+}=\text{min}(f_+/f_-,1)$. Here Eq.~\eqref{bias4} will only not diverge if $f_+/f_- = 1$, meaning that in this case Eq.~\eqref{bias3} is zero throughout its domain, i.e. the multi-ion clusters have no net bias. Now, let us consider the cases where Eq.~\eqref{bias4} diverges to $\pm\infty$. Here, if $f_-/f_+ > 1$ then the second term will diverge first with increasing $p_{-+}$, meaning Eq.~\eqref{bias4} diverges to $-\infty$. This result means that Eq.~\eqref{bias3} will decrease as $p_{-+}$ increases, i.e. Eq.~\eqref{bias3} will be less than zero throughout its domain besides at $p_{-+}$=0. This finding means that for $f_-/f_+ > 1$ multi-ion clusters will have a net positive bias. Now, let us consider if $f_-/f_+ < 1$ then the first term will diverge first with increasing $p_{-+}$, meaning Eq.~\eqref{bias4} diverges to $\infty$. This result means that Eq.~\eqref{bias3} will increase as $p_{-+}$ increases, i.e. Eq.~\eqref{bias3} will be greater than zero throughout its domain besides at $p_{-+}$=0. This finding means that for $f_-/f_+ < 1$ multi-ion clusters will have a net negative bias. From this we can conclude,
    \begin{equation}
    \label{pref4}
        \text{mulit-ion clusters charge bias}=\begin{cases} 
          \text{net positive,} & \frac{f_-}{f_+}>1 \\
          \text{net neutral,} & \frac{f_-}{f_+}=1 \\
          \text{net negative,} & \frac{f_-}{f_+}<1 
       \end{cases}
    \end{equation}

Therefore, as $f_+=4$ and $f_-=3$ in the main paper, the distribution beyond the hydrated lithium and the free TFSI$^-$ is expected to be marginally biased towards net negative clusters. Note in our proof for the bulk under the sticky-cation formalism there are two main edge cases from limiting conditions. Both cases result in multi-ion clusters being ill-defined or neutral bias; if one asserts no multi-ion clusters being present means they have net neutral bias. First, if any ionic species functionality is zero, then there are no multi-cluster ions, so this measure is ill-defined. Second, if $p_{-+}$=0, it would appear that there is no net bias. However considering $p_{-+}$=0's implications, this means there are no multi-ion clusters present, which means this measure is ill-defined once again. In both of these edge cases, they occur when no associations are present, which means for all enlightening parameter choices for our theory, this measure is informative and applicable. 

\subsection{Short-range correlation parameter}

Incorporating the $\alpha$-parameter from Ref.~\citenum{goodwin2017mean} is straight-forward, as it arrives through the excess chemical potential. The resulting changes to the overall system of equations are minor, only the general cluster chemical potential equation is shown here,
    \begin{align}
    \label{alphaChemPot}
        \beta \bar{\mu}_{lms} =& (l-m) \beta e \alpha \Phi - \ln\left( \frac{\text{Sinh}(\beta P |\nabla \Phi|)}{\beta P |\nabla \Phi|} \right)\delta_{l,0}\delta_{m,0}\delta_{s,1} + 1 + \ln(\bar{\phi}_{lms}) + \beta\Delta_{lms} \nonumber \\
        & - (\xi_+l+\xi_-m+s)\Lambda + (\xi_+l+\xi_-m+s)\beta\bar{d}'
    \end{align}

\noindent where $\bar{d}'=\bar{c}_+^{gel}\partial\bar{\Delta}_+^{gel}+\bar{c}_-^{gel}\partial\bar{\Delta}_-^{gel}+\bar{c}_0^{gel}\partial\bar{\Delta}_0^{gel}$, with the derivative being with respect to $\bar{\phi}_{lms}$. This modification can be carried through the subsequent steps to arrive at the modified system of equations. The only change to the system of equations will be $\alpha \Phi$ in place of $\Phi$. Note this means $\nabla\Phi$ is unaffected for the fluctuating Langevin dipole terms. This extension can be directly incorporated into the general and sticky cation system of equations. 

\subsection{System of Equations}\label{SoE}

As discussed in the main text, to connect the bulk equations to quantities within the EDL and the modified Poisson-Boltzmann equation, we need to solve a system of equations. Generally, these equations consist of the Boltzmann closure relations, of which there are 3, described in the main text, which connect the free species in the bulk to the free species in the EDL, to the electrostatic potential and its derivative, in a consistent way. As we do not restrict species to be the same size, we also enforce incompressibility, which is another equation that is introduced through a Lagrange multiplier and appears through $\tau = \exp{(\Lambda)}$. These free species concentrations are intrinsically linked to the conservation of associations and the mass action laws, both of which are also required to solve the EDL system of equations. To remove the dependence of the equations on the free species, and incorporate the conservation of associations and the mass action laws,  we can substitute $\phi_{100} = \phi_{+}(1 - p_{+-} - p_{+0})^{f_{+}}$ [or $\phi_{10f_+} = \phi_{+}(1+f_{+}/\xi_+)(1 - p_{+-})^{f_{+}}$ for the sticky case], $\phi_{010} = \phi_-(1 - p_{-+})^{f_-}$, and $\phi_{001} = \phi_0(1 - p_{0+})$. The conservation of associations and the mass action laws introduce another 4 equations. In total, we have 8 unknown variables and 8 equations, and therefore the system can be solved. For the non-sticky case, we have
    \begin{equation}
        \label{NS_phipb}
        \bar{\phi}_{+} = \frac{\phi_{100}\exp(-e\beta\Phi)\tau^{\xi_+}}{(1-\bar{p}_{+-}-\bar{p}_{+0})^{f_+}} = \frac{\xi_+}{f_+}\bar{\gamma}_{100} \tau^{\xi_+} \frac{1}{{(1-\bar{p}_{+-}-\bar{p}_{+0})^{f_+}}}
    \end{equation}
    \begin{equation}
        \label{NS_phimb}
        \bar{\phi}_{-} = \frac{\phi_{010}\exp(\beta e \Phi) \bar{\tau}^{\xi_-}}{(1-\bar{p}_{-+})^{f_-}} = \frac{\xi_-}{f_-}\bar{\gamma}_{010} \tau^{\xi_-} \frac{1}{(1-\bar{p}_{-+})^{f_-}}
    \end{equation}
    \begin{equation}
        \label{NS_phi0b}
        \bar{\phi}_{0} =  \frac{\text{Sinh}(\beta e P |\nabla \Phi|)}{\beta e P |\nabla \Phi|}\frac{\phi_{001} \tau}{(1-\bar{p}_{0+})} = \bar{\gamma}_{001} \tau \frac{1}{(1-\bar{p}_{0+})}
    \end{equation}
    \begin{equation}
        \label{Incomp1}
       \bar{\phi}_+ + \bar{\phi}_- + \bar{\phi}_{0} = 1
    \end{equation}
    \begin{equation}
        \label{CAC}
        \frac{f_+\phi_+p_{+-}}{\xi_+} = \frac{f_-\phi_-p_{-+}}{\xi_-} = \zeta
    \end{equation}
    \begin{equation}
        \label{CSC}
        \frac{f_+\phi_+p_{+0}}{\xi_+} = \phi_0p_{0+} = \Gamma
    \end{equation}
    \begin{equation}
        \label{CAL}
        \lambda_{+-}\zeta = \frac{p_{+-}p_{-+}}{(1-p_{+-}-p_{+0})(1-p_{-+})}
    \end{equation}
    \begin{equation}
        \label{CSL}
        \lambda_{+0}\Gamma = \frac{p_{+0}p_{0+}}{(1-p_{+-}-p_{+0})(1-p_{0+})}
    \end{equation}
where $\bar{\gamma}_j$ has been introduced and can be considered the number of association sites per lattice site in the EDL \textit{unadjusted by the Lagrange multiplier} for the bare cations (fully hydrated cations), anions, and solvent species. For the sticky-cation formulation,
    \begin{equation}
        \label{S_phipb}
        \bar{\phi}_{+} = \frac{\phi_{10f_+}\exp(-\beta e \Phi) \tau^{(\xi_++f_+)}}{\left(1+f_+/\xi_+\right)\bar{p}_{+0}^{f_+}} = \frac{\xi_++f_+}{f_+}\bar{\gamma}_{10f_+} \tau^{\xi_++f_+} \frac{1}{\left(1+f_+/\xi_+\right)\bar{p}_{+0}^{f_+}}
    \end{equation}
    \begin{equation}
        \label{S_phimb}
        \bar{\phi}_{-} = \frac{\phi_{010}\exp(\beta e \Phi) \tau^{\xi_-}}{(1-\bar{p}_{-+})^{f_-}} = \frac{\xi_-}{f_-}\bar{\gamma}_{010} \tau^{\xi_-} \frac{1}{(1-\bar{p}_{-+})^{f_-}}
    \end{equation}
    \begin{equation}
        \label{S_phi0b}
        \bar{\phi}_{0} =  \frac{\text{Sinh}(\beta e P |\nabla \Phi|)}{\beta e P |\nabla \Phi|}\frac{\phi_{001} \tau}{(1-\bar{p}_{0+})} = \bar{\gamma}_{001} \tau \frac{1}{(1-\bar{p}_{0+})}
    \end{equation}
    \begin{equation}
        \label{Incomp2}
       \bar{\phi}_+ + \bar{\phi}_- + \bar{\phi}_{0} = 1 \nonumber
    \end{equation}
    \begin{equation}
        \label{S_ppm}
        p_{+-} = \frac{\psi_0 - \psi_+ + {\lambda}(\psi_+ + \psi_-) - \sqrt{({\lambda} (\psi_- - \psi_+)+\psi_0+\psi_+)^2 + 4 ({\lambda}-1) \psi_0 \psi_+}}{2 ({\lambda}-1) \psi_+}
    \end{equation}
    \begin{equation}
        \label{S_pmp}
        p_{-+} = \frac{\psi_0 - \psi_+ + {\lambda}(\psi_+ + \psi_-) - \sqrt{({\lambda} (\psi_- - \psi_+)+\psi_0+\psi_+)^2 + 4 ({\lambda}-1) \psi_0 \psi_+}}{2 ({\lambda}-1) \psi_-}
    \end{equation}
    \begin{equation}
        \label{S_pp0}
        p_{+0} = 1 - \frac{\psi_0 - \psi_+ + \lambda(\psi_+ + \psi_-)}{2 (\lambda-1) \psi_+} + \frac{\sqrt{(\lambda (\psi_- - \psi_+)+\psi_0+\psi_+)^2 + 4 (\lambda-1) \psi_0 \psi_+}}{2 (\lambda-1) \psi_+}
    \end{equation}
    \begin{equation}
        \label{S_p0p}
        p_{0+} = \frac{\psi_+}{\psi_0} - \frac{\psi_0 - \psi_+ + {\lambda}(\psi_+ + \psi_-)}{2 ({\lambda}-1) \psi_0} + \frac{\sqrt{({\lambda} (\psi_- - \psi_+)+\psi_0+\psi_+)^2 + 4 ({\lambda}-1) \psi_0 \psi_+}}{2 ({\lambda}-1) \psi_0}
    \end{equation}

    Where the last four equations come from the conservation of associations, the law of mass action on the number of associations, and the sticky cation approximation $p_{+-} + p_{+0} = 1$, we can obtain explicit expressions of our association probabilities in terms of the $\psi_i$ and $\lambda$. Additionally, in both sets of equations, the last five equations hold in a nontrivial fashion for both the bulk and EDL quantities.

    The above system of equations tends to be the clearest way to understand the underlying physics of the system and, hence, its predictions. However, one can reduce these systems of 8 equations down to a system of 2 equations, which can provide some insight into the system of equations being solved and an alternative way of obtaining the roots. This reduction of equations was motivated by and achieved through writing the unknowns in terms of the association probabilities. Initially, we will derive the relationship for the general case followed by the sticky cation approximation. For brevity, the application of algebraic manipulations of $\bar{\lambda}$ is utilized implicitly:
    \begin{equation}
        \label{LBar}
        \bar{\lambda} = \frac{\bar{p}_{-+}(1-\bar{p}_{0+})}{\bar{p}_{0+}(1-\bar{p}_{-+})}.
    \end{equation}
    
    In order to derive these equations, we construct the Boltzmann closure relationships between the bare anions and the free solvent and between the bare cations and the free solvent. These kinds of relationships were first introduced in Ref.~\citenum{Goodwin2022EDL},
        \begin{equation}
        \label{usBCR_AW}
        \frac{\bar{\phi}_-(1-\bar{p}_{-+})^{f_-}}{\bar{\phi}_0(1-\bar{p}_{0+})} = \frac{\xi_-\bar{\gamma}_{010}}{f_-\bar{\gamma}_{001}}\tau^{\xi_--1},
    \end{equation}
    \begin{equation}
        \label{usBCR_CW}
        \frac{\bar{\phi}_+(1-\bar{p}_{+-}-\bar{p}_{+0})^{f_+}}{\bar{\phi}_0(1-\bar{p}_{0+})} = \frac{\xi_+\bar{\gamma}_{100}}{f_+\bar{\gamma}_{001}}\tau^{\xi_+-1}.
    \end{equation}
    These expressions can be further reduced to be explicitly in terms of probabilities, the compressibility constraint, and $\bar{\gamma}_i$'s through using the conservation of association equations, Eq.~\eqref{CAC} \&~\eqref{CSC}, to convert the volume ratios into probability ratios. The idea of this is to utilize substitutions and transformations where possible to reduce the number of unknowns and equations needed to be solved. Note this idea will be used multiple times to simplify our system of equations. Applying this procedure and using Eq.~\eqref{LBar} to simplify our equations we find:
    \begin{equation}
        \label{BCR_AW}
        \frac{1}{\bar{\lambda}}\frac{\bar{p}_{+-}}{\bar{p}_{+0}}(1-\bar{p}_{-+})^{f_--1} = \frac{\bar{\gamma}_{010}}{\bar{\gamma}_{001}}\tau^{\xi_--1},
    \end{equation}
    \begin{equation}
        \label{BCR_CW}
        \frac{\bar{p}_{0+}}{1-\bar{p}_{0+}} \frac{(1-\bar{p}_{+-}-\bar{p}_{+0})^{f_+}}{\bar{p}_{+0}} = \frac{\bar{\gamma}_{100}}{\bar{\gamma}_{001}}\tau^{\xi_+-1}.
    \end{equation}

    To further reduce these equations, one can note that the formulas for $\bar{p}_{ij}$ in terms of $\bar{\eta} = (1-\bar{p}_{+-}-\bar{p}_{+0})$ (the probability a cation association site is empty), $\tau$, and $\bar{\gamma}_i$ can all be written down in terms of the two unknowns $\bar{\eta}$ and $\tau$. Thus, as we will show this allows one to collapse the 8 equations into 2 equations. This will be accomplished through a series of substitutions where we show all the probabilities can be written in terms of $\bar{\eta}$ and $\tau$. This finding is sufficient to show the reduction to 2 equations as we already know the volume fractions can be written in terms of the probabilities and $\tau$, i.e. 8 equations to 5 immediately and with these manipulations to 2. One can start by substituting the cation form of $\zeta$ from Eq.~\eqref{CAC} into Eq.~\eqref{CAL} and using Eq.~\eqref{NS_phipb} for $\bar{\phi}_+$ as well as algebraic manipulation,
    \begin{equation}
        \label{NS_p0p_eta}
        \bar{p}_{0+} = \frac{\lambda_{+-}\bar{\gamma}_{100}\tau^{\xi_+}}{\bar{\lambda}\bar{\eta}^{f_+-1}+\lambda_{+-}\bar{\gamma}_{100}\tau^{\xi_+}},
    \end{equation}

    \noindent which by conversion through the algebraic manipulations of $\bar{\lambda}$ becomes,
    \begin{equation}
        \label{NS_pmp_eta}
        \bar{p}_{-+} = \frac{\lambda_{+-}\bar{\gamma}_{100}\tau^{\xi_+}}{\bar{\eta}^{f_+-1}+\lambda_{+-}\bar{\gamma}_{100}\tau^{\xi_+}}.
    \end{equation}

    By substituting Eq.~\eqref{NS_p0p_eta} into Eq.~\eqref{BCR_CW} we obtain,
    \begin{equation}
        \label{NS_pp0_eta}
        \bar{p}_{+0} = \bar{\lambda}_{+0}\bar{\gamma}_{001}\bar{\eta}\tau,
    \end{equation}

    \noindent which by the definition of $\bar{\eta}$ means,
    \begin{equation}
        \label{NS_ppm_eta}
        \bar{p}_{+-} = 1 -\left(1+ \bar{\lambda}_{+0}\bar{\gamma}_{001}\tau\right)\bar{\eta}. 
    \end{equation}
  
    We produce our new polynomial in $\bar{\eta}$ whose roots contains the solutions by substituting Eq.~\eqref{NS_pmp_eta}-\eqref{NS_ppm_eta} into Eq.~\eqref{BCR_AW},
    \begin{align}
        \label{NS_eta}
        \left(1+\lambda_{+-}\bar{\gamma}_{010}\tau^{\xi_-}+\bar{\lambda}_{+0}\bar{\gamma}_{001}\tau\right)&\bar{\eta}^{(f_+ - 1)(f_- - 1) + 1} - \bar{\eta}^{(f_+ - 1)(f_- - 1)} \nonumber \\
        + &\bar{\eta}\sum_{k=0}^{f_- - 2} \binom{f_- - 1}{k} (\lambda_{+-}\bar{\gamma}_{100}\tau^{\xi_+})^{(f_- - 1) - k} \bar{\eta}^{k (f_+ - 1)}= 0.
    \end{align}

    Lastly, an analogous equation for the incompressibility constraint is required; here, $\bar{p}_{ij}$ will kept as functions of $\bar{\eta}$ and $\tau$ for simplicity. This is accomplished by substituting Eq.~\eqref{NS_phipb}-\eqref{NS_phi0b} into Eq.~\eqref{Incomp1},
    \begin{equation}
        \label{NS_tau}
         \frac{\xi_+}{f_+}\bar{\gamma}_{100} \tau^{\xi_+} \frac{1}{{(1-\bar{p}_{+-}-\bar{p}_{+0})^{f_+}}} + \frac{\xi_-}{f_-}\bar{\gamma}_{010} \tau^{\xi_-} \frac{1}{(1-\bar{p}_{-+})^{f_-}} + \bar{\gamma}_{001} \tau \frac{1}{(1-\bar{p}_{0+})} = 1.
    \end{equation}

    Hence, Eqs.~\eqref{NS_eta}-\eqref{NS_tau}, can be solved in their reduced form for $\bar{\eta}$ and $\tau$, which uniquely determine the composition of the WiSE in the EDL. It's important to note the unique property belonging to Eq.~\eqref{NS_eta} as it is a polynomial, which means it will have $(f_+-1)(f_--1)+1$ roots for a given $\tau$. This information can be further refined by noting for $\bar{\eta}$ to be meaningful that it must take on a value between zero and one inclusively. Thus by utilizing Descartes' rule of sign~\cite{Descartes} and noting that any valid $\tau$ must be a finite positive value, one can prove that for $f_+,f_- > 2$, this polynomial will have at most two positive roots and has at least one zero root which appears to corresponds to sticky cation approximation's case. The zeros correspond to the sticky cation approximation as that solution corresponds to singularities that are introduced from $\eta=0$, which alternatively can be written as $\bar{p}_{+-}+\bar{p}_{+0}=1$ which is the sticky cation approximation. Therefore when zero roots occur, one needs to use the sticky cation version of the polynomial to account for the singularity in the equations introduced with the zero root. 
    
    For the lower functionality cases, it's more case-specific: (1) when either $f_+=1$ or $f_-=1$ there is one positive root, (2) when $f_{+}=2$ and $f_{-} > 2$, there are at most 2 positive roots and at least one zero root, (3) when $f_{+}=f_{-}=2$ there is one positive root and one zero root, and (4) when $f_-=2$ and $f_{+} > 2$ there are at most 2 positive roots and at least 1 zero root. It is important to note that the negative coefficient term in this polynomial will remain negative outside of limiting cases where it goes to zero. Additionally when $f_+ = 2$ and $f_- > 1$, the coefficient that is subtracted from one in the summation leads to a strictly negative net coefficient in the non-limiting or non-enforced sticky cases. We obtain this conclusion as for the coefficient to greater or equal to zero, then (1-$\bar{p}_{-+}$)$^{f_--2}\bar{p}_{+-}\bar{p}_{-+}$-1 $\geq$ 0. This exact result also holds for the stick cation case, but only here does the limiting cases have the capabilities to satisfy the previous inequality. 

    This analysis can be repeated for the sticky cation approximation. The approximation dictates that $\bar{p}_{+-} = 1 - \bar{p}_{+0}$, using this expression helps to reduce the complexity of the intermediate expressions. To derive the reduced equations here, one must construct the Boltzmann closure relationships. In this case, the expressions are between 1) the bare anions and the free solvent and 2) between the fully hydrated cations and the free solvent.         
    \begin{equation}
        \label{usSBCR_AW}
        \frac{\bar{\phi}_-(1-\bar{p}_{-+})^{f_-}}{\bar{\phi}_0(1-\bar{p}_{0+})} = \frac{\xi_-\bar{\gamma}_{010}}{f_-\bar{\gamma}_{001}}\tau^{\xi_--1},
    \end{equation}
    \begin{equation}
        \label{usSBCR_CW}
        \frac{\bar{\phi}_+\bar{p}_{+0}^{f_+}}{\bar{\phi}_0(1-\bar{p}_{0+})} = \frac{\xi_+\bar{\gamma}_{10f_+}}{f_+\bar{\gamma}_{001}}\tau^{\xi_++f_+-1}.
    \end{equation}
    These expressions can be further reduced to be explicitly in terms of probabilities, the compressibility constraint, and $\bar{\gamma}_i$'s through using the conservation of association equations, Eq.~\eqref{CAC} \&~\eqref{CSC}, to convert the volume ratios into probability ratios. Here we also invoke $\bar{p}_{+-} = 1 - \bar{p}_{+0}$. This derivation follows a similar style as the general formalism. Applying the procedure and using Eq.~\eqref{LBar} to simplify our equations we find:
    \begin{equation}
        \label{SBCR_AW}
        \frac{1}{\bar{\lambda}}\frac{1-\bar{p}_{+0}}{\bar{p}_{+0}}\left(1-\bar{p}_{-+}\right)^{f_- - 1} = \frac{\bar{\gamma}_{010}}{\bar{\gamma}_{001}}\tau^{\xi_- - 1},
    \end{equation}
    \begin{equation}
        \label{SBCR_CA}
        \frac{\bar{p}_{0+}}{1-\bar{p}_{0+}} \bar{p}_{+0}^{f_+ - 1} = \frac{\bar{\gamma}_{10f_+}}{\bar{\gamma}_{001}}\tau^{\xi_+ + f_+ - 1}.
    \end{equation}

    To reduce this system of equations, one can note that formulating $\bar{p}_{ij}$ in terms of $\bar{p}_{+0}$, $\tau$, and our $\bar{\gamma}_i$ can collapse the system from 8 to 2 equations and unknowns. This procedure is accomplished in a more direct fashion but similar style to the general case, as Eq.~\eqref{SBCR_CA} can be directly manipulated to give an equation for $\bar{p}_{0+}$ in terms of $\bar{p}_{+0}$, $\tau$, and our $\bar{\gamma}_i$'s,
    \begin{equation}
        \label{S_p0p_pp0}
        \bar{p}_{0+} = \frac{\bar{\gamma}_{10f_+}\tau^{\xi_+ + f_+ - 1}}{\bar{\gamma}_{001}\bar{p}_{+0}^{f_+-1}+\bar{\gamma}_{10f_+}\tau^{\xi_+ + f_+ - 1}},
    \end{equation}

    \noindent which by conversion through the algebraic manipulations of $\bar{\lambda}$ produces a formula for $\bar{p}_{-+}$,
    \begin{equation}
        \label{S_pmp_pp0}
        \bar{p}_{-+} = \frac{\bar{\lambda}\bar{\gamma}_{10f_+}\tau^{\xi_+ + f_+ - 1}}{\bar{\gamma}_{001}\bar{p}_{+0}^{f_+-1} + \bar{\lambda}\bar{\gamma}_{10f_+}\tau^{\xi_+ + f_+ - 1}}.
    \end{equation}

    By substituting Eq.~\eqref{S_pmp_pp0} into Eq.~\eqref{SBCR_AW} we obtain,
    \begin{equation}
        \label{S_pp0_implict}
        \frac{1}{\bar{\lambda}}\frac{1-\bar{p}_{+0}}{\bar{p}_{+0}}\left(\frac{\bar{\gamma}_{001} \bar{p}_{+0}^{f_+ - 1}}{\bar{\gamma}_{001} \bar{p}_{+0}^{f_+ - 1} + \bar{\lambda} \bar{\gamma}_{10f_+} \tau^{\xi_+ + f_+ - 1}}\right)^{f_- - 1} = \frac{\bar{\gamma}_{010}}{\bar{\gamma}_{001}}\tau^{\xi_- - 1}.
    \end{equation}

    This can be algebraically manipulated to produce our new polynomial whose roots are the solutions to $\bar{p}_{+0}$,
    \begin{align}
        \label{S_pp0r}
        (\bar{\lambda}\bar{\gamma}_{010}\tau^{\xi_-}+&\bar{\gamma}_{001}\tau)(\bar{\gamma}_{001}\tau)^{f_- -1}\bar{p}_{+0}^{(f_+ - 1)(f_- - 1) + 1} - \bar{\gamma}_{001}^{f_-}\tau^{f_-}\bar{p}_{+0}^{(f_+ - 1)(f_- - 1)} \nonumber \\
        + &\bar{\lambda}\bar{\gamma}_{010}\tau^{\xi_-}\bar{p}_{+0}\sum_{k=0}^{f_- - 2} \binom{f_- - 1}{k} (\bar{\gamma}_{001}\tau)^{k}(\bar{\lambda}\bar{\gamma}_{10f_{+}}\tau^{\xi_+ + f_+})^{(f_- - 1) - k} \bar{p}_{+0}^{k (f_+ - 1)} = 0.
    \end{align}

    Lastly, an analogous equation enforcing incompressibility must be constructed keeping $\bar{p}_{ij}$ as functions of $\bar{p}_{+0}$ and $\tau$ for simplicity. This is accomplished by substituting Eq.~\eqref{S_phipb}-\eqref{S_phi0b} into Eq.~\eqref{Incomp1},
    \begin{equation}
        \label{S_tau}
        \frac{\xi_+}{f_+}\bar{\gamma}_{10f_+} \tau^{\xi_++f_+} \frac{1}{\bar{p}_{+0}^{f_+}} + \frac{\xi_-}{f_-}\bar{\gamma}_{010} \tau^{\xi_-} \frac{1}{(1-\bar{p}_{-+})^{f_-}} + \bar{\gamma}_{001} \tau \frac{1}{(1-\bar{p}_{0+})} = 1.
    \end{equation}

    Here Eq.~\eqref{S_pp0r}-\eqref{S_tau} can be solved in their reduced form for the $\bar{p}_{+0}$ and $\tau$ which uniquely determines the composition of the WiSE in the EDL. This new polynomial for $\bar{p}_{+0}$, Eq.~\eqref{S_pp0r} brings with it the useful properties discussed earlier which can aid in solving and understanding WiSEs.

    For this and the prior case, the solution roots can be found, and each is tested to find a valid solution. Here to support computational efficiency, this was done by using the previous root solutions as the root to test as the valid solution. For the work presented here, this closest root on a sufficiently fine grid produced a valid solution. Therefore, the other potentially valid solutions were neglected. Most importantly, the predictions from this reduced system were tested in all of the previous equations to validate sufficient convergence against the initial system of equations.

\subsection{EDL Calculations}

In order to solve the sticky cation system of equations with the modified Poisson-Boltzmann (PB) equation, we utilize the following procedure. Initially, we calculate the bulk properties using Eqs.~\eqref{S_ppm}-\eqref{S_p0p} with the cation association ratio. In Section~\ref{SoE}, the system of 8 equations, Eqs.~\eqref{S_phipb}-\eqref{S_p0p}, needs to be solved to obtain the 8 unknowns, i.e., the volume fractions, association probabilities and the Lagrange multiplier. Therefore, we obtain the relationships between $\bar{\phi}_+,\bar{\phi}_-,\bar{\phi}_{001}$ and $\Phi$,$\nabla\Phi$, on a regular grid of $\Phi$ \& $\nabla\Phi$, which can then be used to solve the modified PB equation. For quicker computational, we created additional mappings of $\Phi$ \& $\nabla\Phi$ to all the unknown variables and composite variables, such as $\rho_e$ \& $\epsilon$, to solve the modified PB equation numerically. This procedure is similar to that of Ref.~\citenum{Markiewitz2024}, but with 2D maps depending on $\Phi$ \& $\nabla\Phi$.

Similarly for the non-sticky system of equations with the modified PB equation, we can utilize an analogous procedure. Here, we calculate the bulk properties using Eqs.~\eqref{CAC}-\eqref{CSL} with the association constants. Again, in Section~\ref{SoE}, the system of 8 equations, Eqn.~\eqref{NS_phipb}-\eqref{CSL}, are stated which need to be solved to obtain the 8 unknowns Once again one can establish mappings between $\bar{\phi}_+,\bar{\phi}_-,\bar{\phi}_{001}$ and $\Phi$,$\nabla\Phi$, on a regular grid of $\Phi$ \& $\nabla\Phi$, which can then be used to solve the modified PB equation. Additional mappings of $\Phi$ \& $\nabla\Phi$ to all the unknown variables and composite variables, such as $\rho_e$ \& $\epsilon$, were conducted to simplify numerically solving the modified PB equation numerically. 

\begin{figure}[h!]
     \centering
     \includegraphics[width= 0.9\textwidth]{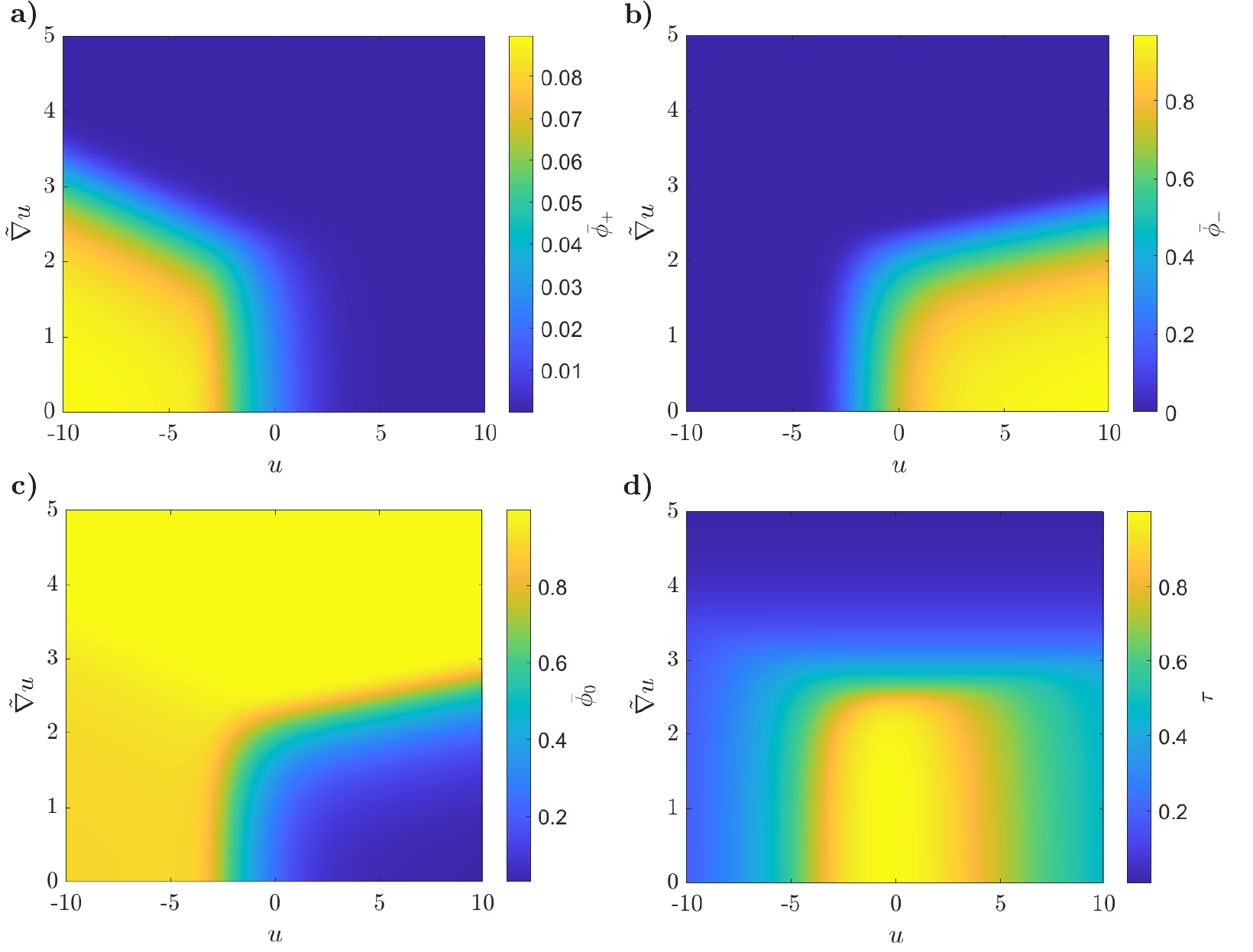}
     \caption{Numerical solution map for the species volume fractions ($\bar{\phi}_i$) and exponential of the Lagrange multiplier ($\tau$) in our system of equations under the sticky-cation formalism for 15m WiSE. a) Volume fraction of cations ($\bar{\phi}_+$). b) Volume fraction of anions ($\bar{\phi}_-$). c) Volume fraction of solvent ($\bar{\phi}_0$). d) Exponential of the Lagrange multiplier ($\tau$). Shown here is how these unknowns vary with the dimensionless electrostatic potential ($u$) and dimensionless electric field strength ($\tilde{\nabla}u$). Here we use $f_+=4$, $f_-=3$, $\xi_{0}=1$, $\xi_+=0.4$, $\xi_-=10.8$, $\epsilon_r = 10.1$, $\lambda = 0.231$, $P$ = 4.995 Debye, and $v_0 = 22.5$ \AA$^3$.}
     \label{fig:nsm_phisp}
\end{figure}

\begin{figure}[h!]
     \centering
     \includegraphics[width= 0.9\textwidth]{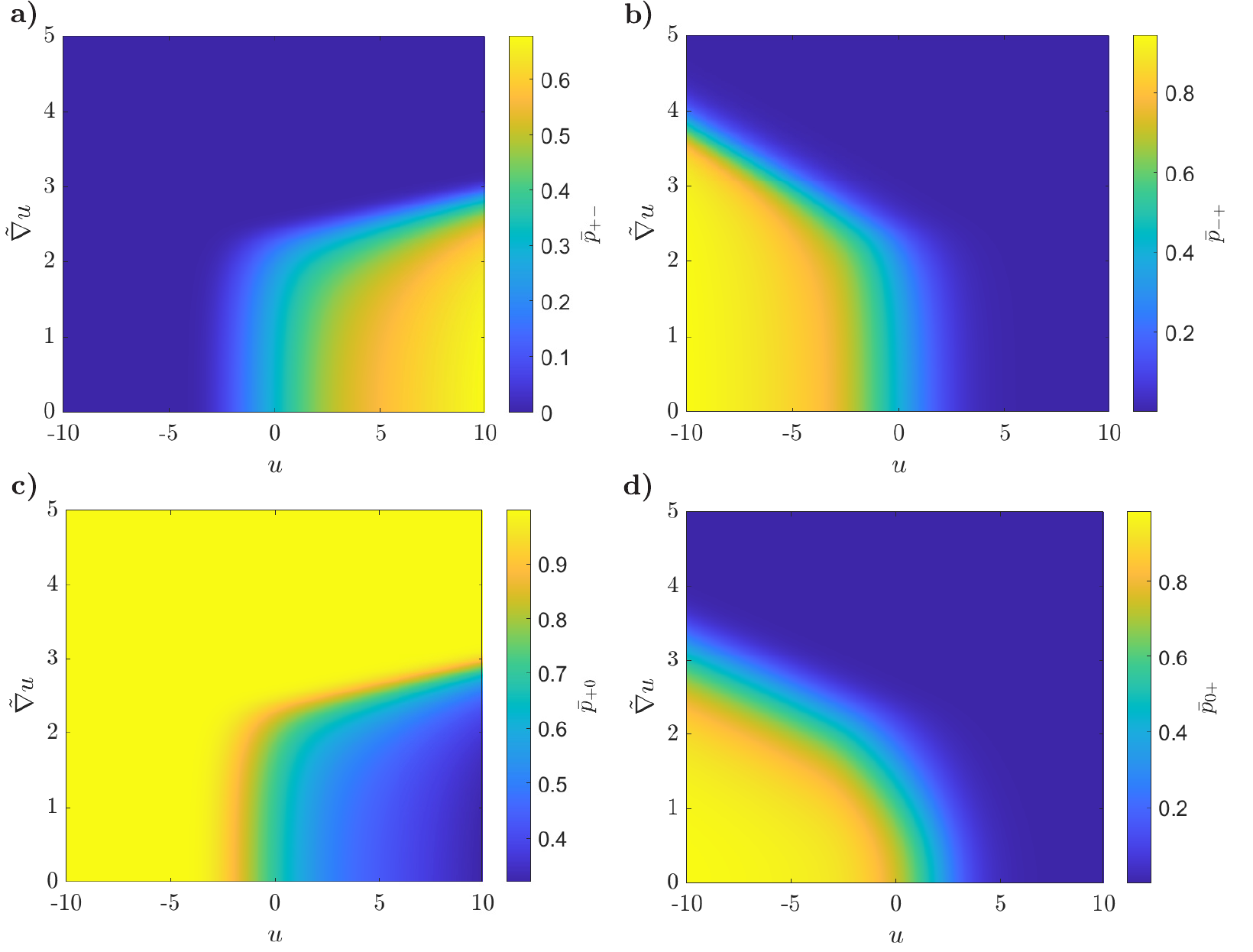}
     \caption{Numerical solution map for the association probabilities ($\bar{p}_{ij}$) in our system of equations under the sticky-cation formalism for 15m WiSE. a) Association probability of cations being bound to anion ($\bar{p}_{+-}$). b) Association probability of anions being bound to cation ($\bar{p}_{-+}$). c) Association probability of cations being bound to water ($\bar{p}_{+0}$). d) Association probability of water being bound to cations ($\bar{p}_{0+}$). Shown here is how the association probabilities vary with the dimensionless electrostatic potential ($u$) and dimensionless electric field strength ($\tilde{\nabla}u$). Here we use $f_+=4$, $f_-=3$, $\xi_{0}=1$, $\xi_+=0.4$, $\xi_-=10.8$, $\epsilon_r = 10.1$, $\lambda = 0.231$, $P$ = 4.995 Debye, and $v_0 = 22.5$ \AA$^3$.}
     \label{fig:nsm_ps}
\end{figure}

\begin{figure}[h!]
     \centering
     \includegraphics[width= 0.9\textwidth]{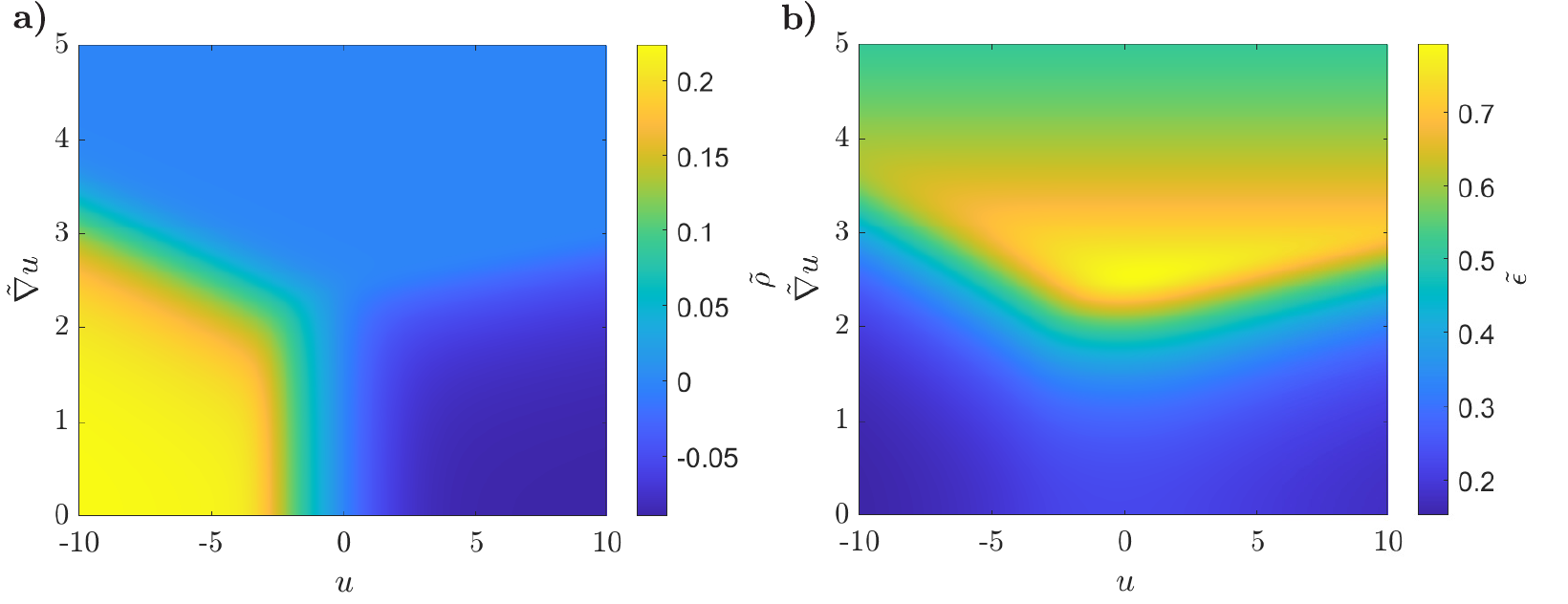}
     \caption{Numerical solution map for dimensionless charge density ($\tilde{\rho}$) and dimensionless dielectric constant ($\tilde{\epsilon}$) in our system of equations under the sticky-cation formalism for 15m WiSE. a) Dimensionless charge density ($\tilde{\rho}$). b) Dimensionless dielectric constant ($\tilde{\epsilon}$). Shown here is how these key unknowns for solving the modified-PB equation vary with the dimensionless electrostatic potential ($u$) and dimensionless electric field strength ($\tilde{\nabla}u$). Here we use $f_+=4$, $f_-=3$, $\xi_{0}=1$, $\xi_+=0.4$, $\xi_-=10.8$, $\epsilon_r = 10.1$, $\lambda = 0.231$, $P$ = 4.995 Debye, and $v_0 = 22.5$ \AA$^3$.}
     \label{fig:nsm_mpb}
\end{figure}

To solve for various profiles in the EDL as well as the screening length, differential capacitance, excess surface concentrations, and interfacial concentration of water, we use the following steps similar to the methodology utilized in Ref.~\citenum{Goodwin2022EDL} \&~\citenum{Markiewitz2024}:

\begin{enumerate}
    \item First, we numerically solve the system of equations in this work. The polynomial formulation was used for $\bar{\phi}_+$, $\bar{\phi}_-$, $\bar{\phi}_0$, dimensionless $\rho_e$, $\bar{p}_{+-}$, $\bar{p}_{+0}$, $\bar{p}_{-+}$, $\bar{p}_{0+}$, $\tau$, and dimensionless $\epsilon$ over a range of electrostatic potential and electric field strength values. This was done for a grid of dimensionless electrostatic potential ($\Phi e\beta$) and dimensionless electric field strength ($e\beta\lambda_D\nabla\Phi$) to create a refined mesh; for our purposes here, a spacing of 0.01 was used. After solving the WiSE at a set composition and $\lambda$, the refined maps for $\bar{\phi}_+$, $\bar{\phi}_-$, $\bar{\phi}_0$, dimensionless $\rho_e$, $\bar{p}_{+-}$, $\bar{p}_{+0}$, $\bar{p}_{-+}$, $\bar{p}_{0+}$, $\tau$, and $\epsilon$ were saved, allowing one to interpolated solution for these quantities from these maps for a given electrostatic potential and electric field strength. Sample maps are shown at the end of this section in Fig.~\ref{fig:nsm_phisp}-\ref{fig:nsm_mpb}.

    \item Using the dimensionless $\rho_e$ and dimensionless $\epsilon$ maps (shown in (Fig.~\ref{fig:nsm_mpb}), we can then numerically solve the modified PB equation to get a solution for the electrostatic potential and electric field profile in the EDL. Our boundary condition for a charge surface is,
    \begin{equation}
    \label{BC_MPB_s}
         \left.(\epsilon\nabla\Phi)\right|_s = -q_s \textbf{n}.
    \end{equation}

    The boundary condition for the bulk is,
    \begin{equation}
    \label{BC_MPB_b}
         \Phi(\textbf{r}\rightarrow\infty) = 0.
    \end{equation}
    
    \item The electrostatic potential and electric field profile in the EDL along with our interpolation maps allows us to predict profiles of the various quantities of interest in the EDL: dimensionless charge density, total volume fractions of each species, volume fraction of free (or fully hydrated) cations, volume fraction free anions, volume fraction of free water, volume fraction aggregates as well as individual clusters volume fractions, association probabilities, and the product of the ionic association probabilities. Lastly from the individual clusters' volume fractions, one can convert into the dimensionless cluster concentrations and, hence, numerically evaluate the simplified form of Eq.~\eqref{LSAgg} till the evaluation converges to determine the length scale of the aggregates. Here, we summed over all valid Sticky Cation Cayley tree clusters containing up to 100 cations and up to 100 anions. Additionally, using the individual clusters' volume fractions, one can obtain the dimensionless concentrations, which allows one to create the cluster distribution plots for $l+m>0$.

    \item To determine the screening length ($\lambda_s$), we applied a $\pm$ 0.001 V electrostatic potential boundary condition at the surface. We obtained the screening length from the electrostatic potential profile in the EDL by fitting an exponential decay to the profile and extracting the exponential decay constant for a range of molalities as shown in Fig.~\ref{fig:ScreenL}. In Fig.~\ref{fig:ScreenL}, it was constructed using a 0.001 V electrostatic potential boundary condition. It is worth noting that the screening lengths obtained by $\pm$ 0.001 V solutions are very similar. Additionally, these profiles are within the pre-gel regime for the screening length plot in Fig.~\ref{fig:ScreenL}. As $\lambda_s$ is plotted against molality and the system's MD parameters fluctuate slightly with the molality, the average value of $v_0$ and $\lambda$ was used to generate this figure. Lastly, to determine the bulk WiSE composition, we utilized the following formula to determine the $\phi_+$ from the molality (m) and the molar mass of the solvent (M$_s$), as from bulk electroneutrality and incompressibility this fully determines the bulk composition:
    \begin{equation}
        \label{Molarity2VF}
        \phi_+ = \left(1+\frac{1}{\xi_+\text{M}_s \text{m}}+\frac{\xi_-}{\xi_+}\right)^{-1}
    \end{equation}
    
    \item To obtain our differential capacitance predictions, we introduce the $\alpha$-parameter and set it to 0.1 to bring the voltage range of the profile more in line with real-world systems. Following this and similar to previous steps, we solved the new system to obtain solution maps to the modified system of equations. From these maps, one could solve the modified PB equation. Then, we solved for the potential at the interface ($\Phi_s$) over a range of surface charge densities (q$_s$) for our boundary conditions, here we used a fine grid spacing of approximately 0.0001 C/m$^2$. Note the solution to the system impacts the value of the dimensional surface charge density; hence, this spacing for q$_s$ was found numerically. From this map, we next constructed splines to calculate how q$_s$ depends on $\Phi_s$. Using these splines, we calculated the differential capacitance by numerically taking the derivative of q$_s$ with respect to $\Phi_s$ using finite differences.

    \item The excess surface concentrations can be obtained in our current theory via numerical integration of the difference between the species' concentration and its bulk concentration. Note to obtain the concentration, one can convert the volume fractions to the dimensionless concentration as $c_i = \phi_i/\xi_i$ then divide by $v_0$ to obtain the concentration. In this work, the numerical integration was conducted on the fine spacing produced by the numerical solver for the boundary value problem.
    
    \item Obtaining the interfacial concentration of water is more intricate as the depletion region is important for correctly evaluating this quantity. Hence to make the predicted values from the theory more analogous to the predictions from the MD simulations, the depletion region found for the MD cases was used to shift the theory's dimensionless concentration before integration. As this region has zero value, one can reduce the upper integration bound accordingly, which was done here. Note for the zero charge surface, the depletion region was not symmetric; hence, the average value was used. With the properly shifted dimensionless concentration of water profiles, which are obtained by converting from volume fraction, one can numerically integrate these curves up from 0 to $\ell_w$ (here 5 \AA) and divide by $\ell_w$ as well as the bulk value of total water. Additionally, as one needs to have it end at a specific distance, the water curves were used to construct splines, which can be numerically integrated over the specific domain with a refined spacing.  
\end{enumerate}

\begin{figure}[h!]
     \centering
     \includegraphics[width= 0.6\textwidth]{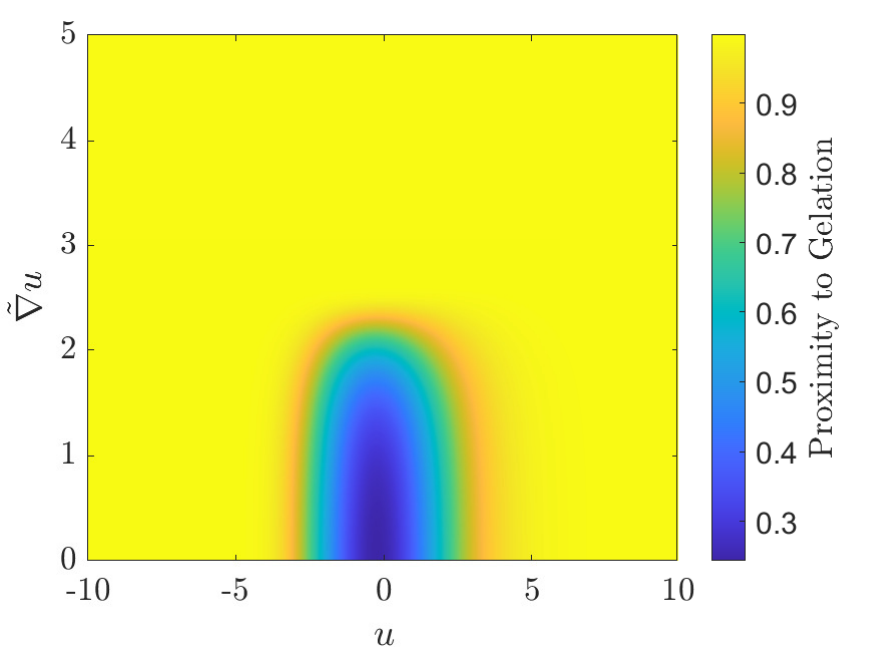}
     \caption{Numerical solution map for $1-(f_+-1)(f_--1)\bar{p}_{+-}\bar{p}_{-+}$ in our system of equations under the sticky-cation formalism for 15m WiSE. Shown here is how the proximity to gelation varies with the dimensionless electrostatic potential ($u$) and the negative dimensionless electric field strength ($\tilde{\nabla}u$). Here we use $f_+=4$, $f_-=3$, $\xi_{0}=1$, $\xi_+=0.4$, $\xi_-=10.8$, $\epsilon_r = 10.1$, $\lambda = 0.231$, $P$ = 4.995 Debye, and $v_0 = 22.5$ \AA$^3$.}
     \label{fig:nsm_Prox}
\end{figure}

As just discussed earlier, the first step towards obtaining our theory's predictions is to generate numerical solutions to the system of equations. Examples of these maps for 15m water-in-LiTFSI under the sticky-cation formalism are shown below in Fig.~\ref{fig:nsm_phisp}-\ref{fig:nsm_mpb}, along with a Fig.~\ref{fig:nsm_Prox}, which highlights how close to the gelation the system is where negative implies gelation has occurred.

\clearpage

\subsection{Additional Theory Predictions}
In the main text, we highlighted the ability of our theory to predict the increase in aggregation due to electric field-induced enhancement in associations. Shown below in Fig.~\ref{fig:15msZoom} is a magnified view of this enhancement in 15m water-in-LiTFSI under the sticky cation approximation near a negatively charged electrode.
\begin{figure}[h!]
     \centering
     \includegraphics[width= 0.7\textwidth]{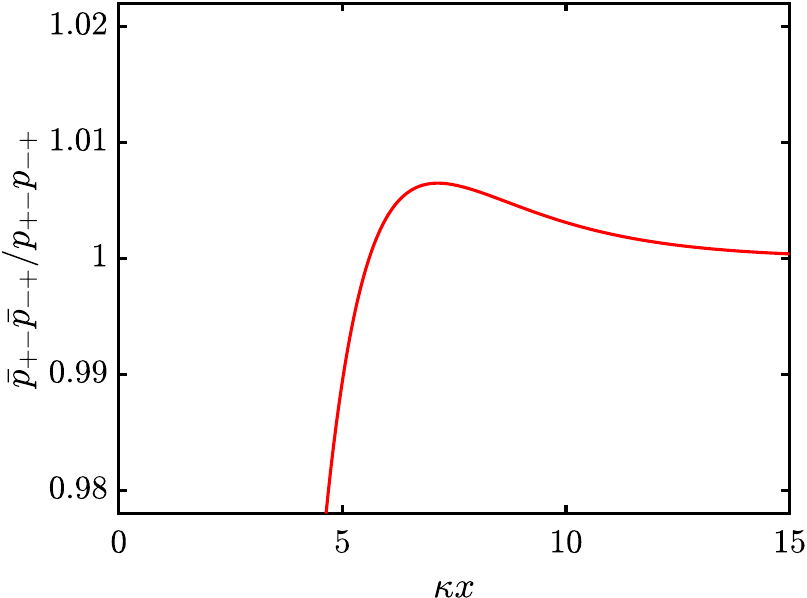}
     \caption{Electric field induced enhancement in associations in 15m WiSE. Shown is the product of the ionic association probabilities, $\bar{p}_{+-}\bar{p}_{-+}$, normalized by its bulk value $p_{+-}p_{-+}$. This figure is plotted only close to the charged surface to highlight the effect but was solved over the same regime as the figures comparing the MD simulation's spatial EDL profiles to the theory's predictions in the main text. Here we use $f_+=4$, $f_-=3$, $\xi_{0}=1$, $\xi_+=0.4$, $\xi_-=10.8$, $\epsilon_r = 10.1$, $\lambda = 0.231$, $P$ = 4.995 Debye, $v_0 = 22.5$ \AA$^3$, and q$_s$ = -0.2 C/m$^2$.}
     \label{fig:15msZoom}
\end{figure}

\begin{figure}[h!]
     \centering
     \includegraphics[width= .9\textwidth]{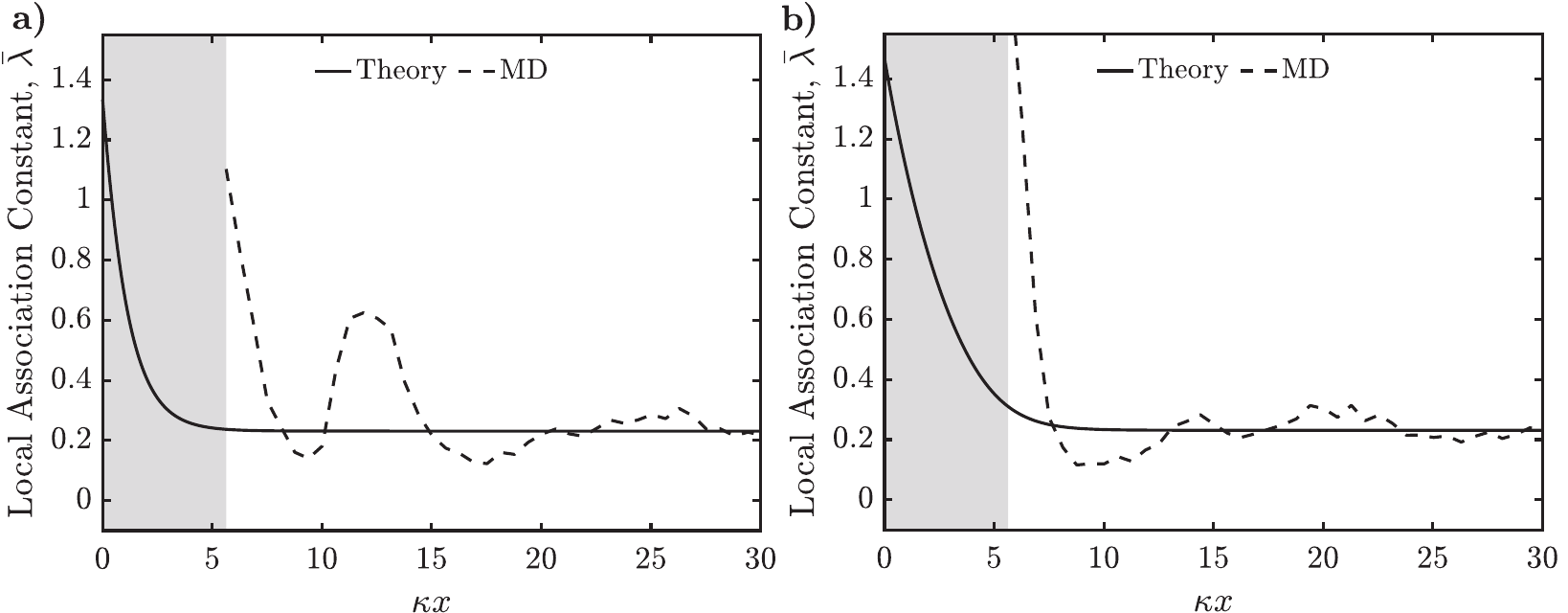}
     \caption{Local association constant ($\bar{\lambda}$) through the EDL in WiSEs. a) $\bar{\lambda}$ of 15m WiSE at q$_s$ = -0.2 C/m$^2$ as a function of distance from the interface in dimensionless units, where $\kappa$ is the inverse Debye length. b) $\bar{\lambda}$ of 15m WiSE at q$_s$ = 0.2  C/m$^2$ as a function of distance from the interface in dimensionless units. Here we use $f_+=4$, $f_-=3$, $\xi_{0}=1$, $\xi_+=0.4$, $\xi_-=10.8$, $\epsilon_r = 10.1$, $\lambda = 0.231$, $P$ = 4.995 Debye, and $v_0 = 22.5$ \AA$^3$.}
     \label{fig:Lambda_bar}
\end{figure}

Next, let us discuss how the local association constant ($\bar{\lambda}$) varies through the EDL. As highlighted in the main text, the general trends in the $\bar{\lambda}$'s spatial profiles are consistent between the MD and the theory, with even the magnitudes agreeing to a reasonable degree. This result is shown for 15m LiTFSI at a surface charge density of $\mp$0.2  C/m$^2$ in Fig.~\ref{fig:Lambda_bar}. This agreement further supports the utility of our theory. Two main deviations between the theory and MD results can be seen in Fig.~\ref{fig:Lambda_bar}. First, at the condensed layer, the MD appears to diverge or become undefined. This deviation is expected given the structure of the condensed layer, as co-ion concentrations vanish which leads to ill-defined association probabilities. Second, there are oscillations in the MD's $\bar{\lambda}$ not seen in the theory's prediction. Once again, this deviation is expected as the theory is a local formulation limiting the model from capturing oscillations in $\bar{\lambda}$ that are likely caused by electric field oscillations or species layering. Nonetheless, the model's ability to capture the trends in $\bar{\lambda}$'s spatial profile and order of magnitude is a promising result.

\begin{figure}[h!]
     \centering
     \includegraphics[width= .9\textwidth]{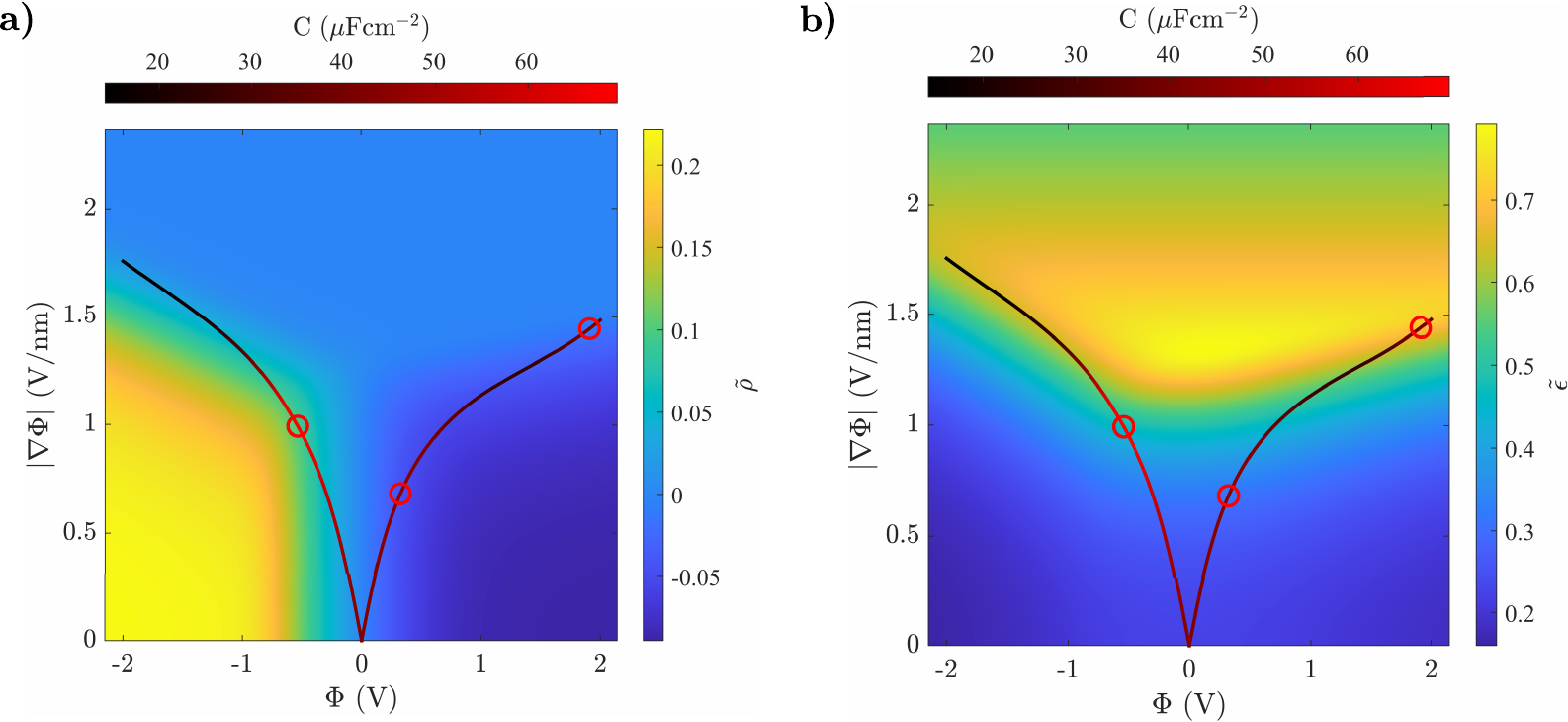}
     \caption{Differential capacitance profile (C) embedded into dimensionless charge density ($\tilde{\rho}$) and dimensionless dielectric constant ($\tilde{\epsilon}$) meshes for WiSE. This projection was constructed for 15m WiSE using the sticky-cation formalism. Local maximums in the differential capacitance profiles are marked by a red circle: a) Differential capacitance profile embedded in the dimensionless charge density ($\tilde{\rho}$). b) Differential capacitance profile embedded in the dimensionless dielectric constant ($\tilde{\epsilon}$). The maps are shown for variations in the electrostatic potential ($\Phi$) and electric field strength ($\nabla\Phi$). Here we use $f_+=4$, $f_-=3$, $\xi_{0}=1$, $\xi_+=0.4$, $\xi_-=10.8$, $\epsilon_r = 10.1$, $\lambda = 0.231$, $P$ = 4.995 Debye, $v_0 = 22.5$ \AA$^3$, and $\alpha$ = 0.1.}
     \label{fig:Cap_Peak_Asso}
\end{figure}

One can embed theory results like the differential capacitance into constructed meshes such as those shown in Fig.~\ref{fig:nsm_phisp}-\ref{fig:nsm_Prox} gaining insight into the system. For example in Fig.~\ref{fig:Cap_Peak_Asso}, the embedding provides insight into the factors that give rise to the local peaks in the differential capacitance profile. This embedding can be done with all the meshes for results with the same parametrization and for results that depend on the electrostatic profile. In this case, we focus on how the non-dimensional charge density (Fig.~\ref{fig:Cap_Peak_Asso}.a) and dielectric constant (Fig.~\ref{fig:Cap_Peak_Asso}.b) vary throughout the differential capacitance profile. First consider the peak at moderate negative potential, it occurs in a region of cation enrichment and has some dielectric enhancement that leads to the large peak we see at the same location in the differential capacitance. Second consider the peak at moderate positive potential, it occurs in a region of anion enrichment but with less significant dielectric enhancement leading to the smaller peak we see at this same location in the differential capacitance. Third consider the peak at large positive potential, it occurs in a region of decreasing non-dimensional charge density and strongly increasing non-dimensional dielectric constant suggesting that this is a region of free water enrichment with significant dielectric enhancement. Additionally, these observed features can be further supported by considering other meshes similar to those shown in Fig.~\ref{fig:nsm_phisp}-\ref{fig:nsm_mpb}. One can also utilize these meshes by constructing additional local quantity maps from these fundamental ones. For example, one could construct maps of the concentration of free water or ion pairs to provide additional insight into these systems. Moreover, one could also use these new meshes to develop intuition on key features in predicted profiles. This concept is demonstrated here as we embedded the differential capacitance profile into the non-dimensional charge density (Fig.~\ref{fig:Cap_Peak_Asso}.a) and dielectric constant (Fig.~\ref{fig:Cap_Peak_Asso}.b) meshes.

\begin{figure}[h!]
     \centering
     \includegraphics[width= .75\textwidth]{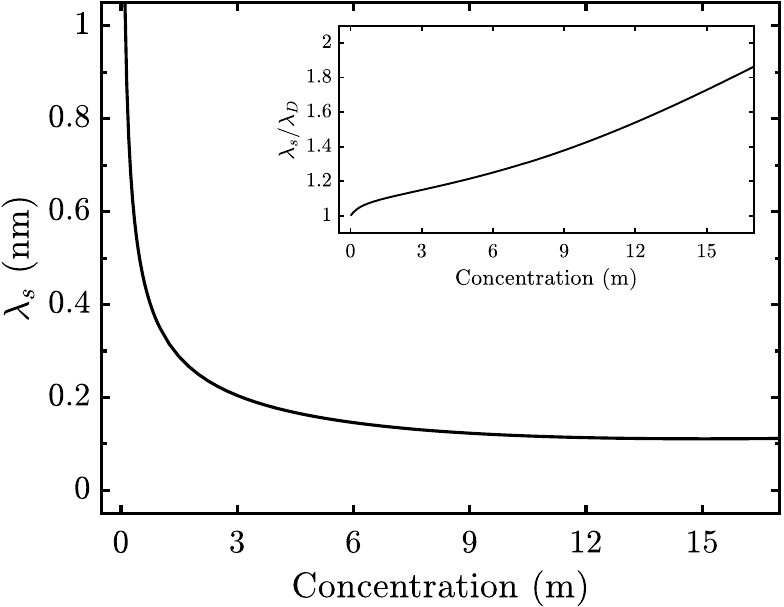}
     \caption{Screening length prediction for WiSEs. Screening length of WiSE as a function of molality. The inset shows the screening length normalized by the formal Debye length, $\lambda_D = \sqrt{v_0\epsilon/e^2\beta(c_++c_-)}$, as a function of molality. Here we use $f_+=4$, $f_-=3$, $\xi_{0}=1$, $\xi_+=0.4$, $\xi_-=10.8$, $\epsilon_r = 10.1$, $\lambda = 0.228$, $P$ = 4.995 Debye, and $v_0 = 22.7$ \AA$^3$.}
     \label{fig:ScreenL}
\end{figure}

Now turning to the screening length, $\lambda_s$, varies with the molality in the pre-gel regime. In Fig.~\ref{fig:ScreenL}, we can observe that the screening length decrease as the molality of LiTFSI increases up to 9~m, after which it is relatively constant. The enhancement of the screening length due to the associations can be seen in the inset of Fig.~\ref{fig:ScreenL} where $\lambda_s/\lambda_D$ is always greater than, or equal to, one. Additionally, we can observe around 9m that the contributions to the screening length from the associations begin to increase, leading to the screening length increasing more strongly with concentration.

\section{Extended Simulation Section}
Presented here is an in-depth explanation and discussion of the analysis of the molecular dynamic (MD) simulations as well as the implementation of the sticky-cation formalism. For the MD simulations presented here, we utilized the same molecular dynamics procedure and force fields used in Ref.~\citenum{mceldrew2018}. 

\subsection{Molecular Dynamics Simulation Methodology}

Here we performed classical atomistic MD simulations using LAMMPS~\cite{plimpton1995}, following the methodology outlined in Ref.~\citenum{mceldrew2018}, which we will briefly recap. Our simulations were for LiTFSI in water at concentrations of 12~m and 15~m in a slit geometry in contact with charged interfaces.

We simulated the EDL of this WiSE in the NVT ensemble at 300 K, where the geometry of the cell was taken to be $33\times33\times266$ \AA$^3$, with two $33\times33\times33$ \AA$^3$ electrodes sandwiching the electrolyte region was made up of fixed Lennard Jones (LJ) spheres arranged in an fcc lattice (100). For 15~m, the box contained 636 ion pairs, 2356 water molecules, and 4096 electrode atoms. For 12~m, the box contained 588 ion pairs, 2725 water molecules, and 4096 electrode atoms. The initial configurations for all simulations were generated using the open-source software, PACKMOL~\cite{martinez2009packmol}. Surface charges of $\pm 0.2$ C/m$^2$ were applied by placing partial charges on the first layer of the electrode atoms. Additionally identical simulations were conducted for 15~m with surface charge of $\pm 0.15$ C/m$^2$, $\pm 0.1$ C/m$^2$, $\pm 0.05$ C/m$^2$, and $0$ C/m$^2$.

For all Li$^+$ and TFSI$^-$ we employed the CL$\&$P force field~\cite{lopes2012}. For water, we employed the spc/e force field. Inter-atomic interactions are determined using Lorentz-Berthelot mixing rules. For the electrode, we did not explicitly model the dynamics, omitting the need for an `electrode'-`electrode' force field. The electrode only interacts with the fluid through coulomb and Lennard-Jones interactions, which were made to be the same no matter what atom is interacting with the electrode atom with LJ well depth $\varepsilon = 0.001\text{eV}$ and LJ well distance $\sigma = 3$ \AA. Long-range electrostatic interactions were computed using the Particle-Particle Particle-Mesh (PPPM) solver with cut-off of 12 \AA, which maps particle charge to a 2D mesh in the transverse direction for the nano-slit simulation\cite{hockney1988}. 

Equilibration runs of about 12 ns (1 fs time steps) were performed initially with no applied potential/charge. Then the surface charge was ramped up from zero, allowing for 12 ns of equilibration, and 4 (2)~ns of production at each electrode surface charge for 15~m (12~m) collecting frames every 4 (2)~ps, giving a total of 1000 frames in each case.

\subsection{Parameters for Theory and Analysis of Simulation Data}\label{ASD}

\begin{figure}[h!]
     \centering
     \includegraphics[width= .63\textwidth]{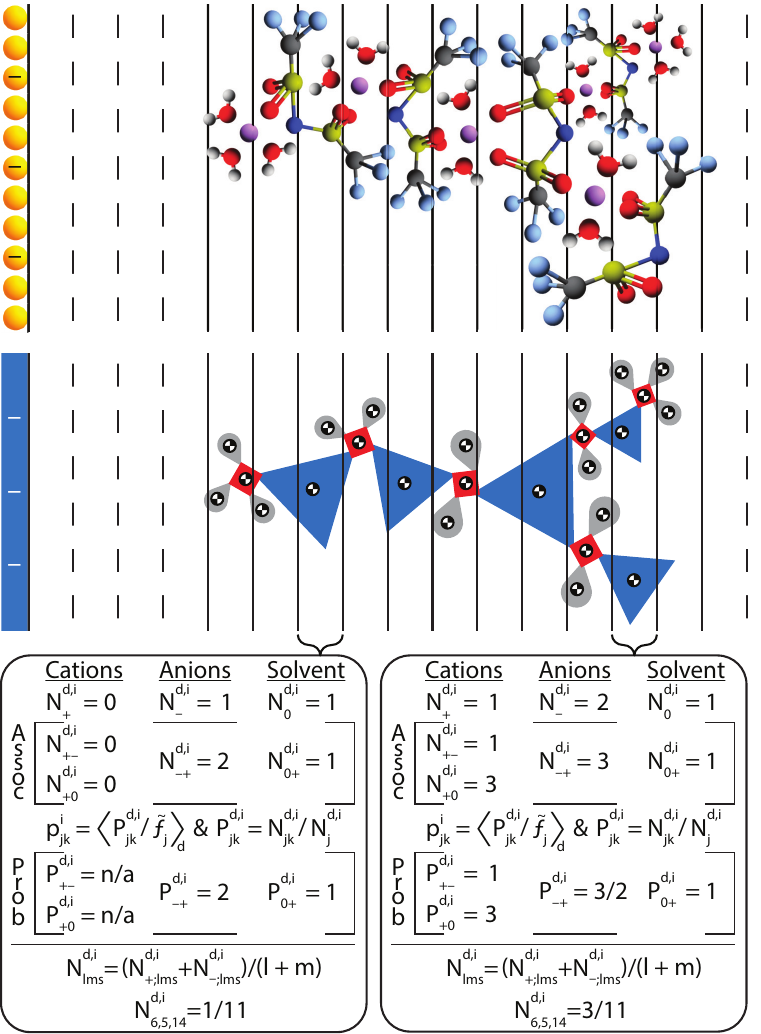}
     \caption{Schematic of Sample Molecular Dynamics Simulation Analysis. Here the top layer shows what a single cluster would look in our MD simulation. The second layer highlights how we decomposed the cluster, by localizing associations and positions to their center of masses. N$_j^{d,i}$ indicates the number of species type $j$ that are found in the $i^{th}$ single partition/bin for the $d^{th}$ data frame. N$_{jk}^{d,i}$ represents the number of associations from species $j$ to species $k$ localized to the center of mass of species $j$. The number of species and associations can then be used by N$_j^{d,i}$/N$_{jk}^{d,i}$ to obtain the average association number P$_{jk}^{d,i}$. This quantity is used to find the average association number per bin later by averaging it overall data frames and dividing by its functionality, $\tilde{f}_j$. Here we use $\tilde{f}_j$ since it can depart from the functionality when analyzing under different approximations like the sticky-cation formalism. Lastly, we calculate the approximate number density per bin of a $lms$-ranked cluster N$_{lms}^{d,i}$ via the fraction of that clusters ionic backbone present in the bin.}
     \label{fig:MD_analysis}
\end{figure}

To model a given WiSE, one needs to obtain predictions for the volume ratios ($\xi_+$ \& $\xi_-$), the functionality of the cations and anions ($f_+$ \& $f_-$), and association constants ($\lambda_{+-}$ \& $\lambda_{+0}$). We will discuss how these are calculated, and then explain how quantities within the EDL are computed, before moving to screening lengths and integrated quantities. A detailed depiction of the analysis of the MD data is shown in Fig.~\ref{fig:MD_analysis}.

The volume ratios can be determined by summing over the van der Waals spheres of the elements that comprise the molecules, ensuring not to double count for overlapping contributions, and normalizing this volume by the volume of water ($v_0$). In this work for the volume fractions, we use the same values found in Ref.~\citenum{mceldrew2021ion}, $\xi_+ = 0.4$ \& $\xi_- = 10.8$.

To compute the associations in this MD simulation, we first identified the threshold distance for which an association would be classified. This can also be accomplished via studying the spatial distribution functions of the associating molecules and counting the number of ``hot-spots" that are present~\cite{mceldrew2020corr}. Alternatively, one can utilize kinetic criteria~\cite{feng2019free} or machine learning methods~\cite{jones2021bayesian} to define associations between species in the electrolyte. Following the previous analysis of this type of system~\cite{mceldrew2018}, with the same simulation procedure and force fields, the association threshold was determined to be 2.7 \AA. This means that if Li$^+$ is within 2.7 \AA\ of an oxygen atom belonging to TFSI$^-$ or H$_2$O, an association was present. Note we do not consider any association between water and TFSI$^-$ as these are rare in WiSEs considered here~\cite{mceldrew2018,mceldrew2021ion}.

The coordination number of each Li$^+$ was obtained by counting the number of Li-O (Water) associations as well as Li-O (TFSI$^-$). Here, if multiple oxygens' from a TFSI$^-$ associate to a single Li$^+$, it was only counted once, i.e., bi-dentate or multi-dentate associations are only counted as one. This kind of procedure was also used to evaluate the number of associations each Li$^+$ had. Similarly, the number of Li$^+$ associated to a single water molecule's oxygen was extracted. Lastly, we repeated this procedure for the TFSI$^-$, but here, we only counted the associations by the number of unique Li$^+$ associated to the oxygen atoms belonging to the TFSI$^-$. In all these cases, the associations were assigned to the molecule's center of mass, which was calculated directly from the position data. This approximation was made to treat the individual molecules discretely and at one point in space. By doing this, however, it means the associations that are spread out over space are localized to the center of mass.

\begin{figure}[h!]
     \centering
     \includegraphics[width= .6\textwidth]{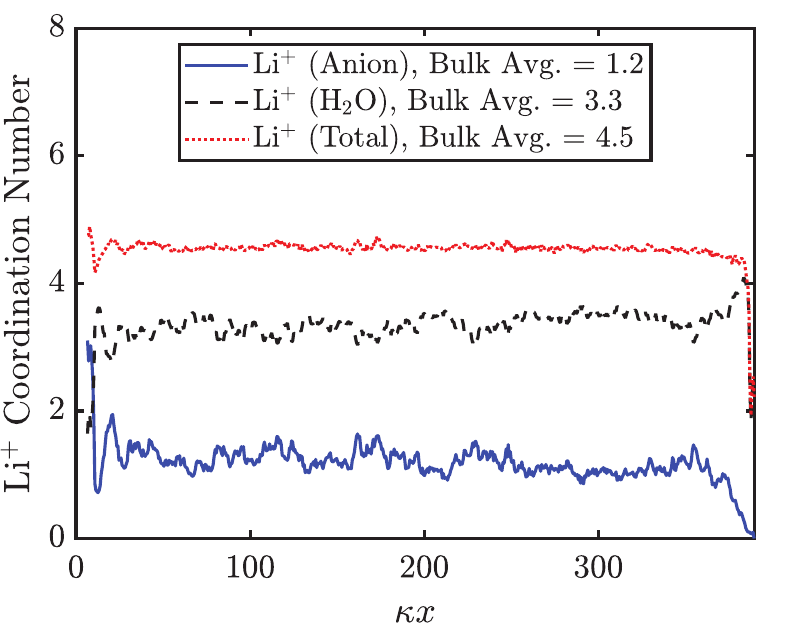}
     \caption{Lithium Coordination Number for 12~m water-in-LiTFSI for electrodes with $q_s =\pm$0.2 C/m$^2$ surface charge. Here the left side of the plot corresponds to the interface with $q_s = $~0.2 C/m$^2$ and the right side with $q_s =-$0.2 C/m$^2$.}
     \label{Coord12m}
\end{figure}

\begin{figure}[h!]
     \centering
     \includegraphics[width= .6\textwidth]{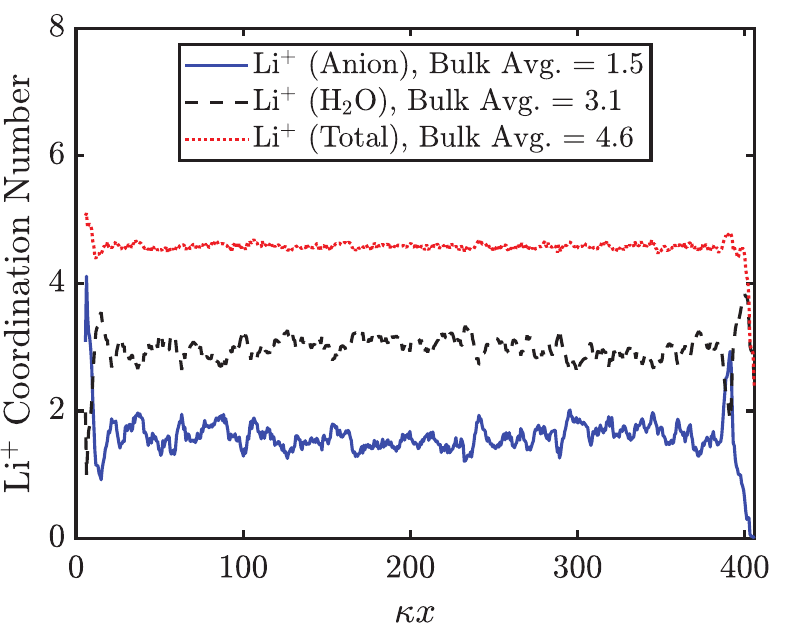}
     \caption{Lithium Coordination Number for 15~m water-in-LiTFSI for electrodes with $q_s =\pm$0.2 C/m$^2$ surface charge. Here the left side of the plot corresponds to the interface with $q_s = $~0.2 C/m$^2$ and the right side with $q_s =-$0.2 C/m$^2$.}
     \label{Coord15m}
\end{figure}

Considering the total coordination number of Li$^+$ in the bulk, for 12~m and 15~m water-in-LiTFSI with electrodes with $\pm$0.2 C/m$^2$ surface charge in Fig.~\ref{Coord12m} and Fig.~\ref{Coord15m} respectively, one can note on average that they form more than four associations at both molalities. From this finding the Li$^+$ functionality should be $f_+$ = \{4,5\}. Previously in Ref.~\citenum{mceldrew2021ion}, they found that Li$^+$ in water-in-LiTFSI for a range of molalities had a coordination number slightly greater than four, leading to them choosing $f_+$ = 4. The functionality of the TFSI anions has been previously studied by analyzing their spatial distribution function and how their moieties interact with the lithium ions; it is set as $f_-=3$~\cite{mceldrew2021ion}. This comes from the partial negative charge distributed among the oxygen atoms of the TFSI$^-$ leading to the association being formed between Li-O. However, typically, when three Li$^+$ are associated with TFSI$^-$, it leads to two oxygen atoms being associated on a single Li$^+$. However, there are rare cases where four Li$^+$ can be associated with a single TFSI$^-$, but since they are rare, it is justifiable to set $f_-$=3.

Following Ref.~\citenum{mceldrew2021ion}, we calculate the association probabilities through
\begin{equation}
    \label{sim_prob}
    p_{ij} = \left<\frac{\# \text{ of associations of type } ij}{f_i \cdot \# \text{ of molecules of type } i}\right>
\end{equation}
where there is an average temporally, and over space if in the bulk region. We also compute these probabilities as a function of position, where the normalization has to occur over only those bins/partitions that have species present. The process of extracting the association probabilities spatially is highlighted in Fig.~\ref{fig:MD_analysis}. Note as shown a partition (or bin) can only produce a sample if the respective species are present, i.e., if no Li$^+$ is present, one cannot obtain a sampled value for $\bar{p}_{+-}$ and $\bar{p}_{+0}$. This makes sampling the association probabilities of the counter-ions in the EDL more challenging given the lower statistics available given their rare appearance close to the interface. In the sticky cation approximation, $f_+$ is replaced with the bulk coordination number of Li$^+$, to ensure the association probabilities are bounded with 0 and 1. Note, it is still possible to have a probability greater than one as uncommon coordination structures can exist in the EDL. An effect of this modification is that it slightly adjusts the extracted $\lambda$ used for the sticky cation approximation.

From these association probabilities one can directly calculate the product of the ionic association probabilities, $\bar{p}_{+-}\bar{p}_{-+}$. Additionally, one can extract the variance of these sampled values over the production period. From this calculation and using the propagation of error, one can calculate the standard deviation for $\bar{p}_{+-}\bar{p}_{-+}$.

As discussed in the work of McEldrew \textit{et al.}~\cite{mceldrew2020theory,mceldrew2021ion}, the association constants can be obtained from Eqs.~\eqref{CAL}-\eqref{CSL} using the previously computed ensemble average association probabilities. An equivalent procedure was utilized for the sticky case to extract $\lambda$ using Eq.~\eqref{lambdaeqSM} shown below,
    \begin{equation}
    \label{lambdaeqSM}
        \lambda = \frac{\lambda_{+-}}{\lambda_{+0}} = \frac{p_{-+}(1-p_{0+})}{p_{0+}(1-p_{-+})}.
    \end{equation} 
From this analysis we found that for 12~m $\lambda_{+-} = 24.1\pm0.66$ \& $\lambda_{+0} = 106\pm1.7$ and 15~m $\lambda_{+-} =  36.9\pm1.7$ \& $\lambda_{+0} = 159\pm5.18$. Additionally under the sticky cation approximation, we found that 12~m $\lambda = 0.226\pm0.0066$, 15~m $\lambda = 0.231\pm0.0066$. Alternatively, one can obtain predictions for these association constants without fitting from MD simulations by using integral equations and Wertheim's formalism~\cite{wertheim1984fluids1,wertheim1984fluids2,wertheim1986fluids1,wertheim1986fluids2,laria1990cluster,blum1995general,simonin1999ionic,sciortino2007self}. The main drawback to this fitting-free approach is that can only be applied to certain simple cases~\cite{Goodwin2023}.

To calculate the aggregates, we construct an adjacency matrix. This matrix represents the connectivity of the ionic species by labeling the species' local associations. Analyzing this adjacency matrix allows us to establish the $lms$ rank of the cluster to which the ionic species and water belong. This information allows one to label each ionic species with its cluster identity; a similar treatment could be done for water or solvent molecules.

Next, we turn to describing in more detail how we calculate EDL quantities. As described above, we have outlined how the coordination numbers of Li$^+$ are dealt with, through localizing the associations on a Li$^+$. In what follows, we describe how we partition species and associations into bins as a function from the interface, which can then be compared against our theory.

Firstly, the volume fraction of each species was obtained by first calculating the average number of each species in a bin/partition. From this, we invoked the incompressibility compressibility constraint to normalize the total species volume by the total volume from the number density profiles to obtain the spatial species volume fractions profiles in the EDL. Here we selected bins to minimize the numerical artifacts of spatial profiles typically seen in larger bins. This leads to small bins with widths around 0.3 \AA\ selected as a result of the overly small Debye length in these concentrated electrolytes. This partitioning is in line with previous studies~\cite{mceldrew2018}.

An equivalent method was used to find the volume fraction of each cluster, where first, we obtained the average number of each type of cluster in each partition. Here, the cluster was segmented by the cations and anions that composed them as they can exist across multiple partitions; hence, accounting for the fractional amount of the cluster in each partition allowed for a finer partition to be utilized. This is still an approximation as the bound water molecules are effectively shrink-wrapped to the ionic backbone in this method. Following the previous method, the volume of each cluster in the partition is normalized by the total volume in the partition, producing the volume fraction of each cluster.

To find the volume fraction of each cluster, where first, we obtained the average number of each type of cluster in each partition/bin. Here, the cluster was segmented by the cations and anions that composed them as they can exist across multiple partitions; hence, accounting for the fractional amount of the cluster in each partition allowed for a finer partition to be utilized. We achieved this here by counting what fraction of the ionic species that made up a cluster is found in that partition, highlighted in Fig.~\ref{fig:MD_analysis}. If we required the partitions to contain the entire clusters, then only coarse partitioning of the MD simulation cell could be used. This is still an approximation as the bound water molecules are effectively shrink-wrapped to the ionic backbone of the cluster in this method, i.e. not explicitly counted for ion-containing clusters. Following the previous method, the volume of each cluster in the partition is normalized by the total volume in the partition, producing the volume fraction of each cluster. Note a more exacting way to analyze these clusters would be to forgo extracting the number density of the clusters initially. Through this approach, one would label all the species with the cluster to which they belong. Then, one can calculate what fraction of the volume in the partition they contribute from the total species volume fractions. Hence, this approach directly provides the volume fraction of each cluster in a partition without needing the number of densities. One can still extract the number densities by back-calculating from $\bar{\phi}_{lms}$ to $\bar{c}_{lms}$. 

One needs to extract from the simulations the site size, which in the theory is set to the size of a water molecule ($v_0$). To determine this value, one first calculates the volume of the electrolyte simulation cell ($V$) from the initialization of the simulation cell. Then, by utilizing the incompressibility criteria, one can find, 
\begin{equation}
    \label{Incomp_exp}
    1 = \phi_+ + \phi_- + \phi_0 = \frac{v_0}{V}\left(\xi_+ N_+ + \xi_- N_- + N_0\right).
\end{equation}
Since the number of molecules is known for each simulation's molality, and the electrolyte volume can be extracted, one can explicitly solve for $v_0$ from Eq.~\eqref{Incomp_exp}. Using the volume ratios, as they are independent of the MD, the number of each species, and the volume of a single site ($v_0$), we can define the bulk species volume fractions for the simulations. This composition represents the theory-equivalent bulk solution that can be tested against the simulation results. 

From these volume fractions and the bulk dielectric constant found previously to be $\epsilon_r=$10.1 for water-in-LiTFSI~\cite{mceldrew2018}, one can calculate the Debye length ($\lambda_D$) for the simulations and the theory for which the distance from the charged interface is normalized by the inverse Debye length ($\kappa$). Performing this analysis for the 12~m and 15~m cases and comparing against theory, we found that for 12~m $v_0$ = 22.9 \AA$^3$ with $\phi_+$ $\approx$ 0.0253 and for 15~m $v_0$ = 22.5 \AA$^3$ with $\phi_+$ $\approx$ 0.0268. These values for the site volume roughly align with the volume of a single water molecule and fall roughly around that of the volume a bulk water molecule would take up, albeit slightly smaller, which is consistent with the more concentrated nature of the WiSEs.

To calculate the length scale of the aggregates, one can utilize the average volume fraction of each type of cluster in each partition to obtain $\bar{c}_{lms}$ in these partitions. As MD is restricted to finite domains, this restricts any potential gelation to existing as finite sized clusters, albeit with a large number of loops. Therefore we can explicitly use the pre-gel regime version of the length scale of the aggregates equation:
    \begin{equation}
        \label{LSAggSM}
        \ell_{A}^3 = \frac{v_0\sum_{lms}(\xi_+l+\xi_-m+s)^2 \bar{c}_{lms}}{\sum_{lms}(\xi_+l+\xi_-m+s) \bar{c}_{lms}}= v_0\sum_{lms}(\xi_+l+\xi_-m+s)^2 \bar{c}_{lms}.
    \end{equation}

Thus by explicitly doing these summations, the length scale of the aggregates in each partition can be calculated. For simplicity, these calculations were conducted using the pre-gel regime formula, although the oscillations in the EDL predict the segments of the system will gel. This treatment simplifies the comparison against the theory and experimental data.

Turning our attention to the integrated quantities. The excess surface concentrations, shown in Eq.~\eqref{exsurfcSM} where $\infty$ is the middle of the simulation cell, can be found by integrating the difference between the concentration of each species by their theoretical bulk reservoir, from the surface of each charged interface to the center of the simulation box. This analysis is done through spline interpolation and then integration of our numerical concentrations. Additionally, we exclude the depletion region where no species are present. This methodology allows us to compare these results directly against the theory without needing to implement a depletion region shift into the theory to make the extracted measurements have an equivalent meaning. Note that this treatment means that the left and right electrodes at no charge can produce different excess surface concentrations, and thus, these two values are averaged to obtain the reported one.
    \begin{equation}
        \label{exsurfcSM}
        \Gamma_i(q_s) = \frac{1}{v_0}\int_0^{\infty} \left(\bar{c}_i (x,q_s)-c_i^{\text{bulk}}\right)\,dx,
    \end{equation}

Lastly to calculate the interfacial concentration of water, shown in Eq.~\eqref{wsurfSM} where $\ell_w$=5 \AA\, one needs the bulk value of total water to normalize this quantity and the spatial profiles to integrate over. The bulk value was obtained from the middle third of the simulation cell, where we calculated the concentration of the total water molecules. Splines interpolate the profile to allow for numerical integration from the surface of the charged interface to 5 \AA\ away from it. Enabling an analogous comparison against the MD prediction in the theory, the same depletion region in the MD is used in the theory, i.e. the region where no species are found near the charged interface. Since the theory was not developed to account for steric effects at the interfaces, which would create a deletion region, a synthetic one from the MD was incorporated. This shift was performed to enable a more direct comparison of the simulations and the theory's predictions. Lastly, as the shift values extracted at zero surface charge were not equal, their values were averaged.
  \begin{equation}
        \label{wsurfSM}
        \tilde{\rho}_{w,n}^{ads}(q_s) = \frac{\int_0^{\ell_w} \bar{c}_n(x,q_s)\,dx}{\ell_wc_0^{\text{bulk}}}.
    \end{equation}

\subsection{Stick Cation Testing and Validation for 12~m Water-in-LiTFSI}
 
\begin{figure}[h!]
     \centering
     \includegraphics[width= .89\textwidth]{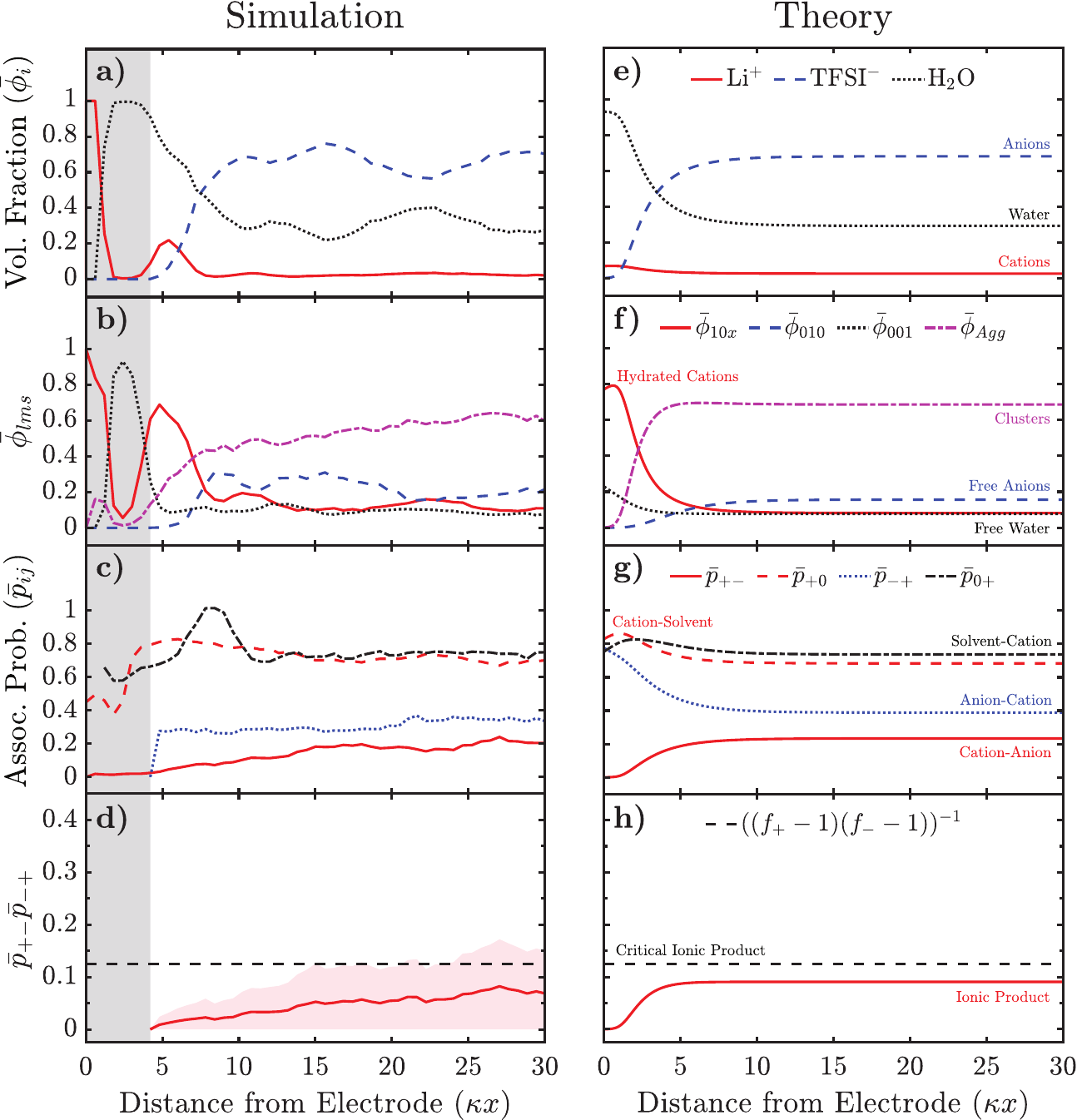}
     \caption{Distributions of properties of general 12m WiSEs in the EDL as a function from the interface, in dimensionless units, where $\kappa$ is the inverse Debye length. a-d) are the results from MD simulations, and e-h) are the corresponding predictions from theory. The gray region indicates the minimum distance from the electrode at which a species was never found. a,e) Total volume fraction of each species. b,f) Volume fractions of hydrated cations, free anions, free water, and aggregates. c,g) Association probabilities. d,h) Product of the ionic association probabilities, $\bar{p}_{+-}\bar{p}_{-+}$, where the dashed line indicates the critical line for gelation. Here we use $f_+=5$, $f_-=3$, $\xi_{0}=1$, $\xi_+=0.4$, $\xi_-=10.8$, $\epsilon_r = 10.1$, $\lambda_{+-} = 24.1$, $\lambda_{+0} = 106$, $P$ = 4.995 Debye, $v_0 = 22.9$ \AA$^3$, and q$_s$ = -0.2 C/m$^2$.}
     \label{fig:12mnsneg}
\end{figure}

\begin{figure}[h!]
     \centering
     \includegraphics[width= .89\textwidth]{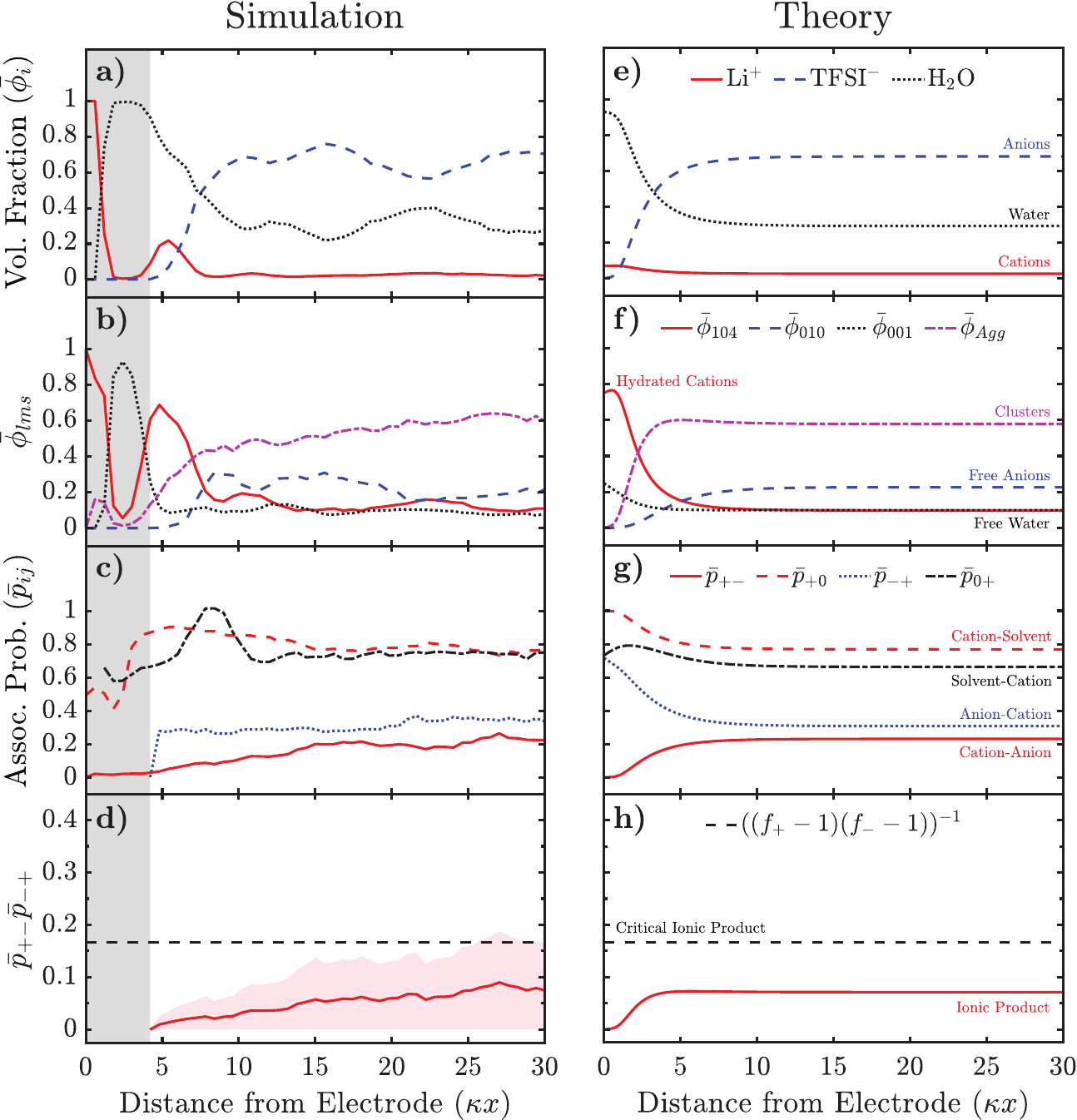}
     \caption{Distributions of properties of 12m WiSEs in the EDL using the sticky cation approximation as a function from the interface, in dimensionless units, where $\kappa$ is the inverse Debye length. a-d) are the results from MD simulations, and e-h) are the corresponding predictions from theory. The gray region indicates the minimum distance from the electrode at which a species was never found. a,e) Total volume fraction of each species. b,f) Volume fractions of hydrated cations, free anions, free water, and aggregates. c,g) Association probabilities. d,h) Product of the ionic association probabilities, $\bar{p}_{+-}\bar{p}_{-+}$, where the dashed line indicates the critical line for gelation. Here we use $f_+=4$, $f_-=3$, $\xi_{0}=1$, $\xi_+=0.4$, $\xi_-=10.8$, $\epsilon_r = 10.1$, $\lambda = 0.226$, $P$ = 4.995 Debye, $v_0 = 22.9$ \AA$^3$, and q$_s$ = -0.2 C/m$^2$.}
     \label{fig:12msneg}
\end{figure}

\begin{figure}[h!]
     \centering
     \includegraphics[width= .89\textwidth]{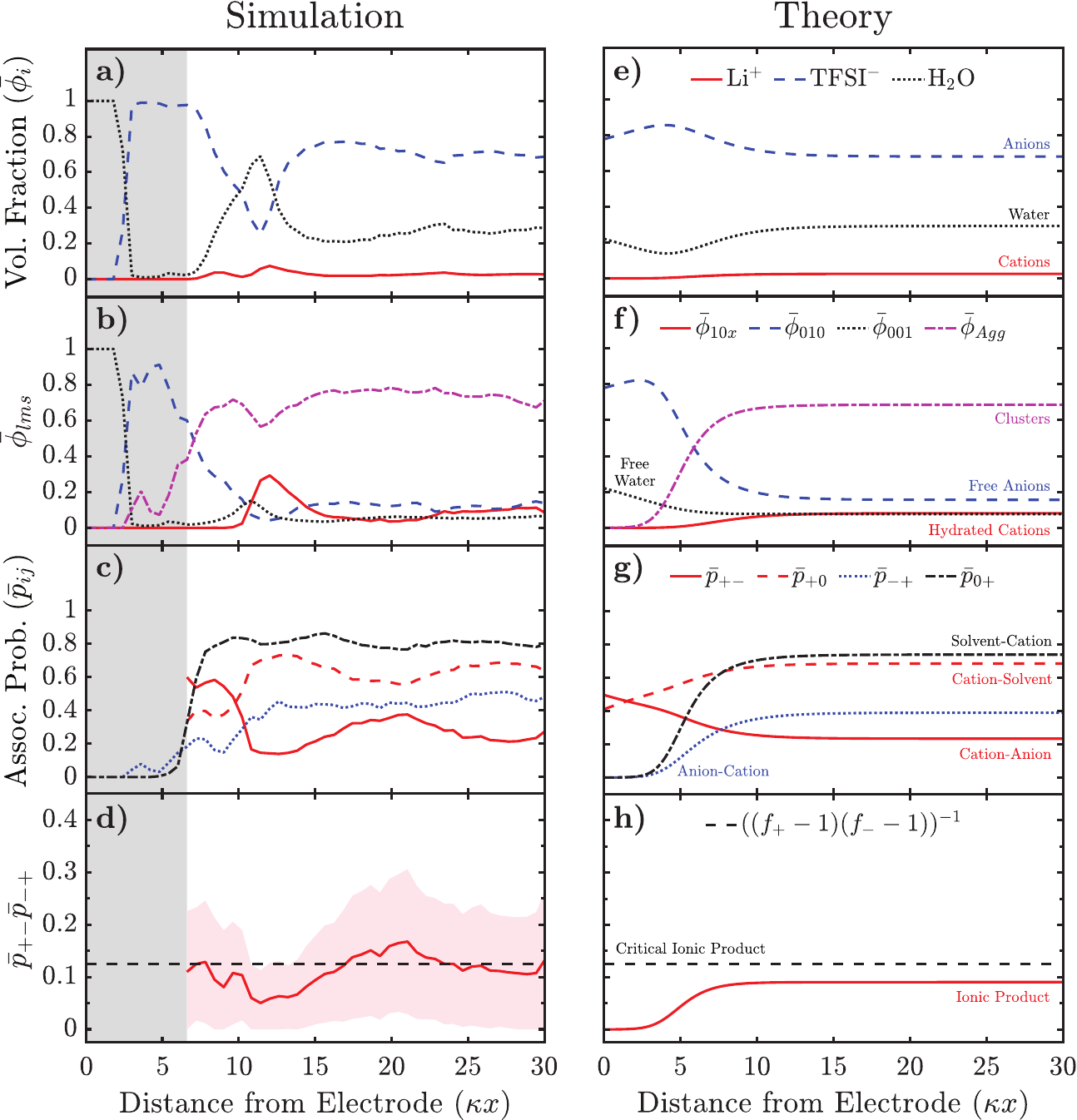}
     \caption{Distributions of properties of general 12m WiSEs in the EDL as a function from the interface, in dimensionless units, where $\kappa$ is the inverse Debye length. a-d) are the results from MD simulations, and e-h) are the corresponding predictions from theory. The gray region indicates the minimum distance from the electrode at which a species was never found. a,e) Total volume fraction of each species. b,f) Volume fractions of hydrated cations, free anions, free water, and aggregates. c,g) Association probabilities. d,h) Product of the ionic association probabilities, $\bar{p}_{+-}\bar{p}_{-+}$, where the dashed line indicates the critical line for gelation. Here we use $f_+=5$, $f_-=3$, $\xi_{0}=1$, $\xi_+=0.4$, $\xi_-=10.8$, $\epsilon_r = 10.1$, $\lambda_{+-} = 24.1$, $\lambda_{+0} = 106$, $P$ = 4.995 Debye, $v_0 = 22.9$ \AA$^3$, and q$_s$ = 0.2 C/m$^2$.}
     \label{fig:12mnspos}
\end{figure}

\begin{figure}[h!]
     \centering
     \includegraphics[width= .89\textwidth]{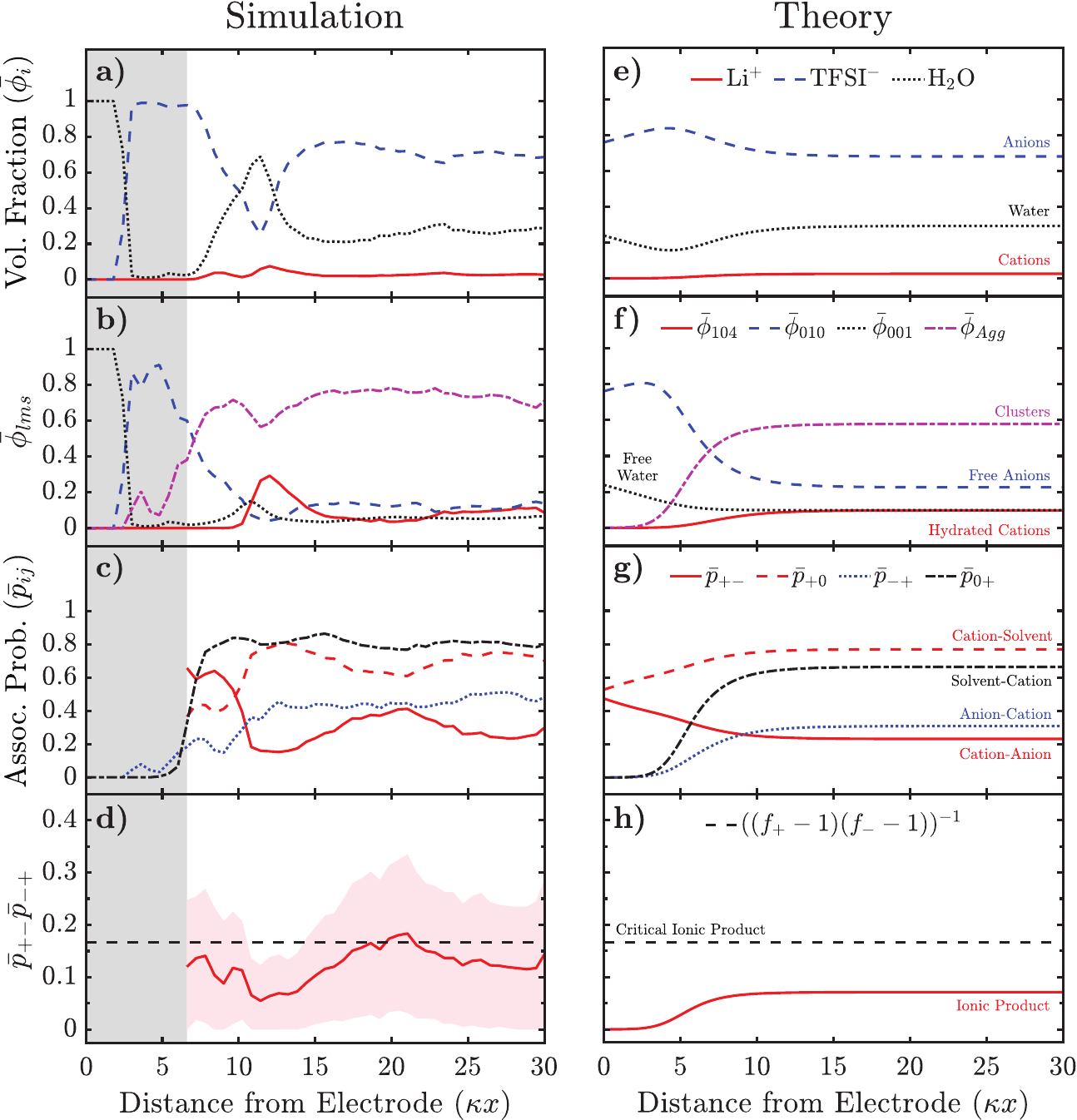}
     \caption{Distributions of properties of 12m WiSEs in the EDL using the sticky cation approximation as a function from the interface, in dimensionless units, where $\kappa$ is the inverse Debye length. a-d) are the results from MD simulations, and e-h) are the corresponding predictions from theory. The gray region indicates the minimum distance from the electrode at which a species was never found. a,e) Total volume fraction of each species. b,f) Volume fractions of hydrated cations, free anions, free water, and aggregates. c,g) Association probabilities. d,h) Product of the ionic association probabilities, $\bar{p}_{+-}\bar{p}_{-+}$, where the dashed line indicates the critical line for gelation. Here we use $f_+=4$, $f_-=3$, $\xi_{0}=1$, $\xi_+=0.4$, $\xi_-=10.8$, $\epsilon_r = 10.1$, $\lambda = 0.226$, $P$ = 4.995 Debye, $v_0 = 22.9$ \AA$^3$, and q$_s$ = 0.2 C/m$^2$.}
     \label{fig:12mspos}
\end{figure}

This section primarily focuses on the implications of making the sticky cation approximation and how it influences the theory and simulation results. For this reason, the discussion focuses mainly on the general trends and implications observed without going into fine details on the individual profiles in each of the figures as conducted in the main text. Presented here are the main EDL profiles at q$_s=\pm0.2\ $C/m$^2$ for 12~m water-in-LiTFSI, the general case plots are shown in Fig.~\ref{fig:12mnsneg} and Fig.~\ref{fig:12mnspos}, and the sticky cation approximation plots are shown in Fig.~\ref{fig:12msneg} and Fig.~\ref{fig:12mspos}.

Initially, let us discuss what is expected to change between the general case and the sticky cation approximation. In the analysis of the MD simulation's data, the only change introduced by the sticky cation approximation is replacing $f_+$ with the bulk coordination number of Li$^+$, as discussed in the previous subsection. While this does not affect the extracted species volume fractions ($\bar{\phi}_i$) or the cluster volume fractions ($\bar{\phi}_{lms}$), it does impact the association probabilities of cations to anions ($\bar{p}_{+-}$) and cations to solvent ($\bar{p}_{+0}$). For this reason, it is also expected to impact the product of the ionic association probabilities, $\bar{p}_{+-}\bar{p}_{-+}$, as well as $\lambda$, since it is calculated as a direct average instead of from the averages of $\lambda_{+-}$ and $\lambda_{+0}$. In these three cases, it is expected to scale the curves by a factor of roughly $5/4.6 \approx 1.1$, which comes from the ratio of the general functionality $f_+=5$ to the bulk coordination number of Li$^+$ 4.6. Additionally, this change from the general case to sticky cation approximation will shift the critical $\bar{p}_{+-}\bar{p}_{-+} = (f_+ - 1)(f_- - 1)$ at which gelation is predicted to occur, owing to the different functionalities. In regards to the expected changes in the theory between these two cases, it is less clear, as it is governed by the solution to a system of polynomials with large order. However, one can loosely expect the trends and profiles predicted by the theory between the two cases to be roughly similar. However, the exact values may differ slightly as $f_+$ and $\lambda$ are changing in addition to the enforcement of $\bar{p}_{+-}+\bar{p}_{+0} = 1$. The place of greatest change is expected to be in the association probabilities ($\bar{p}_{ij}$) profiles as in the sticky cation case $\bar{p}_{+-}+\bar{p}_{+0} = 1$ whereas in the general case this is not enforced. This can additional lead to changes in $\bar{p}_{+-}\bar{p}_{-+}$ if $\bar{p}_{+-}$ is strongly impacted. Similar to the changes in the simulations critical $\bar{p}_{+-}\bar{p}_{-+}$ threshold, the theory's threshold will change in the same manner.

Let us now consider the simulations for the negatively charged electrode. This can be done by comparing the simulation general case results shown in the left column of Fig.~\ref{fig:12mnsneg}) against the sticky case results in the left column of Fig.~\ref{fig:12msneg}). As expected, the species volume fraction profiles match exactly in Fig.~\ref{fig:12mnsneg}.a) and Fig.~\ref{fig:12msneg}.a). The cluster volume fraction profiles shown in Fig.~\ref{fig:12mnsneg}.b) and Fig.~\ref{fig:12msneg}.b) also match. Considering now Fig.~\ref{fig:12mnsneg}.c) and Fig.~\ref{fig:12msneg}.c) one can observe that the profiles for $\bar{p}_{-+}$ and $\bar{p}_{0+}$ are an exact match; however $\bar{p}_{+-}$ and $\bar{p}_{+0}$ in Fig.~\ref{fig:12msneg}.c) appear to be the same profiles seen in Fig.~\ref{fig:12mnsneg}.c) but scaled up by a constant. This difference is exactly as expected from the change in the MD data analysis. Lastly, one can note that the product of the ionic association probabilities, $\bar{p}_{+-}\bar{p}_{-+}$, shown in Fig.~\ref{fig:12msneg}.d) is a scaled-up by a constant factor version of $\bar{p}_{+-}\bar{p}_{-+}$ in Fig.~\ref{fig:12mnsneg}.d). Here, one can also note between Fig.~\ref{fig:12mnsneg}.d) and Fig.~\ref{fig:12msneg}.d) the critical threshold for $\bar{p}_{+-}\bar{p}_{-+}$ shifting up from the general case to the sticky cation case. Overall, the changes in the results and their profiles in the EDL do not appear to differ drastically between the general case and the sticky cation case in the MD simulations.

Next, let us consider how the theory's predictions change between the general case and the sticky cation case for the negatively charged electrode. This can be done by comparing the theory's general prediction shown in the right column of Fig.~\ref{fig:12mnsneg}) to the right column of Fig.~\ref{fig:12msneg}, which contains the theory's prediction for the sticky cation approximation. Viewing Fig.~\ref{fig:12mnsneg}.e) and Fig.~\ref{fig:12msneg}.e), one can compare the species volume fractions. In both Fig.~\ref{fig:12mnsneg}.e) and Fig.~\ref{fig:12msneg}.e), the Li$^+$ volume fraction profile as it approaches the interface appears to slowly increase before increasing faster then plateauing near the interface in nearly identical fashions. For the TFSI$^-$ volume fraction profile in both Fig.~\ref{fig:12mnsneg}.e) and Fig.~\ref{fig:12msneg}.e) appears to slowly decrease before rapidly decreasing close to the interface to a near zero value. In the case of H$_2$O volume fraction as it approaches the interface, it appears to slowly increase before rapidly increasing and plateauing near the interface in both cases as seen in Fig.~\ref{fig:12mnsneg}.e) and Fig.~\ref{fig:12msneg}.e). For the species volume fractions, both cases produced nearly identical predictions.

Considering the cluster volume fractions, one can note in both Fig.~\ref{fig:12mnsneg}.f) and Fig.~\ref{fig:12msneg}.f) that the strictly hydrated cation clusters slowly increase before rapidly increasing close to the interface before slightly decreasing from its peak value. Here while the trends are the same, the exact values close to the interface are a little different with the general case taking on slightly larger values compared to the sticky cation case. The volume fraction of bare anions ($\bar{\phi}_{010}$) appears to present the same trends in Fig.~\ref{fig:12mnsneg}.f) and Fig.~\ref{fig:12msneg}.f) where it is slowly decreasing before decreasing faster closer to the interface. Here, $\bar{\phi}_{001}$ appears to start at a notably higher value in the bulk in the sticky cation case compared to the general case, leading the sticky case's $\bar{\phi}_{010}$ profile appearing as a scaled-up version of the general case's $\bar{\phi}_{010}$ profile. The volume fraction of free water molecules ($\bar{\phi}_{001}$) seems to present the same trends in both Fig.~\ref{fig:12mnsneg}.f) and Fig.~\ref{fig:12msneg}.f) where it slow increases before quickly increasing near the interface. For the general case's $\bar{\phi}_{001}$, the profile seems to obtain a slightly lower value near the interface than in the sticky cation case. For the volume fraction of aggregates ($\bar{\phi}_{Agg}$) profile in the EDL, the two cases appear to have the same trends as displayed in Fig.~\ref{fig:12mnsneg}.f) and Fig.~\ref{fig:12msneg}.f) where it slow increases before achieving a local maximum after which it quickly decreases near the interface. For $\bar{\phi}_{Agg}$, one can note that the general case appears to start at a somewhat higher value in the bulk than in the sticky cation case. Additionally, the local maximum in $\bar{\phi}_{Agg}$ is slightly more notable in the sticky cation case compared to the general case. The trends and predictions for the cluster volume fractions appear to be similar between the general and the sticky cation cases, with minor deviations in the precise values. 

Turning to the association probabilities displayed in Fig.~\ref{fig:12mnsneg}.g) and Fig.~\ref{fig:12msneg}.g) in both cases $\bar{p}_{+-}$ appears to slowly decrease before quickly decreasing towards zero close to the interface. The trends and values seen in the $\bar{p}_{+-}$ profile appear to agree in both the general and sticky cation cases. Considering now $\bar{p}_{+0}$ in the general case shown in Fig.~\ref{fig:12mnsneg}.g), $\bar{p}_{+0}$ initially increases before quickly increasing close to the interface, then obtains its maximum value before decreasing slightly. In Fig.~\ref{fig:12msneg}.g), $\bar{p}_{+0}$ in the sticky case also initially slowly increasing before rapidly increasing near the interface before plateauing around 1. This deviation between the general case and the sticky cation case is expected as $\bar{p}_{+-}+\bar{p}_{+0} = 1$ is enforced in the sticky cation case but not the general case, allowing for the more sophisticated profile in $\bar{p}_{+0}$ to exist independent of $\bar{p}_{+-}$ in the general case. Additionally, $\bar{p}_{+0}$ in the general case appears to take on a slightly lower value in the bulk compared to the sticky cation case. As seen in both Fig.~\ref{fig:12mnsneg}.g) and Fig.~\ref{fig:12msneg}.g), $\bar{p}_{-+}$ slowly increase before quickly increasing at a decelerating rate near the interface. While the trends match, the exact profiles appear to differ by scaling factor, i.e. the sticky case seems to be a slightly scaled-down version of the general case. Last for the association probabilities, $\bar{p}_{0+}$ demonstrated the same trends in its EDL profile as shown in both Fig.~\ref{fig:12mnsneg}.g) and Fig.~\ref{fig:12msneg}.g), where it is initially slowly increasing before increasing faster near the interface and achieving a maximum value before slightly decreasing. Once again while both the general and sticky cation trends are similar, here the sticky cation $\bar{p}_{0+}$ profile appears to be a scaled-down version of the general $\bar{p}_{0+}$ profile with a slightly more noticeable maximum. Overall, even with slight deviations between the association probability profiles in the general and sticky cation cases, they appear to produce similar trends and predictions.

Finally let us compare the product of the ionic association probabilities, $\bar{p}_{+-}\bar{p}_{-+}$, in the general and sticky cation cases shown in Fig.~\ref{fig:12mnsneg}.h) and Fig.~\ref{fig:12msneg}.h), respectively. In both cases $\bar{p}_{+-}\bar{p}_{-+}$ very slightly increases till obtaining a maximum after which is gradually decays to zero close to the interface. In Fig.~\ref{fig:12mnsneg}.h) and Fig.~\ref{fig:12msneg}.h), the trends for $\bar{p}_{+-}\bar{p}_{-+}$ are the same, but the exact value is slightly different with the general case taking on a slightly larger value in the bulk than the sticky cation case. Considering that trends in the predicted EDL quantities are rather close between the general and the sticky cation cases and that the slight deviations in values or trends do not significantly change the overarching predictions, this narrow difference suggests that the sticky cation approximation is reasonable for negatively charged electrodes.

Now, let us consider the simulation measurements for the positively charged electrode. This can be done by comparing the simulation general case results shown in the left column of Fig.~\ref{fig:12mnspos}) against the sticky case results in the left column of Fig.~\ref{fig:12mspos}). Similar to the negatively charged cases, the species volume fraction profiles in Fig.~\ref{fig:12mnspos}.a) and Fig.~\ref{fig:12mspos}.a) are an exact match. Comparing Fig.~\ref{fig:12mnspos}.b) and Fig.~\ref{fig:12mspos}.b), the cluster volume fractions are also an exact match. Turning to the association probabilities are shown in Fig.~\ref{fig:12mnspos}.c) and Fig.~\ref{fig:12mspos}.c) one can note that the profiles for $\bar{p}_{-+}$ and $\bar{p}_{0+}$ as in the negatively charged case are exact matches; however as expected $\bar{p}_{+-}$ and $\bar{p}_{+0}$ in Fig.~\ref{fig:12mspos}.c) appear to be scaled up by a constant factor compared to the $\bar{p}_{+-}$ and $\bar{p}_{+0}$ curves in Fig.~\ref{fig:12mnspos}.c). This change was expected to occur when the MD data analysis was changed from the general case to the sticky cation case. Lastly, considering the product of the ionic association probabilities, $\bar{p}_{+-}\bar{p}_{-+}$, displayed in Fig.~\ref{fig:12mspos}.d) one can note the curve is a scaled-up version of $\bar{p}_{+-}\bar{p}_{-+}$ seen in Fig.~\ref{fig:12mnspos}.d). Once again, one can also note between Fig.~\ref{fig:12mnspos}.d) and Fig.~\ref{fig:12mspos}.d) the critical threshold for $\bar{p}_{+-}\bar{p}_{-+}$ shifted up from the general case to the sticky cation case. As with the negatively charged electrode, the changes in the results and their profiles in the EDL do not appear to differ significantly between the general case and the sticky cation case in the MD simulations.

Let us consider how the theory's predictions change between the general case and the sticky cation case for the positively charged electrode. As before, we compare the theory's general prediction shown in the right column of Fig.~\ref{fig:12mnspos}) to the right column of Fig.~\ref{fig:12mspos}, which contains the theory's prediction for the sticky cation approximation. Through Fig.~\ref{fig:12mnspos}.e) and Fig.~\ref{fig:12mspos}.e), the predictions for species volume fractions can be compared. The Li$^+$ volume fraction ($\bar{\phi}_+$) profile is shown in Fig.~\ref{fig:12mnspos}.e) and Fig.~\ref{fig:12mspos}.e), in both cases when approaching the interface $\bar{\phi}_+$ appears to slow decrease before decreasing faster near the interface. For the TFSI$^-$ volume fraction ($\bar{\phi}_-$) profile in both Fig.~\ref{fig:12mnspos}.e) and Fig.~\ref{fig:12mspos}.e) it appears to slowly increase before quickly increasing closer to the interface and obtaining a maximum value around 4$\lambda_D$ from the electrode, before moderately decreases. In the general case, $\bar{\phi}_-$ appears to have a slightly higher surface value than in the sticky cation case. In both cases, the H$_2$O volume fraction ($\bar{\phi}_0$), as it approaches the interface, appears to slowly decrease before quickly decreasing closer to the interface and obtaining a minimum around 4$\lambda_D$ from the electrode before it moderately increases as seen in Fig.~\ref{fig:12mnspos}.e) and Fig.~\ref{fig:12mspos}.e). Here, we see that in the general case, the $\bar{\phi}_0$ appears to take on a surface value just marginally less than in the sticky cation case. For the species volume fractions, both cases produce similar trends with only slight deviations between the general case and the sticky cation approximation predictions.

For the cluster volume fractions, one can observe in both Fig.~\ref{fig:12mnspos}.f) and Fig.~\ref{fig:12mspos}.f) that the strictly hydrated cation clusters slowly decrease before rapidly decreasing towards zero close to the interface. Here, the strictly hydrated cation clusters display the same trend and nearly identical profiles. The volume fraction of bare anions ($\bar{\phi}_{010}$) appears to present the same trends in Fig.~\ref{fig:12mnspos}.f) and Fig.~\ref{fig:12mspos}.f) where it slowly increases before increasing faster near the interface where it achieves a maximum before decaying slightly to the interface. Here $\bar{\phi}_{001}$ appears to start at a higher value in the bulk and end at a slightly lower surface value in the sticky cation case than in the general case. 
The volume fraction of free water molecules ($\bar{\phi}_{001}$) seems to present the same trends in both Fig.~\ref{fig:12mnspos}.f) and Fig.~\ref{fig:12mspos}.f) where it slow increases before quickly increasing near the interface. While the trends are the same, the exact surface $\bar{\phi}_{001}$ appears slightly larger in the sticky cation case compared to the general case. Lastly, for the volume fraction of aggregates ($\bar{\phi}_{Agg}$) profile in the EDL, the two cases appear to have the same trends as displayed in Fig.~\ref{fig:12mnspos}.f) and Fig.~\ref{fig:12mspos}.f) where it slow decreases quickly decreasing towards zero closer to the interface. For $\bar{\phi}_{Agg}$, one can note that the general case appears to start at a somewhat higher value in the bulk than in the sticky cation case. The trends and predictions for the cluster volume fractions appear similar between the general and the sticky cation cases with only minor deviations. 

The association probabilities are displayed in Fig.~\ref{fig:12mnspos}.g) and Fig.~\ref{fig:12mspos}.g) in both cases for $\bar{p}_{+-}$ it appears to slowly increase before quickly increasing closer to the interface then increase slightly slower around 4$\lambda_D$ from the interface. The trends seen in the $\bar{p}_{+-}$ profile appear to agree in both the general and sticky cation cases. The exact surface $\bar{p}_{+-}$ in the general case appears slightly larger than in the sticky cation approximation. In Fig.~\ref{fig:12mnspos}.g), $\bar{p}_{+0}$ shown in the general case initially decreases slowly before decreasing quickly close to the interface. In Fig.~\ref{fig:12mspos}.g), $\bar{p}_{+0}$ displayed in the sticky case also initially slowly decreases before quickly decreasing closer to the interface and then decreasing slightly slower around 4$\lambda_D$ from the interface. This deviation between the general and the sticky cation cases is expected as $\bar{p}_{+-}+\bar{p}_{+0} = 1$ is enforced in the sticky cation case but not the general case allowing for the structure of $\bar{p}_{+0}$ to not be a direct transformation of $\bar{p}_{+-}$ in the general case. Additionally, $\bar{p}_{+0}$ in the general case appears to take on a lower value in the bulk compared to the sticky cation case. Shown in both Fig.~\ref{fig:12mnspos}.g) and Fig.~\ref{fig:12mspos}.g) $\bar{p}_{-+}$ slowly decreasing before quickly decreasing towards zero near the interface. Here, the trends appear to match; however, the exact profiles appear to differ by scaling factor, i.e. the sticky case seems to be a slightly scaled-down version of the general case. Last for the association probabilities, $\bar{p}_{0+}$ demonstrated the same trends in its EDL profile as shown in both Fig.~\ref{fig:12mnspos}.g) and Fig.~\ref{fig:12mspos}.g) where it is initially slowly decreasing before quickly decreasing towards zero near the interface. Once again while both the general and sticky cation trends are similar, here the sticky cation $\bar{p}_{0+}$ profile appears to be a scaled-down version of the general case's $\bar{p}_{0+}$ profile. Even with slight deviations between the association probability profiles in the general and sticky cation cases, they appear to produce similar trends and predictions.

Finally let us compare the product of the ionic association probabilities, $\bar{p}_{+-}\bar{p}_{-+}$, in the general and sticky cation cases shown in Fig.~\ref{fig:12mnspos}.h) and Fig.~\ref{fig:12mspos}.h), respectively. In both cases, $\bar{p}_{+-}\bar{p}_{-+}$ very slightly decreases before gradually decays to zero close to the interface. In Fig.~\ref{fig:12mnspos}.h) and Fig.~\ref{fig:12mspos}.h), the trends for $\bar{p}_{+-}\bar{p}_{-+}$ are the same, but the exact value is slightly different with the general case taking on a slightly larger value in the bulk than the sticky cation case. Given that the trends in the predicted EDL quantities are rather close between the general and the sticky cation cases and that the slight deviations in values or trends do not significantly change the overarching predictions, this finding suggests that the sticky cation approximation is reasonable for positively charged electrodes.

In conclusion as the results from MD simulations and the theory's predictions in the general case and under the sticky cation approximation do not vary significantly, this finding validates the usage of the approximation for this work for both negatively and positively charged electrodes. From the deviations observed, one can note that the accuracy of using this approximation may worsen at close proximity to the interface. This is expected as close to the interface, the system approaches limiting conditions for different species, which can significantly impact certain predictions from the theory, such as the association probabilities. Additionally, the theory is expected to deviate from the MD simulations as one closes the distance to the interface. In this region, one would expect to have a condensed layer in which the kinds of interactions and the form of species associations would vary significantly from the diffuse double layer.

\section{Extended Experimental Methods Section}

\subsection{Experimental Methods}

Bis(trifluoromethane)sulfonimide lithium salt (LiTFSI, $\geq$99.0\% ($^{19}$F-NMR), Sigma-Aldrich) was stored in a vacuum desiccator and dried in a vacuum oven at 95 $^{\circ}$C for 24 h before use. Aqueous solutions of 1 m, 12 m, 15 m, and 21 m LiTFSI were prepared by dissolving salt in Milli-Q water (18.2 M$\Omega$ cm$^{-1}$) inside a N$_2$-filled anaerobic chamber (relative humidity controlled below 3\% RH). Electrochemical measurements were performed in a sealed three-electrode cell with a gold disk electrode (2 mm diameter, CH Instruments) as working electrode, a gold wire (MSE Supplies) as counter electrode, and a silver wire (Sigma-Aldrich) as reference electrode. The gold disk electrode was polished mechanically on a microcloth polishing pad (Buehler) in 0.05 $\mu$m alumina particle (CH Instruments) slurry, rinsed with Deionized water and sonicated for 15 min, and blow-dried by ultrapure N$_2$ before use. 

\begin{figure}[h!]
     \centering
     \includegraphics[width= .75\textwidth]{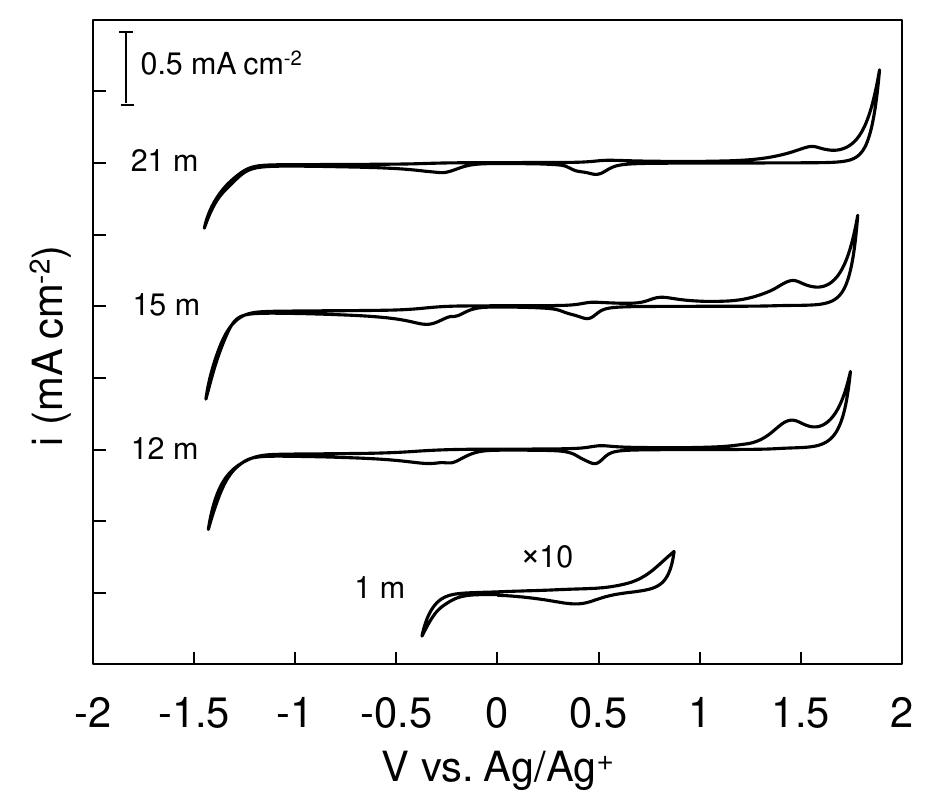}
     \caption{Cyclic voltammograms of 1 m, 12 m, 15 m, and 21 m LiTFSI on gold electrode measured after EIS with a scan rate of 10 mV s$^{-1}$. The current density for 1 m LiTFSI is plotted with ×10 magnification.}
     \label{fig:CV}
\end{figure}

Electrochemical impedance spectroscopy measurements (EIS) were performed using a Gamry Reference 620 potentiostat. The impedance spectra were collected by applying an AC sinusoidal potential (10 mV) over a DC voltage in the frequency range from 0.1 Hz to 1 MHz. To eliminate hysteresis, the sequence of applied DC potential with 0.1 V step was chosen to start from open circuit potential (OCP) to positive or negative potential limits. Cyclic voltammetry measurements were taken with a scan rate of 10 mV s$^{-1}$ after EIS measurements to confirm that the measurements were performed within the electrochemical stability window shown in Fig.~\ref{fig:CV}. In our experiments, no unexpected or unusually high safety hazards were encountered.

\subsection{Methods for Determining Differential Capacitance}

\begin{figure}[h!]
     \centering
     \includegraphics[width= 1\textwidth]{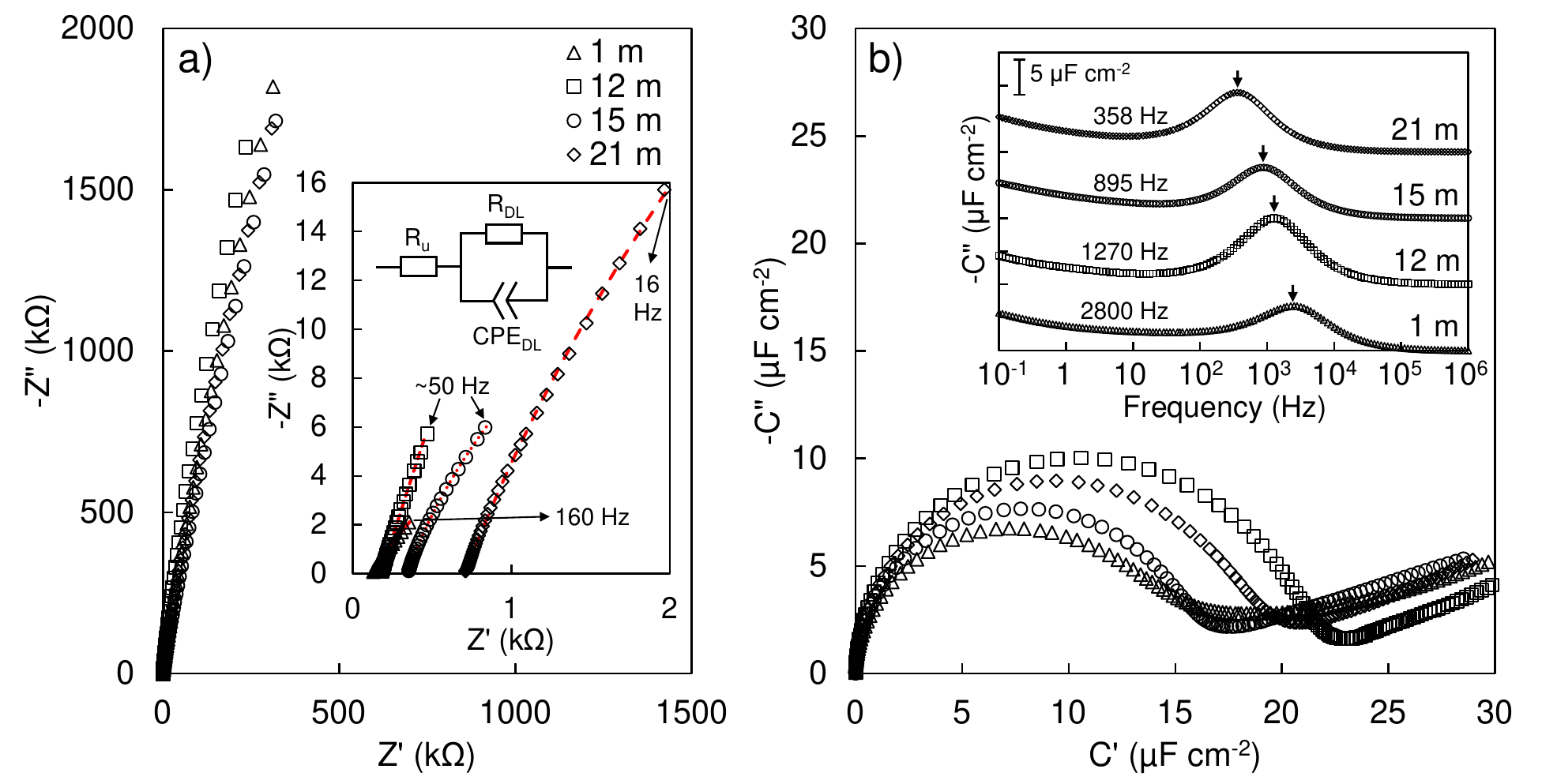}
     \caption{Nyquist plot and Cole-Cole (complex capacitance plane) plot for water-in-LiTFSI at open circuit potential. a) Nyquist plot and (inset) electric circuit fitting in the frequency range of double layer charging. The fitting circuit model is shown in the inset and the lower frequency limit for the fittings are labeled on the plot. The fitting curves are plotted in red. b) Cole-Cole plot and (inset) imaginary component of capacitance vs. frequency. A peak in the imaginary capacitance vs. frequency plot is associated with the capacitive process and the peak frequencies are noted by an arrow ($\downarrow$) for each concentration on the plot.}
     \label{fig:NC}
\end{figure}

Differential capacitance data were obtained by fitting the EIS data at each potential. Three methods were implemented based on the Nyquist plot and Cole-Cole plot; see Fig.~\ref{fig:NC}. By examining the Nyquist plot in Fig.~\ref{fig:NC}a, we found that a circuit model with only a pure capacitor does not adequately fit the impedance data. Therefore, an electric circuit model with a constant phase element (CPE) was considered~\cite{brug1984analysis}. The impedance of a CPE element (Z$_{CPE}$) can be described as,
\begin{equation}
    \label{exp1}
    Z_{CPE} = \frac{1}{\text{Y}_0(j\omega)^\alpha} = \frac{\text{cos}(\alpha\pi/2)}{\text{Y}_0\omega^\alpha} - j\frac{\text{sin}(\alpha\pi/2)}{\text{Y}_0\omega^\alpha},
\end{equation}

\noindent where $\omega$ is angular frequency, Y$_0$[F s$^{\alpha-1}$] and $\alpha$ are the CPE parameters. $\alpha$ has a value such that $0 \le \alpha \le 1$; if $\alpha$ = 1, the behavior is identical to a pure capacitor. 

The equivalent circuit for fitting the Nyquist plot is shown in the inset of Fig.~\ref{fig:NC}a. Here, R$_u$ represents the bulk electrolyte resistance, R$_{DL}$ represents the double layer resistance, and CPE$_{DL}$ describes the non-ideal double layer impedance. Only the double layer charging regime of the impedance spectra is selected for the fitting. The average $\alpha$ values of the fitted CPE element for each electrolyte across all the potentials are 0.930$\pm$0.004 for 1 m, 0.961$\pm$0.012 for 12 m, 0.965$\pm$0.008 for 15 m, and 0.960$\pm$0.010 for 21 m. 

A more straightforward approach to determine the capacitive behavior is based on the complex capacitance plane shown in Fig.~\ref{fig:NC}b. The complex capacitance can be derived from the impedance data,
\begin{equation}
    \label{exp2}
    C = C' + iC'' = \frac{1}{i\omega Z} = \frac{-Z''}{\omega\left(\left(Z'\right)^2+\left(-Z''\right)^2\right)}-i\frac{Z'}{\omega\left(\left(Z'\right)^2+\left(-Z''\right)^2\right)}.
\end{equation}

The experimental results show a distinctive capacitive process (slightly suppressed semicircle) at higher frequencies followed by a non-ideal behavior at lower frequencies. Capacitive behavior can also be indicated by a prominent peak in imaginary capacitance vs. frequency, see inset in Fig.~\ref{fig:NC}b; non-ideal slower processes lead to the increase of the imaginary capacitance at lower frequencies. The fast process (first semicircle in Cole-Cole plot/peak in -C'') has been attributed to double layer charging~\cite{druschler2011capacitive,klein2019potential}. The physical origin of the non-ideal behavior at lower frequencies is still not fully understood, but it is often associated with molecular mechanisms such as the reconstruction of the electrode surface, ion reorientation, and ion adsorption~\cite{druschler2011capacitive,roling2012slow,reichert2018molecular}. We selected the double layer charging frequency range for fitting as this experimental investigation focuses on its capacitance. 

\noindent \textit{Method 1. Determining the capacitance at a selected frequency.} As seen from Eq.~\eqref{exp1}, a CPE element contributes to both the real and imaginary parts of impedance. However, only the imaginary part originates from the capacitive behavior. The capacitance can thus be derived from the imaginary component of the CPE~\cite{lockett2008differential}:  
\begin{equation}
    \label{exp3}
    C(\omega) = \frac{1}{j\omega Z_{CPE,j}} = \frac{\text{Y}_0\omega^{\alpha-1}}{\text{sin}(\alpha\pi/2)}.
\end{equation}

The results display a frequency-dispersion of the capacitance, and therefore, a careful selection of the frequency to determine the capacitance is needed. Because the frequency range of double layer charging shifts significantly with water-in-LiTFSI electrolyte concentration, see inset in Fig.~\ref{fig:NC}b, using the same frequency to calculate the double layer capacitance across different concentrations of electrolytes is not ideal. Here, we select the peak-C'' frequency to calculate the capacitance at each concentration. 

\noindent \textit{Method 2. Deriving effective capacitance from circuit fitting parameters.} This approach was first developed by Brug \textit{et al.}~\cite{brug1984analysis} and explained in more detail by Hirschorn \textit{et al.}~\cite{hirschorn2010determination}. Their methodology applies to cases with a distribution of surface time-constants where additive contributions are considered for each site of the electrode surface. The effective capacitance extracted is associated with a CPE and is expressed as,
\begin{equation}
    \label{exp4}
    C_{eff} = Y_{0}^{1/\alpha}\left(\frac{1}{R_{u}}+\frac{1}{R_{DL}}\right)^{(\alpha-1)/\alpha}= Y_{0}^{1/\alpha}\left(\frac{R_u R_{DL}}{R_u+R_{DL}}\right)^{(1-\alpha)/\alpha}.
\end{equation}

\noindent When R$_{DL}$ $\gg$ R$_{u}$, Eq.~\eqref{exp4} simplifies to,
\begin{equation}
    \label{exp5}
    C_{eff} = Y_{0}^{1/\alpha}\left(R_{u}\right)^{(1-\alpha)/\alpha}.
\end{equation}

\noindent \textit{Method 3. Cole-Cole fit.} Instead of analyzing data in the impedance plane, the capacitance plane is used for fitting. In the complex capacitance plane, the fitting can be performed using the Cole-Cole function: 
\begin{equation}
    \label{exp6}
    C(\omega) = \frac{C}{1+(i\omega\tau)^\alpha}
\end{equation}

\noindent Where $\tau$ denotes the relaxation time and $\alpha$ indicates the ideality of the capacitive process. 

If multiple capacitive processes are observed, each process \textit{i} is assumed to occur in parallel and the total capacitive process can be described as follows,
\begin{equation}
    \label{exp7}
    C(\omega) = \sum_i\frac{C_i}{1+(i\omega\tau)^{\alpha_i}}.
\end{equation}

Here, we only fitted the fast capacitive process (1$^{st}$ semicircle) attributed to the double layer charging. 

\begin{figure}[h!]
     \centering
     \includegraphics[width= 1\textwidth]{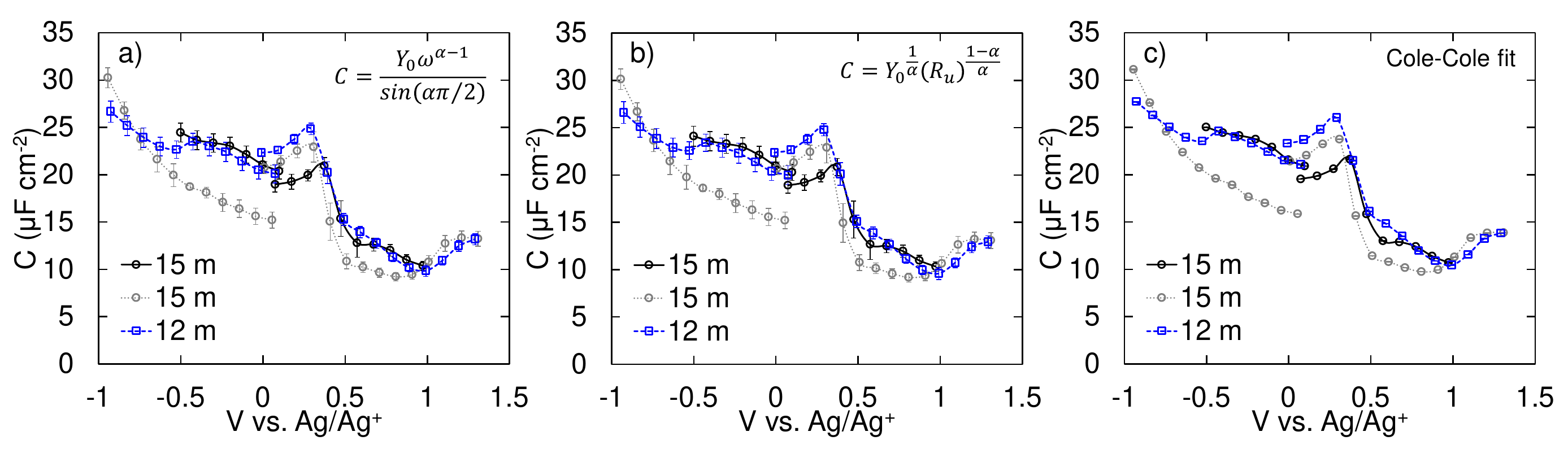}
     \caption{Differential capacitance of 15 m and 12 m water-in-LiTFSI as a function of the applied voltage from EIS measurements. The differential capacitance was obtained using three different fitting methods: a) Capacitance at a selected frequency, b) Capacitance derived from circuit fitting parameters, and c) Capacitance determined by the Cole-Cole method. Two measurements were taken for both the positive and negative branches at the concentration of 15 m LiTFSI, respectively, which are shown in the three plots for comparison.}
     \label{fig:1215exp}
\end{figure}

\begin{figure}[h!]
     \centering
     \includegraphics[width= 1\textwidth]{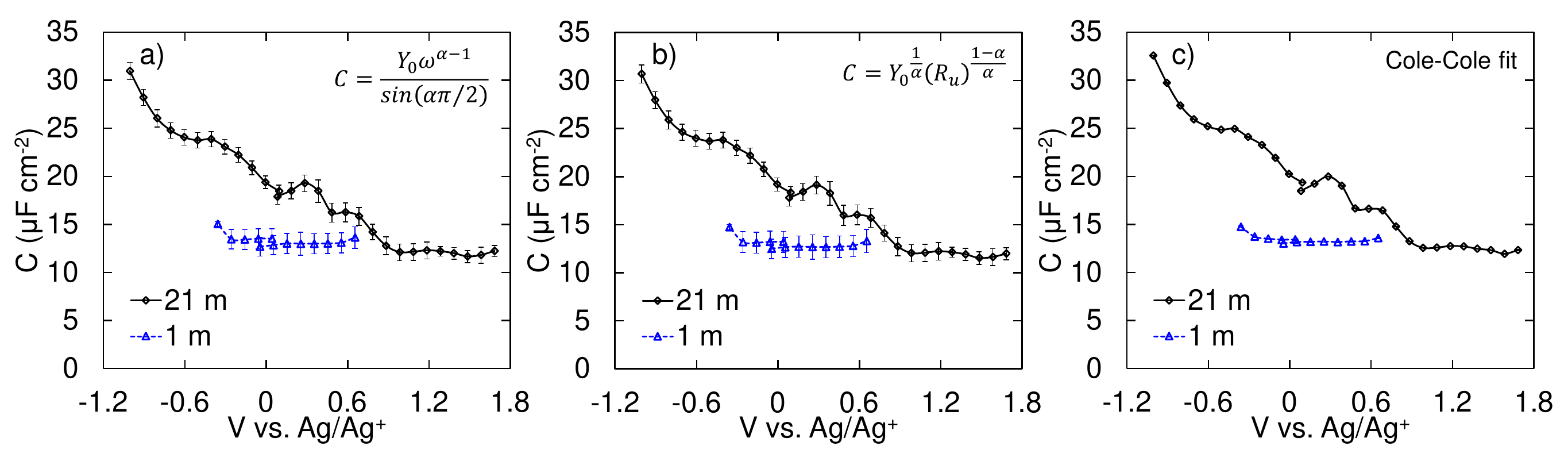}
     \caption{Differential capacitance of 21 m and 1 m aqueous LiTFSI as a function of the applied voltage from EIS measurements. The differential capacitance was obtained using three different fitting methods: a) Capacitance at a selected frequency, b) Capacitance derived from circuit fitting parameters, and c) Capacitance fitted by the Cole-Cole method.}
     \label{fig:121exp}
\end{figure}

The extracted differential capacitances are shown in Fig.~\ref{fig:1215exp} and Fig.~\ref{fig:121exp}. The capacitance values derived from Cole-Cole methods are slightly higher than from the other two methods based on fitting the Nyquist plot, which is consistent with the comparison of analysis methods reported by Small \textit{et al.}~\cite{small2014influence}. But overall, the three methods deliver very similar results. 

Our current experimental findings highlight the function of electrode material and surface roughness that may play a key role in the measured differential capacitance. In Zhang \textit{et al}.~\cite{zhang2020potential}, their differential capacitance measurements for 21m water-in-LiTFSI are much higher at negative potentials than these results with a similar structure as shown here in Fig.~\ref{fig:121exp}.b. The difference in measurements could be due to the surface preparation or EIS protocol and fitting methods used. Since Zhang \textit{et al}.~\cite{zhang2020potential} displays similar trends, the different methodology used to analyze the data could explain the quantitative differences.

\subsection{Additional Theory-Experimental Comparison}

\begin{figure}[h!]
     \centering
     \includegraphics[width= .9\textwidth]{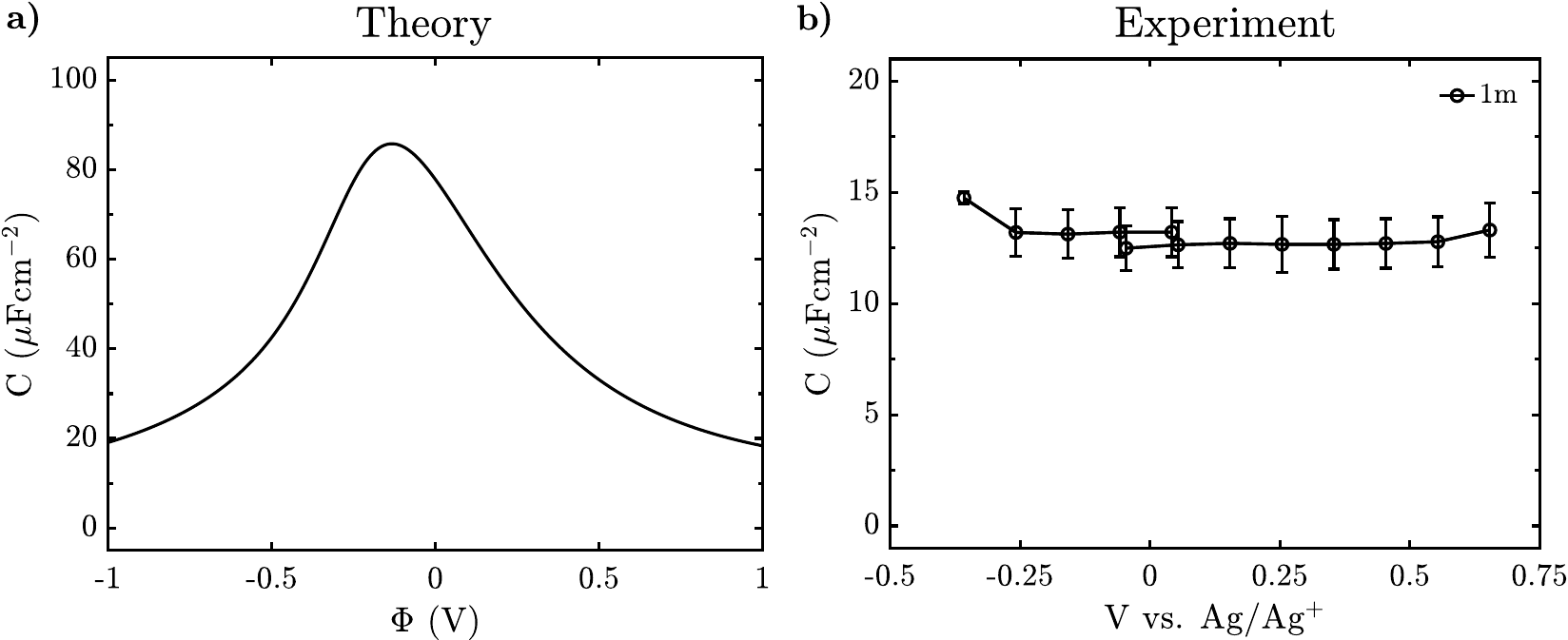}
     \caption{Differential Capacitance Comparison for 1m water-in-LiTFSI. a) Theory prediction for the differential capacitance of water-in-LiTFSI as a function of the electrostatic potential using $f_+=4$, $f_-=3$, $\xi_{0}=1$, $\xi_+=0.4$, $\xi_-=10.8$, $\epsilon_r = 10.1$, $\lambda = 0.228$, $P$ = 4.995 Debye, $v_0 = 22.7$ \AA$^3$, and $\alpha$ = 0.1. b) Experimental measurement of the differential capacitance of water-in-LiTFSI derived from circuit fitting parameters as a function of the applied electrostatic potential.}
     \label{fig:1mDC}
\end{figure}

\begin{figure}[h!]
     \centering
     \includegraphics[width= .9\textwidth]{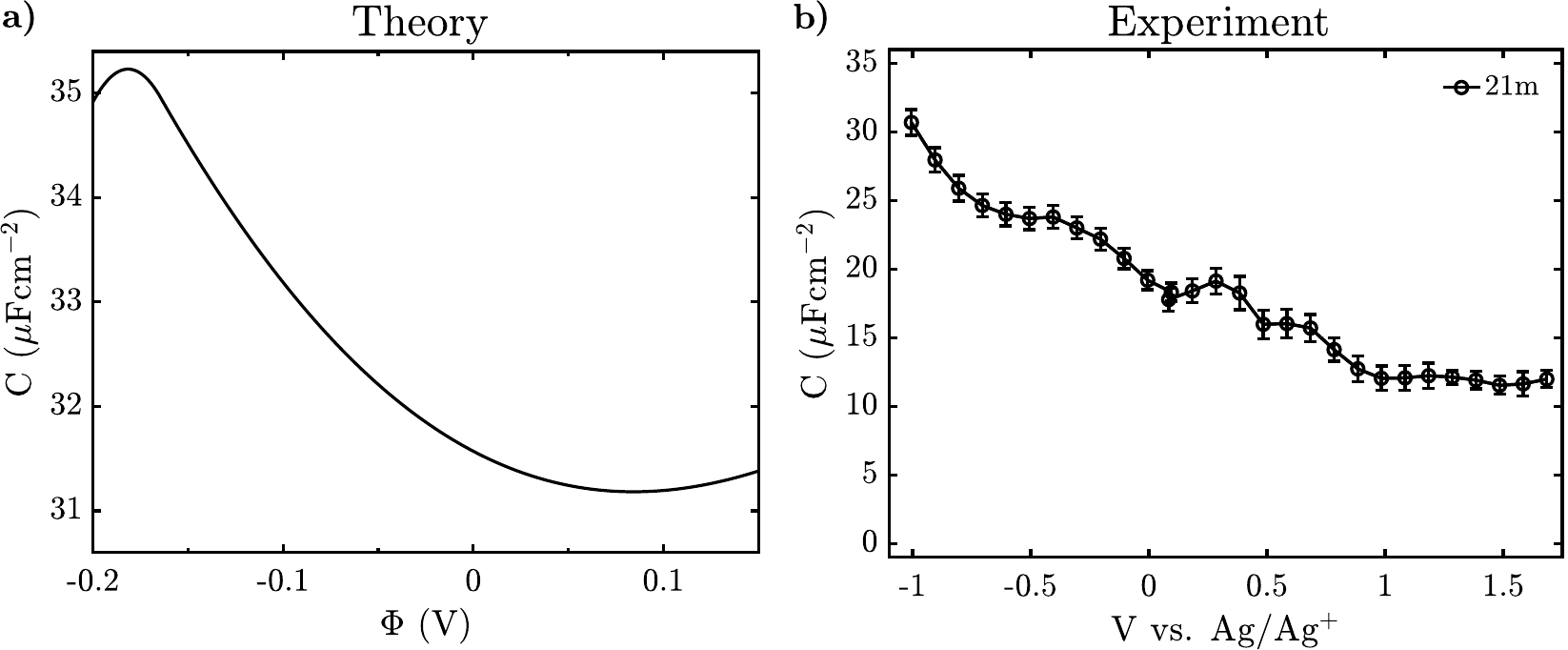}
     \caption{Differential Capacitance Comparison for 21m water-in-LiTFSI. a) Theory prediction for the differential capacitance of water-in-LiTFSI as a function of the electrostatic potential using $f_+=4$, $f_-=3$, $\xi_{0}=1$, $\xi_+=0.4$, $\xi_-=10.8$, $\epsilon_r = 10.1$, $\lambda = 0.228$, $P$ = 4.995 Debye, $v_0 = 22.7$ \AA$^3$, and $\alpha$ = 0.1. b) Experimental measurement of the differential capacitance of water-in-LiTFSI derived from circuit fitting parameters as a function of the applied electrostatic potential.}
     \label{fig:21mDC}
\end{figure}

\clearpage

\bibliography{WiSE}

\providecommand{\latin}[1]{#1}
\makeatletter
\providecommand{\doi}
  {\begingroup\let\do\@makeother\dospecials
  \catcode`\{=1 \catcode`\}=2 \doi@aux}
\providecommand{\doi@aux}[1]{\endgroup\texttt{#1}}
\makeatother
\providecommand*\mcitethebibliography{\thebibliography}
\csname @ifundefined\endcsname{endmcitethebibliography}
  {\let\endmcitethebibliography\endthebibliography}{}
\begin{mcitethebibliography}{167}
\providecommand*\natexlab[1]{#1}
\providecommand*\mciteSetBstSublistMode[1]{}
\providecommand*\mciteSetBstMaxWidthForm[2]{}
\providecommand*\mciteBstWouldAddEndPuncttrue
  {\def\EndOfBibitem{\unskip.}}
\providecommand*\mciteBstWouldAddEndPunctfalse
  {\let\EndOfBibitem\relax}
\providecommand*\mciteSetBstMidEndSepPunct[3]{}
\providecommand*\mciteSetBstSublistLabelBeginEnd[3]{}
\providecommand*\EndOfBibitem{}
\mciteSetBstSublistMode{f}
\mciteSetBstMaxWidthForm{subitem}{(\alph{mcitesubitemcount})}
\mciteSetBstSublistLabelBeginEnd
  {\mcitemaxwidthsubitemform\space}
  {\relax}
  {\relax}

\bibitem[Suo \latin{et~al.}(2013)Suo, Hu, Li, Armand, and Chen]{Suo2013}
Suo,~L.; Hu,~Y.-S.; Li,~H.; Armand,~M.; Chen,~L. {A new class of
  Solvent-in-Salt electrolyte for high-energy rechargeable metallic lithium
  batteries}. \emph{Nat. Commun.} \textbf{2013}, \emph{4}, 1481\relax
\mciteBstWouldAddEndPuncttrue
\mciteSetBstMidEndSepPunct{\mcitedefaultmidpunct}
{\mcitedefaultendpunct}{\mcitedefaultseppunct}\relax
\EndOfBibitem
\bibitem[Suo \latin{et~al.}(2015)Suo, Borodin, Gao, Olguin, Ho, Fan, Luo, Wang,
  and Xu]{Suo2015}
Suo,~L.; Borodin,~O.; Gao,~T.; Olguin,~M.; Ho,~J.; Fan,~X.; Luo,~C.; Wang,~C.;
  Xu,~K. {"Water-in-salt" electrolyte enables high-voltage aqueous lithium-ion
  chemistries.} \emph{Science} \textbf{2015}, \emph{350}, 938--43\relax
\mciteBstWouldAddEndPuncttrue
\mciteSetBstMidEndSepPunct{\mcitedefaultmidpunct}
{\mcitedefaultendpunct}{\mcitedefaultseppunct}\relax
\EndOfBibitem
\bibitem[Smith and Dunn(2015)Smith, and Dunn]{Smith2015}
Smith,~L.; Dunn,~B. {Opening the window for aqueous electrolytes}.
  \emph{Science} \textbf{2015}, \emph{350}, 918--918\relax
\mciteBstWouldAddEndPuncttrue
\mciteSetBstMidEndSepPunct{\mcitedefaultmidpunct}
{\mcitedefaultendpunct}{\mcitedefaultseppunct}\relax
\EndOfBibitem
\bibitem[Wang \latin{et~al.}(2016)Wang, Yamada, Sodeyama, Chiang, Tateyama, and
  Yamada]{Wang2016}
Wang,~J.; Yamada,~Y.; Sodeyama,~K.; Chiang,~C.~H.; Tateyama,~Y.; Yamada,~A.
  {Superconcentrated electrolytes for a high-voltage lithium-ion battery}.
  \emph{Nat. Commun.} \textbf{2016}, \emph{7}, 12032\relax
\mciteBstWouldAddEndPuncttrue
\mciteSetBstMidEndSepPunct{\mcitedefaultmidpunct}
{\mcitedefaultendpunct}{\mcitedefaultseppunct}\relax
\EndOfBibitem
\bibitem[Wang \latin{et~al.}(2018)Wang, Borodin, Gao, Fan, Sun, Han, Faraone,
  Dura, Xu, and Wang]{Wang2018}
Wang,~F.; Borodin,~O.; Gao,~T.; Fan,~X.; Sun,~W.; Han,~F.; Faraone,~A.;
  Dura,~J.~A.; Xu,~K.; Wang,~C. {Highly reversible zinc metal anode for aqueous
  batteries}. \emph{Nature Materials} \textbf{2018}, \emph{17}\relax
\mciteBstWouldAddEndPuncttrue
\mciteSetBstMidEndSepPunct{\mcitedefaultmidpunct}
{\mcitedefaultendpunct}{\mcitedefaultseppunct}\relax
\EndOfBibitem
\bibitem[Wang and Xu(2018)Wang, and Xu]{Wang2018a}
Wang,~C.; Xu,~K. {Advanced Aqueous Electrolytes for Li-ion Batteries}.
  \emph{Meet. Abstr.} \textbf{2018}, \emph{MA2018-01}, 1199--1199\relax
\mciteBstWouldAddEndPuncttrue
\mciteSetBstMidEndSepPunct{\mcitedefaultmidpunct}
{\mcitedefaultendpunct}{\mcitedefaultseppunct}\relax
\EndOfBibitem
\bibitem[Sun \latin{et~al.}(2017)Sun, Suo, Wang, Eidson, Yang, Han, Ma, Gao,
  Zhu, and Wang]{Sun2017}
Sun,~W.; Suo,~L.; Wang,~F.; Eidson,~N.; Yang,~C.; Han,~F.; Ma,~Z.; Gao,~T.;
  Zhu,~M.; Wang,~C. {“Water-in-Salt” electrolyte enabled LiMn2O4/TiS2
  Lithium-ion batteries}. \emph{Electrochem. Commun.} \textbf{2017}, \emph{82},
  71--74\relax
\mciteBstWouldAddEndPuncttrue
\mciteSetBstMidEndSepPunct{\mcitedefaultmidpunct}
{\mcitedefaultendpunct}{\mcitedefaultseppunct}\relax
\EndOfBibitem
\bibitem[Sodeyama \latin{et~al.}(2014)Sodeyama, Yamada, Aikawa, Yamada, and
  Tateyama]{Sodeyama2014}
Sodeyama,~K.; Yamada,~Y.; Aikawa,~K.; Yamada,~A.; Tateyama,~Y. {Sacrificial
  Anion Reduction Mechanism for Electrochemical Stability Improvement in Highly
  Concentrated Li-Salt Electrolyte}. \emph{J. Phys. Chem. C} \textbf{2014},
  \emph{118}, 14091--14097\relax
\mciteBstWouldAddEndPuncttrue
\mciteSetBstMidEndSepPunct{\mcitedefaultmidpunct}
{\mcitedefaultendpunct}{\mcitedefaultseppunct}\relax
\EndOfBibitem
\bibitem[Yamada \latin{et~al.}(2016)Yamada, Usui, Sodeyama, Ko, Tateyama, and
  Yamada]{Yamada2016}
Yamada,~Y.; Usui,~K.; Sodeyama,~K.; Ko,~S.; Tateyama,~Y.; Yamada,~A.
  {Hydrate-melt electrolytes for high-energy-density aqueous batteries}.
  \emph{Nat. Energy} \textbf{2016}, \emph{1}, 16129\relax
\mciteBstWouldAddEndPuncttrue
\mciteSetBstMidEndSepPunct{\mcitedefaultmidpunct}
{\mcitedefaultendpunct}{\mcitedefaultseppunct}\relax
\EndOfBibitem
\bibitem[K{\"u}hnel \latin{et~al.}(2017)K{\"u}hnel, Reber, and
  Battaglia]{kuhnel2017}
K{\"u}hnel,~R.-S.; Reber,~D.; Battaglia,~C. A high-voltage aqueous electrolyte
  for sodium-ion batteries. \emph{ACS Energy Lett.} \textbf{2017}, \emph{2},
  2005--2006\relax
\mciteBstWouldAddEndPuncttrue
\mciteSetBstMidEndSepPunct{\mcitedefaultmidpunct}
{\mcitedefaultendpunct}{\mcitedefaultseppunct}\relax
\EndOfBibitem
\bibitem[Suo \latin{et~al.}(2017)Suo, Borodin, Wang, Rong, Sun, Fan, Xu,
  Schroeder, Cresce, Wang, \latin{et~al.} others]{suo2017water}
Suo,~L.; Borodin,~O.; Wang,~Y.; Rong,~X.; Sun,~W.; Fan,~X.; Xu,~S.;
  Schroeder,~M.~A.; Cresce,~A.~V.; Wang,~F., \latin{et~al.}
  “Water-in-salt” electrolyte makes aqueous sodium-ion battery safe, green,
  and long-lasting. \emph{Advanced Energy Materials} \textbf{2017}, \emph{7},
  1701189\relax
\mciteBstWouldAddEndPuncttrue
\mciteSetBstMidEndSepPunct{\mcitedefaultmidpunct}
{\mcitedefaultendpunct}{\mcitedefaultseppunct}\relax
\EndOfBibitem
\bibitem[Leonard \latin{et~al.}(2018)Leonard, Wei, Chen, Du, and
  Ji]{leonard2018}
Leonard,~D.~P.; Wei,~Z.; Chen,~G.; Du,~F.; Ji,~X. Water-in-salt electrolyte for
  potassium-ion batteries. \emph{ACS Energy Lett.} \textbf{2018}, \emph{3},
  373--374\relax
\mciteBstWouldAddEndPuncttrue
\mciteSetBstMidEndSepPunct{\mcitedefaultmidpunct}
{\mcitedefaultendpunct}{\mcitedefaultseppunct}\relax
\EndOfBibitem
\bibitem[Thareja and Kumar(2021)Thareja, and Kumar]{thareja2021water}
Thareja,~S.; Kumar,~A. “Water-in-salt” electrolyte-based high-voltage (2.7
  V) sustainable symmetric supercapacitor with superb electrochemical
  performance—an analysis of the role of electrolytic ions in extending the
  cell voltage. \emph{ACS Sustainable Chemistry \& Engineering} \textbf{2021},
  \emph{9}, 2338--2347\relax
\mciteBstWouldAddEndPuncttrue
\mciteSetBstMidEndSepPunct{\mcitedefaultmidpunct}
{\mcitedefaultendpunct}{\mcitedefaultseppunct}\relax
\EndOfBibitem
\bibitem[Park \latin{et~al.}(2022)Park, Lee, and Kim]{park2022redox}
Park,~J.; Lee,~J.; Kim,~W. Redox-active water-in-salt electrolyte for
  high-energy-density supercapacitors. \emph{ACS Energy Letters} \textbf{2022},
  \emph{7}, 1266--1273\relax
\mciteBstWouldAddEndPuncttrue
\mciteSetBstMidEndSepPunct{\mcitedefaultmidpunct}
{\mcitedefaultendpunct}{\mcitedefaultseppunct}\relax
\EndOfBibitem
\bibitem[Haregewoin \latin{et~al.}(2016)Haregewoin, Wotango, and
  Hwang]{haregewoin2016electrolyte}
Haregewoin,~A.~M.; Wotango,~A.~S.; Hwang,~B.-J. Electrolyte additives for
  lithium ion battery electrodes: progress and perspectives. \emph{Energy \&
  Environmental Science} \textbf{2016}, \emph{9}, 1955--1988\relax
\mciteBstWouldAddEndPuncttrue
\mciteSetBstMidEndSepPunct{\mcitedefaultmidpunct}
{\mcitedefaultendpunct}{\mcitedefaultseppunct}\relax
\EndOfBibitem
\bibitem[Dou \latin{et~al.}(2018)Dou, Lei, Wang, Zhang, Xiao, Guo, Wang, Yang,
  Li, Shi, , and Yan]{dou2018safe}
Dou,~Q.; Lei,~S.; Wang,~D.-W.; Zhang,~Q.; Xiao,~D.; Guo,~H.; Wang,~A.;
  Yang,~H.; Li,~Y.; Shi,~S.; ; Yan,~X. Safe and high-rate supercapacitors based
  on an “acetonitrile/water in salt” hybrid electrolyte. \emph{Energy
  Environ. Sci.} \textbf{2018}, \emph{11}, 3212--3219\relax
\mciteBstWouldAddEndPuncttrue
\mciteSetBstMidEndSepPunct{\mcitedefaultmidpunct}
{\mcitedefaultendpunct}{\mcitedefaultseppunct}\relax
\EndOfBibitem
\bibitem[Zhang \latin{et~al.}(2018)Zhang, Ye, Henkensmeier, Hempelmann, and
  Chen]{zhang2018water}
Zhang,~Y.; Ye,~R.; Henkensmeier,~D.; Hempelmann,~R.; Chen,~R. “Water-in-ionic
  liquid” solutions towards wide electrochemical stability windows for
  aqueous rechargeable batteries. \emph{Electrochimica Acta} \textbf{2018},
  \emph{263}, 47--52\relax
\mciteBstWouldAddEndPuncttrue
\mciteSetBstMidEndSepPunct{\mcitedefaultmidpunct}
{\mcitedefaultendpunct}{\mcitedefaultseppunct}\relax
\EndOfBibitem
\bibitem[Yang \latin{et~al.}(2019)Yang, Chen, Ji, Pollard, L{\"u}, Sun, Hou,
  Liu, Liu, Qing, \latin{et~al.} others]{yang2019aqueous}
Yang,~C.; Chen,~J.; Ji,~X.; Pollard,~T.~P.; L{\"u},~X.; Sun,~C.-J.; Hou,~S.;
  Liu,~Q.; Liu,~C.; Qing,~T., \latin{et~al.}  Aqueous Li-ion battery enabled by
  halogen conversion--intercalation chemistry in graphite. \emph{Nature}
  \textbf{2019}, \emph{569}, 245--250\relax
\mciteBstWouldAddEndPuncttrue
\mciteSetBstMidEndSepPunct{\mcitedefaultmidpunct}
{\mcitedefaultendpunct}{\mcitedefaultseppunct}\relax
\EndOfBibitem
\bibitem[Yang \latin{et~al.}(2017)Yang, Suo, Borodin, Wang, Sun, Gao, Fan, Hou,
  Ma, Amine, Xu, and Wang]{yang2017}
Yang,~C.; Suo,~L.; Borodin,~O.; Wang,~F.; Sun,~W.; Gao,~T.; Fan,~X.; Hou,~S.;
  Ma,~Z.; Amine,~K.; Xu,~K.; Wang,~C. {Unique aqueous Li-ion/sulfur chemistry
  with high energy density and reversibility}. \emph{Proceedings of the
  National Academy of Sciences of the United States of America} \textbf{2017},
  \emph{114}\relax
\mciteBstWouldAddEndPuncttrue
\mciteSetBstMidEndSepPunct{\mcitedefaultmidpunct}
{\mcitedefaultendpunct}{\mcitedefaultseppunct}\relax
\EndOfBibitem
\bibitem[Borodin \latin{et~al.}(2020)Borodin, Self, Persson, Wang, and
  Xu]{borodin2020uncharted}
Borodin,~O.; Self,~J.; Persson,~K.~A.; Wang,~C.; Xu,~K. Uncharted Waters:
  Super-Concentrated Electrolytes. \emph{Joule} \textbf{2020}, \emph{4},
  69--100\relax
\mciteBstWouldAddEndPuncttrue
\mciteSetBstMidEndSepPunct{\mcitedefaultmidpunct}
{\mcitedefaultendpunct}{\mcitedefaultseppunct}\relax
\EndOfBibitem
\bibitem[Sayah \latin{et~al.}(2022)Sayah, Ghosh, Baazizi, Amine, Dahbi, Amine,
  Ghamouss, and Amine]{sayah2022super}
Sayah,~S.; Ghosh,~A.; Baazizi,~M.; Amine,~R.; Dahbi,~M.; Amine,~Y.;
  Ghamouss,~F.; Amine,~K. How do super concentrated electrolytes push the
  Li-ion batteries and supercapacitors beyond their thermodynamic and
  electrochemical limits? \emph{Nano Energy} \textbf{2022}, \emph{98},
  107336\relax
\mciteBstWouldAddEndPuncttrue
\mciteSetBstMidEndSepPunct{\mcitedefaultmidpunct}
{\mcitedefaultendpunct}{\mcitedefaultseppunct}\relax
\EndOfBibitem
\bibitem[Vatamanu and Borodin(2017)Vatamanu, and Borodin]{vatamanu2017}
Vatamanu,~J.; Borodin,~O. Ramifications of water-in-salt interfacial structure
  at charged electrodes for electrolyte electrochemical stability. \emph{J.
  Phys. Chem. Lett.} \textbf{2017}, \emph{8}, 4362--4367\relax
\mciteBstWouldAddEndPuncttrue
\mciteSetBstMidEndSepPunct{\mcitedefaultmidpunct}
{\mcitedefaultendpunct}{\mcitedefaultseppunct}\relax
\EndOfBibitem
\bibitem[McEldrew \latin{et~al.}(2018)McEldrew, Goodwin, Kornyshev, and
  Bazant]{mceldrew2018}
McEldrew,~M.; Goodwin,~Z.~A.; Kornyshev,~A.~A.; Bazant,~M.~Z. Theory of the
  double layer in water-in-salt electrolytes. \emph{The journal of physical
  chemistry letters} \textbf{2018}, \emph{9}, 5840--5846\relax
\mciteBstWouldAddEndPuncttrue
\mciteSetBstMidEndSepPunct{\mcitedefaultmidpunct}
{\mcitedefaultendpunct}{\mcitedefaultseppunct}\relax
\EndOfBibitem
\bibitem[McEldrew \latin{et~al.}(2021)McEldrew, Goodwin, Bi, Kornyshev, and
  Bazant]{mceldrew2021ion}
McEldrew,~M.; Goodwin,~Z.~A.; Bi,~S.; Kornyshev,~A.; Bazant,~M.~Z. Ion Clusters
  and Networks in Water-in-Salt Electrolytes. \emph{J. Electrochem. Soc.}
  \textbf{2021}, \emph{168}, 050514\relax
\mciteBstWouldAddEndPuncttrue
\mciteSetBstMidEndSepPunct{\mcitedefaultmidpunct}
{\mcitedefaultendpunct}{\mcitedefaultseppunct}\relax
\EndOfBibitem
\bibitem[Borodin \latin{et~al.}(2017)Borodin, Suo, Gobet, Ren, Wang, Faraone,
  Peng, Olguin, Schroeder, Ding, \latin{et~al.} others]{borodin2017liquid}
Borodin,~O.; Suo,~L.; Gobet,~M.; Ren,~X.; Wang,~F.; Faraone,~A.; Peng,~J.;
  Olguin,~M.; Schroeder,~M.; Ding,~M.~S., \latin{et~al.}  Liquid structure with
  nano-heterogeneity promotes cationic transport in concentrated electrolytes.
  \emph{ACS nano} \textbf{2017}, \emph{11}, 10462--10471\relax
\mciteBstWouldAddEndPuncttrue
\mciteSetBstMidEndSepPunct{\mcitedefaultmidpunct}
{\mcitedefaultendpunct}{\mcitedefaultseppunct}\relax
\EndOfBibitem
\bibitem[Zheng \latin{et~al.}(2018)Zheng, Tan, Shan, Liu, Hu, Feng, Yang,
  Zhang, Chen, Lin, \latin{et~al.} others]{zheng2018understanding}
Zheng,~J.; Tan,~G.; Shan,~P.; Liu,~T.; Hu,~J.; Feng,~Y.; Yang,~L.; Zhang,~M.;
  Chen,~Z.; Lin,~Y., \latin{et~al.}  Understanding thermodynamic and kinetic
  contributions in expanding the stability window of aqueous electrolytes.
  \emph{Chem} \textbf{2018}, \emph{4}, 2872--2882\relax
\mciteBstWouldAddEndPuncttrue
\mciteSetBstMidEndSepPunct{\mcitedefaultmidpunct}
{\mcitedefaultendpunct}{\mcitedefaultseppunct}\relax
\EndOfBibitem
\bibitem[Lim \latin{et~al.}(2018)Lim, Park, Lee, Kim, Kwak, and Cho]{lim2018}
Lim,~J.; Park,~K.; Lee,~H.; Kim,~J.; Kwak,~K.; Cho,~M. Nanometric water
  channels in water-in-salt lithium ion battery electrolyte. \emph{Journal of
  the American Chemical Society} \textbf{2018}, \emph{140}, 15661--15667\relax
\mciteBstWouldAddEndPuncttrue
\mciteSetBstMidEndSepPunct{\mcitedefaultmidpunct}
{\mcitedefaultendpunct}{\mcitedefaultseppunct}\relax
\EndOfBibitem
\bibitem[Choi \latin{et~al.}(2018)Choi, Lee, Choi, and Cho]{choi2018graph}
Choi,~J.-H.; Lee,~H.; Choi,~H.~R.; Cho,~M. Graph theory and ion and molecular
  aggregation in aqueous solutions. \emph{Annual review of physical chemistry}
  \textbf{2018}, \emph{69}, 125--149\relax
\mciteBstWouldAddEndPuncttrue
\mciteSetBstMidEndSepPunct{\mcitedefaultmidpunct}
{\mcitedefaultendpunct}{\mcitedefaultseppunct}\relax
\EndOfBibitem
\bibitem[Han \latin{et~al.}(2020)Han, Yu, Wang, Redfern, Ma, Cheng, Chen, Hu,
  Curtiss, Xu, \latin{et~al.} others]{han2020origin}
Han,~K.~S.; Yu,~Z.; Wang,~H.; Redfern,~P.~C.; Ma,~L.; Cheng,~L.; Chen,~Y.;
  Hu,~J.~Z.; Curtiss,~L.~A.; Xu,~K., \latin{et~al.}  Origin of Unusual Acidity
  and Li+ Diffusivity in a Series of Water-in-Salt Electrolytes. \emph{The
  Journal of Physical Chemistry B} \textbf{2020}, \relax
\mciteBstWouldAddEndPunctfalse
\mciteSetBstMidEndSepPunct{\mcitedefaultmidpunct}
{}{\mcitedefaultseppunct}\relax
\EndOfBibitem
\bibitem[Andersson \latin{et~al.}(2020)Andersson, {\AA}r{\'e}n, Franco, and
  Johansson]{andersson2020ion}
Andersson,~R.; {\AA}r{\'e}n,~F.; Franco,~A.~A.; Johansson,~P. Ion Transport
  Mechanisms via Time-Dependent Local Structure and Dynamics in Highly
  Concentrated Electrolytes. \emph{Journal of the Electrochemical Society}
  \textbf{2020}, \emph{167}, 140537\relax
\mciteBstWouldAddEndPuncttrue
\mciteSetBstMidEndSepPunct{\mcitedefaultmidpunct}
{\mcitedefaultendpunct}{\mcitedefaultseppunct}\relax
\EndOfBibitem
\bibitem[Yu \latin{et~al.}(2020)Yu, Curtiss, Winans, Zhang, Li, and
  Cheng]{yu2020asymmetric}
Yu,~Z.; Curtiss,~L.~A.; Winans,~R.~E.; Zhang,~Y.; Li,~T.; Cheng,~L. Asymmetric
  Composition of Ionic Aggregates and the Origin of High Correlated
  Transference Number in Water-in-Salt Electrolytes. \emph{The Journal of
  Physical Chemistry Letters} \textbf{2020}, \emph{11}, 1276--1281\relax
\mciteBstWouldAddEndPuncttrue
\mciteSetBstMidEndSepPunct{\mcitedefaultmidpunct}
{\mcitedefaultendpunct}{\mcitedefaultseppunct}\relax
\EndOfBibitem
\bibitem[Lewis \latin{et~al.}(2020)Lewis, Zhang, Dereka, Carino, Maginn, and
  Tokmakoff]{lewis2020signatures}
Lewis,~N.~H.; Zhang,~Y.; Dereka,~B.; Carino,~E.~V.; Maginn,~E.~J.;
  Tokmakoff,~A. Signatures of Ion-Pairing and Aggregation in the Vibrational
  Spectroscopy of Super-Concentrated Aqueous Lithium Bistriflimide Solutions.
  \emph{The Journal of Physical Chemistry C} \textbf{2020}, \relax
\mciteBstWouldAddEndPunctfalse
\mciteSetBstMidEndSepPunct{\mcitedefaultmidpunct}
{}{\mcitedefaultseppunct}\relax
\EndOfBibitem
\bibitem[Gonz{\'a}lez \latin{et~al.}(2020)Gonz{\'a}lez, Borodin, Kofu, Shibata,
  Yamada, Yamamuro, Xu, Price, and Saboungi]{gonzalez2020nanoscale}
Gonz{\'a}lez,~M.~A.; Borodin,~O.; Kofu,~M.; Shibata,~K.; Yamada,~T.;
  Yamamuro,~O.; Xu,~K.; Price,~D.~L.; Saboungi,~M.-L. Nanoscale Relaxation in
  “Water-in-Salt” and “Water-in-Bisalt” Electrolytes. \emph{The Journal
  of Physical Chemistry Letters} \textbf{2020}, \emph{11}, 7279--7284\relax
\mciteBstWouldAddEndPuncttrue
\mciteSetBstMidEndSepPunct{\mcitedefaultmidpunct}
{\mcitedefaultendpunct}{\mcitedefaultseppunct}\relax
\EndOfBibitem
\bibitem[Zhang \latin{et~al.}(2020)Zhang, Han, Ta, Madsen, Chen, Zhang,
  Espinosa-Marzal, and Gewirth]{zhang2020potential}
Zhang,~R.; Han,~M.; Ta,~K.; Madsen,~K.~E.; Chen,~X.; Zhang,~X.;
  Espinosa-Marzal,~R.~M.; Gewirth,~A.~A. Potential-dependent layering in the
  electrochemical double layer of water-in-salt electrolytes. \emph{ACS Applied
  Energy Materials} \textbf{2020}, \emph{3}, 8086--8094\relax
\mciteBstWouldAddEndPuncttrue
\mciteSetBstMidEndSepPunct{\mcitedefaultmidpunct}
{\mcitedefaultendpunct}{\mcitedefaultseppunct}\relax
\EndOfBibitem
\bibitem[Han \latin{et~al.}(2021)Han, Zhang, Gewirth, and
  Espinosa-Marzal]{Han2021WiSE}
Han,~M.; Zhang,~R.; Gewirth,~A.~A.; Espinosa-Marzal,~R.~M. Nanoheterogeneity of
  LiTFSI Solutions Transitions Close to a Surface and with Concentration.
  \emph{Nano Lett.} \textbf{2021}, \emph{21}, 2304--2309\relax
\mciteBstWouldAddEndPuncttrue
\mciteSetBstMidEndSepPunct{\mcitedefaultmidpunct}
{\mcitedefaultendpunct}{\mcitedefaultseppunct}\relax
\EndOfBibitem
\bibitem[Liu \latin{et~al.}(2021)Liu, Yu, Sarnello, Qian, Seifert, Winans,
  Cheng, and Li]{liu2021microscopic}
Liu,~X.; Yu,~Z.; Sarnello,~E.; Qian,~K.; Seifert,~S.; Winans,~R.~E.; Cheng,~L.;
  Li,~T. Microscopic understanding of the ionic networks of “water-in-salt”
  electrolytes. \emph{Energy Material Advances} \textbf{2021}, \relax
\mciteBstWouldAddEndPunctfalse
\mciteSetBstMidEndSepPunct{\mcitedefaultmidpunct}
{}{\mcitedefaultseppunct}\relax
\EndOfBibitem
\bibitem[Groves \latin{et~al.}(2021)Groves, Perez-Martinez, Lhermerout, and
  Perkin]{groves2021surface}
Groves,~T.~S.; Perez-Martinez,~C.~S.; Lhermerout,~R.; Perkin,~S. Surface forces
  and structure in a water-in-salt electrolyte. \emph{The Journal of Physical
  Chemistry Letters} \textbf{2021}, \emph{12}, 1702--1707\relax
\mciteBstWouldAddEndPuncttrue
\mciteSetBstMidEndSepPunct{\mcitedefaultmidpunct}
{\mcitedefaultendpunct}{\mcitedefaultseppunct}\relax
\EndOfBibitem
\bibitem[Ichii \latin{et~al.}(2020)Ichii, Ichikawa, Yamada, Murata, Utsunomiya,
  and Sugimura]{ichii2020solvation}
Ichii,~T.; Ichikawa,~S.; Yamada,~Y.; Murata,~M.; Utsunomiya,~T.; Sugimura,~H.
  Solvation structure on water-in-salt/mica interfaces and its molality
  dependence investigated by atomic force microscopy. \emph{Japanese Journal of
  Applied Physics} \textbf{2020}, \emph{59}, SN1003\relax
\mciteBstWouldAddEndPuncttrue
\mciteSetBstMidEndSepPunct{\mcitedefaultmidpunct}
{\mcitedefaultendpunct}{\mcitedefaultseppunct}\relax
\EndOfBibitem
\bibitem[Li \latin{et~al.}(2022)Li, Chen, Liu, Lu, Meng, Yan, Abru{\~n}a, Feng,
  and Lian]{li2022unconventional}
Li,~C.-Y.; Chen,~M.; Liu,~S.; Lu,~X.; Meng,~J.; Yan,~J.; Abru{\~n}a,~H.~D.;
  Feng,~G.; Lian,~T. Unconventional interfacial water structure of highly
  concentrated aqueous electrolytes at negative electrode polarizations.
  \emph{Nature Communications} \textbf{2022}, \emph{13}, 5330\relax
\mciteBstWouldAddEndPuncttrue
\mciteSetBstMidEndSepPunct{\mcitedefaultmidpunct}
{\mcitedefaultendpunct}{\mcitedefaultseppunct}\relax
\EndOfBibitem
\bibitem[Bazant \latin{et~al.}(2011)Bazant, Storey, and Kornyshev]{Bazant2011}
Bazant,~M.~Z.; Storey,~B.~D.; Kornyshev,~A.~A. {Double Layer in Ionic Liquids:
  Overscreening versus Crowding}. \emph{Phys. Rev. Lett.} \textbf{2011},
  \emph{106}, 046102\relax
\mciteBstWouldAddEndPuncttrue
\mciteSetBstMidEndSepPunct{\mcitedefaultmidpunct}
{\mcitedefaultendpunct}{\mcitedefaultseppunct}\relax
\EndOfBibitem
\bibitem[McEldrew \latin{et~al.}(2020)McEldrew, Goodwin, Bi, Bazant, and
  Kornyshev]{mceldrew2020theory}
McEldrew,~M.; Goodwin,~Z.~A.; Bi,~S.; Bazant,~M.~Z.; Kornyshev,~A.~A. Theory of
  Ion Aggregation and Gelation in Super-Concentrated Electrolytes. \emph{J.
  Chem. Phys.} \textbf{2020}, \emph{152}, 234506\relax
\mciteBstWouldAddEndPuncttrue
\mciteSetBstMidEndSepPunct{\mcitedefaultmidpunct}
{\mcitedefaultendpunct}{\mcitedefaultseppunct}\relax
\EndOfBibitem
\bibitem[McEldrew \latin{et~al.}(2021)McEldrew, Goodwin, Zhao, Bazant, and
  Kornyshev]{mceldrew2020corr}
McEldrew,~M.; Goodwin,~Z. A.~H.; Zhao,~H.; Bazant,~M.~Z.; Kornyshev,~A.~A.
  Correlated Ion Transport and the Gel Phase in Room Temperature Ionic Liquids.
  \emph{J. Phys. Chem B} \textbf{2021}, \emph{125}, 2677–2689\relax
\mciteBstWouldAddEndPuncttrue
\mciteSetBstMidEndSepPunct{\mcitedefaultmidpunct}
{\mcitedefaultendpunct}{\mcitedefaultseppunct}\relax
\EndOfBibitem
\bibitem[McEldrew \latin{et~al.}(2021)McEldrew, Goodwin, Molinari, Kozinsky,
  Kornyshev, and Bazant]{McEldrewsalt2021}
McEldrew,~M.; Goodwin,~Z. A.~H.; Molinari,~N.; Kozinsky,~B.; Kornyshev,~A.~A.;
  Bazant,~M.~Z. Salt-in-ionic-liquid electrolytes: Ion network formation and
  negative effective charges of alkali metal cations. \emph{J. Phys. Chem. B}
  \textbf{2021}, \emph{125}, 13752--13766\relax
\mciteBstWouldAddEndPuncttrue
\mciteSetBstMidEndSepPunct{\mcitedefaultmidpunct}
{\mcitedefaultendpunct}{\mcitedefaultseppunct}\relax
\EndOfBibitem
\bibitem[Goodwin \latin{et~al.}(2023)Goodwin, McEldrew, Kozinsky, and
  Bazant]{Goodwin2023}
Goodwin,~Z. A.~H.; McEldrew,~M.; Kozinsky,~B.; Bazant,~M.~Z. Theory of Cation
  Solvation and Ionic Association in Nonaqueous Solvent Mixtures. \emph{PRX
  Energy} \textbf{2023}, \emph{2}, 013007\relax
\mciteBstWouldAddEndPuncttrue
\mciteSetBstMidEndSepPunct{\mcitedefaultmidpunct}
{\mcitedefaultendpunct}{\mcitedefaultseppunct}\relax
\EndOfBibitem
\bibitem[Flory(1942)]{flory1942thermodynamics}
Flory,~P.~J. Thermodynamics of high polymer solutions. \emph{The Journal of
  chemical physics} \textbf{1942}, \emph{10}, 51--61\relax
\mciteBstWouldAddEndPuncttrue
\mciteSetBstMidEndSepPunct{\mcitedefaultmidpunct}
{\mcitedefaultendpunct}{\mcitedefaultseppunct}\relax
\EndOfBibitem
\bibitem[Flory(1953)]{flory1953principles}
Flory,~P.~J. \emph{Principles of polymer chemistry}; Cornell University Press,
  1953\relax
\mciteBstWouldAddEndPuncttrue
\mciteSetBstMidEndSepPunct{\mcitedefaultmidpunct}
{\mcitedefaultendpunct}{\mcitedefaultseppunct}\relax
\EndOfBibitem
\bibitem[Stockmayer(1943)]{stockmayer1943theory}
Stockmayer,~W.~H. Theory of molecular size distribution and gel formation in
  branched-chain polymers. \emph{The Journal of chemical physics}
  \textbf{1943}, \emph{11}, 45--55\relax
\mciteBstWouldAddEndPuncttrue
\mciteSetBstMidEndSepPunct{\mcitedefaultmidpunct}
{\mcitedefaultendpunct}{\mcitedefaultseppunct}\relax
\EndOfBibitem
\bibitem[Stockmayer(1944)]{stockmayer1944theory}
Stockmayer,~W.~H. Theory of molecular size distribution and gel formation in
  branched polymers II. General cross linking. \emph{The Journal of Chemical
  Physics} \textbf{1944}, \emph{12}, 125--131\relax
\mciteBstWouldAddEndPuncttrue
\mciteSetBstMidEndSepPunct{\mcitedefaultmidpunct}
{\mcitedefaultendpunct}{\mcitedefaultseppunct}\relax
\EndOfBibitem
\bibitem[Stockmayer(1952)]{stockmayer1952molecular}
Stockmayer,~W.~H. Molecular distribution in condensation polymers. \emph{J.
  Polym. Sci.} \textbf{1952}, \emph{9}, 69--71\relax
\mciteBstWouldAddEndPuncttrue
\mciteSetBstMidEndSepPunct{\mcitedefaultmidpunct}
{\mcitedefaultendpunct}{\mcitedefaultseppunct}\relax
\EndOfBibitem
\bibitem[Tanaka(1989)]{tanaka1989}
Tanaka,~F. Theory of thermoreversible gelation. \emph{Macromolecules}
  \textbf{1989}, \emph{22}, 1988--1994\relax
\mciteBstWouldAddEndPuncttrue
\mciteSetBstMidEndSepPunct{\mcitedefaultmidpunct}
{\mcitedefaultendpunct}{\mcitedefaultseppunct}\relax
\EndOfBibitem
\bibitem[Tanaka(1990)]{tanaka1990thermodynamic}
Tanaka,~F. Thermodynamic theory of network-forming polymer solutions. 1.
  \emph{Macromolecules} \textbf{1990}, \emph{23}, 3784--3789\relax
\mciteBstWouldAddEndPuncttrue
\mciteSetBstMidEndSepPunct{\mcitedefaultmidpunct}
{\mcitedefaultendpunct}{\mcitedefaultseppunct}\relax
\EndOfBibitem
\bibitem[Tanaka and Stockmayer(1994)Tanaka, and Stockmayer]{tanaka1994}
Tanaka,~F.; Stockmayer,~W.~H. Thermoreversible gelation with junctions of
  variable multiplicity. \emph{Macromolecules} \textbf{1994}, \emph{27},
  3943--3954\relax
\mciteBstWouldAddEndPuncttrue
\mciteSetBstMidEndSepPunct{\mcitedefaultmidpunct}
{\mcitedefaultendpunct}{\mcitedefaultseppunct}\relax
\EndOfBibitem
\bibitem[Tanaka and Ishida(1995)Tanaka, and Ishida]{tanaka1995}
Tanaka,~F.; Ishida,~M. Thermoreversible gelation of hydrated polymers. \emph{J.
  Chem. Soc. Faraday Trans.} \textbf{1995}, \emph{91}, 2663--2670\relax
\mciteBstWouldAddEndPuncttrue
\mciteSetBstMidEndSepPunct{\mcitedefaultmidpunct}
{\mcitedefaultendpunct}{\mcitedefaultseppunct}\relax
\EndOfBibitem
\bibitem[Ishida and Tanaka(1997)Ishida, and Tanaka]{ishida1997}
Ishida,~M.; Tanaka,~F. Theoretical study of the postgel regime in
  thermoreversible gelation. \emph{Macromolecules} \textbf{1997}, \emph{30},
  3900--3909\relax
\mciteBstWouldAddEndPuncttrue
\mciteSetBstMidEndSepPunct{\mcitedefaultmidpunct}
{\mcitedefaultendpunct}{\mcitedefaultseppunct}\relax
\EndOfBibitem
\bibitem[Tanaka(1998)]{tanaka1998}
Tanaka,~F. Thermoreversible gelation of associating polymers. \emph{Physica A:
  Statistical Mechanics and its Applications} \textbf{1998}, \emph{257},
  245--255\relax
\mciteBstWouldAddEndPuncttrue
\mciteSetBstMidEndSepPunct{\mcitedefaultmidpunct}
{\mcitedefaultendpunct}{\mcitedefaultseppunct}\relax
\EndOfBibitem
\bibitem[Tanaka and Ishida(1999)Tanaka, and Ishida]{tanaka1999}
Tanaka,~F.; Ishida,~M. Thermoreversible gelation with two-component networks.
  \emph{Macromolecules} \textbf{1999}, \emph{32}, 1271--1283\relax
\mciteBstWouldAddEndPuncttrue
\mciteSetBstMidEndSepPunct{\mcitedefaultmidpunct}
{\mcitedefaultendpunct}{\mcitedefaultseppunct}\relax
\EndOfBibitem
\bibitem[Tanaka(2002)]{tanaka2002}
Tanaka,~F. Theoretical study of molecular association and thermoreversible
  gelation in polymers. \emph{Polym. J.} \textbf{2002}, \emph{34}, 479\relax
\mciteBstWouldAddEndPuncttrue
\mciteSetBstMidEndSepPunct{\mcitedefaultmidpunct}
{\mcitedefaultendpunct}{\mcitedefaultseppunct}\relax
\EndOfBibitem
\bibitem[Goodwin \latin{et~al.}(2022)Goodwin, McEldrew, de~Souza, Bazant, and
  Kornyshev]{Goodwin2022EDL}
Goodwin,~Z.~A.; McEldrew,~M.; de~Souza,~J.~P.; Bazant,~M.~Z.; Kornyshev,~A.~A.
  Gelation, Clustering and Crowding in the Electrical Double Layer of Ionic
  Liquids. \emph{J. Chem. Phys.} \textbf{2022}, \emph{157}, 094106\relax
\mciteBstWouldAddEndPuncttrue
\mciteSetBstMidEndSepPunct{\mcitedefaultmidpunct}
{\mcitedefaultendpunct}{\mcitedefaultseppunct}\relax
\EndOfBibitem
\bibitem[Goodwin and Kornyshev(2022)Goodwin, and
  Kornyshev]{Goodwin2022Kornyshev}
Goodwin,~Z.~A.; Kornyshev,~A.~A. Cracking Ion Pairs in the Electrical Double
  Layer of Ionic Liquids. \emph{Electrochim. Acta} \textbf{2022}, \emph{434},
  141163\relax
\mciteBstWouldAddEndPuncttrue
\mciteSetBstMidEndSepPunct{\mcitedefaultmidpunct}
{\mcitedefaultendpunct}{\mcitedefaultseppunct}\relax
\EndOfBibitem
\bibitem[Markiewitz \latin{et~al.}(2024)Markiewitz, Goodwin, McEldrew,
  de~Souza, Zhang, Espinosa-Marzal, and Bazant]{Markiewitz2024}
Markiewitz,~D.~M.; Goodwin,~Z.~A.; McEldrew,~M.; de~Souza,~J.~P.; Zhang,~X.;
  Espinosa-Marzal,~R.~M.; Bazant,~M.~Z. Electric field induced associations in
  the double layer of salt-in-ionic-liquid electrolytes. \emph{Faraday
  Discussions} \textbf{2024}, \relax
\mciteBstWouldAddEndPunctfalse
\mciteSetBstMidEndSepPunct{\mcitedefaultmidpunct}
{}{\mcitedefaultseppunct}\relax
\EndOfBibitem
\bibitem[Grimley and Mott(1947)Grimley, and Mott]{grimley1947general}
Grimley,~T.; Mott,~N. I. General and theoretical. The contact between a solid
  and a liquid electrolyte. \emph{Discussions of the Faraday Society}
  \textbf{1947}, \emph{1}, 3--11\relax
\mciteBstWouldAddEndPuncttrue
\mciteSetBstMidEndSepPunct{\mcitedefaultmidpunct}
{\mcitedefaultendpunct}{\mcitedefaultseppunct}\relax
\EndOfBibitem
\bibitem[Borukhov \latin{et~al.}(1997)Borukhov, Andelman, and
  Orland]{borukhov1997steric}
Borukhov,~I.; Andelman,~D.; Orland,~H. Steric effects in electrolytes: A
  modified Poisson-Boltzmann equation. \emph{Physical review letters}
  \textbf{1997}, \emph{79}, 435\relax
\mciteBstWouldAddEndPuncttrue
\mciteSetBstMidEndSepPunct{\mcitedefaultmidpunct}
{\mcitedefaultendpunct}{\mcitedefaultseppunct}\relax
\EndOfBibitem
\bibitem[Kornyshev(2007)]{Kornyshev2007}
Kornyshev,~A.~A. {Double-Layer in Ionic Liquids:  Paradigm Change?} \emph{J.
  Phys. Chem. B} \textbf{2007}, \emph{111}, 5545--5557\relax
\mciteBstWouldAddEndPuncttrue
\mciteSetBstMidEndSepPunct{\mcitedefaultmidpunct}
{\mcitedefaultendpunct}{\mcitedefaultseppunct}\relax
\EndOfBibitem
\bibitem[Bazant \latin{et~al.}(2009)Bazant, Kilic, Storey, and
  Ajdari]{Bazant2009a}
Bazant,~M.~Z.; Kilic,~M.~S.; Storey,~B.~D.; Ajdari,~A. Towards an understanding
  of induced-charge electrokinetics at large applied voltages in concentrated
  solutions. \emph{Advances in colloid and interface science} \textbf{2009},
  \emph{152}, 48--88\relax
\mciteBstWouldAddEndPuncttrue
\mciteSetBstMidEndSepPunct{\mcitedefaultmidpunct}
{\mcitedefaultendpunct}{\mcitedefaultseppunct}\relax
\EndOfBibitem
\bibitem[Hughes(1996)]{hughes1996random}
Hughes,~B.~D. \emph{Random walks and random environments}; Oxford University
  Press, 1996\relax
\mciteBstWouldAddEndPuncttrue
\mciteSetBstMidEndSepPunct{\mcitedefaultmidpunct}
{\mcitedefaultendpunct}{\mcitedefaultseppunct}\relax
\EndOfBibitem
\bibitem[Abrashkin \latin{et~al.}(2007)Abrashkin, Andelman, and
  Orland]{abrashkin2007}
Abrashkin,~A.; Andelman,~D.; Orland,~H. Dipolar Poisson-Boltzmann equation:
  ions and dipoles close to charge interfaces. \emph{Phys. Rev. Lett.}
  \textbf{2007}, \emph{99}, 077801\relax
\mciteBstWouldAddEndPuncttrue
\mciteSetBstMidEndSepPunct{\mcitedefaultmidpunct}
{\mcitedefaultendpunct}{\mcitedefaultseppunct}\relax
\EndOfBibitem
\bibitem[Gongadze \latin{et~al.}(2013)Gongadze, van Rienen, Kralj-Igli{\v{c}},
  and Igli{\v{c}}]{gongadze2013spatial}
Gongadze,~E.; van Rienen,~U.; Kralj-Igli{\v{c}},~V.; Igli{\v{c}},~A. Spatial
  variation of permittivity of an electrolyte solution in contact with a
  charged metal surface: A mini review. \emph{Computer methods in biomechanics
  and biomedical engineering} \textbf{2013}, \emph{16}, 463--480\relax
\mciteBstWouldAddEndPuncttrue
\mciteSetBstMidEndSepPunct{\mcitedefaultmidpunct}
{\mcitedefaultendpunct}{\mcitedefaultseppunct}\relax
\EndOfBibitem
\bibitem[de~Souza \latin{et~al.}(2022)de~Souza, Kornyshev, and
  Bazant]{pedro2022polar}
de~Souza,~J.~P.; Kornyshev,~A.~A.; Bazant,~M.~Z. Polar liquids at charged
  interfaces: A dipolar shell theory. \emph{The Journal of Chemical Physics}
  \textbf{2022}, \emph{156}\relax
\mciteBstWouldAddEndPuncttrue
\mciteSetBstMidEndSepPunct{\mcitedefaultmidpunct}
{\mcitedefaultendpunct}{\mcitedefaultseppunct}\relax
\EndOfBibitem
\bibitem[Chen \latin{et~al.}(2018)Chen, Goodwin, Feng, and Kornyshev]{Chen2017}
Chen,~M.; Goodwin,~Z. A.~H.; Feng,~G.; Kornyshev,~A.~A. {On the temperature
  dependence of the double layer capacitance of ionic liquids}. \emph{J.
  Electroanal. Chem.} \textbf{2018}, \emph{819}, 347--358\relax
\mciteBstWouldAddEndPuncttrue
\mciteSetBstMidEndSepPunct{\mcitedefaultmidpunct}
{\mcitedefaultendpunct}{\mcitedefaultseppunct}\relax
\EndOfBibitem
\bibitem[Goodwin and Kornyshev(2017)Goodwin, and
  Kornyshev]{goodwin2017underscreening}
Goodwin,~Z.~A.; Kornyshev,~A.~A. Underscreening, overscreening and double-layer
  capacitance. \emph{Electrochemistry Communications} \textbf{2017}, \emph{82},
  129--133\relax
\mciteBstWouldAddEndPuncttrue
\mciteSetBstMidEndSepPunct{\mcitedefaultmidpunct}
{\mcitedefaultendpunct}{\mcitedefaultseppunct}\relax
\EndOfBibitem
\bibitem[Zhang \latin{et~al.}(2020)Zhang, Ye, Chen, Goodwin, Feng, Huang, and
  Kornyshev]{Yufan2020}
Zhang,~Y.; Ye,~T.; Chen,~M.; Goodwin,~Z.~A.; Feng,~G.; Huang,~J.;
  Kornyshev,~A.~A. Enforced Freedom: Electric-Field-Induced Declustering of
  Ionic-Liquid Ions in the Electrical Double Layer. \emph{Energy Environ.
  Mater.} \textbf{2020}, \emph{3}, 414–420\relax
\mciteBstWouldAddEndPuncttrue
\mciteSetBstMidEndSepPunct{\mcitedefaultmidpunct}
{\mcitedefaultendpunct}{\mcitedefaultseppunct}\relax
\EndOfBibitem
\bibitem[Goodwin \latin{et~al.}(2017)Goodwin, Feng, and
  Kornyshev]{goodwin2017mean}
Goodwin,~Z.~A.; Feng,~G.; Kornyshev,~A.~A. Mean-field theory of electrical
  double layer in ionic liquids with account of short-range correlations.
  \emph{Electrochim. Acta} \textbf{2017}, \emph{225}, 190--197\relax
\mciteBstWouldAddEndPuncttrue
\mciteSetBstMidEndSepPunct{\mcitedefaultmidpunct}
{\mcitedefaultendpunct}{\mcitedefaultseppunct}\relax
\EndOfBibitem
\bibitem[Jitvisate and Seddon(2018)Jitvisate, and Seddon]{Jitvisate2018}
Jitvisate,~M.; Seddon,~J. R.~T. Direct Measurement of the Differential
  Capacitance of Solvent-Free and Dilute Ionic Liquids. \emph{J. Phys. Chem.
  Lett.} \textbf{2018}, \emph{9}, 26--131\relax
\mciteBstWouldAddEndPuncttrue
\mciteSetBstMidEndSepPunct{\mcitedefaultmidpunct}
{\mcitedefaultendpunct}{\mcitedefaultseppunct}\relax
\EndOfBibitem
\bibitem[Guldbrand \latin{et~al.}(1984)Guldbrand, J{\"o}nsson, Wennerstr{\"o}m,
  and Linse]{guldbrand1984electrical}
Guldbrand,~L.; J{\"o}nsson,~B.; Wennerstr{\"o}m,~H.; Linse,~P. Electrical
  double layer forces. A Monte Carlo study. \emph{The Journal of chemical
  physics} \textbf{1984}, \emph{80}, 2221--2228\relax
\mciteBstWouldAddEndPuncttrue
\mciteSetBstMidEndSepPunct{\mcitedefaultmidpunct}
{\mcitedefaultendpunct}{\mcitedefaultseppunct}\relax
\EndOfBibitem
\bibitem[Kjellander and Mar{\v{c}}elja(1986)Kjellander, and
  Mar{\v{c}}elja]{kjellander1986interaction}
Kjellander,~R.; Mar{\v{c}}elja,~S. Interaction of charged surfaces in
  electrolyte solutions. \emph{Chemical physics letters} \textbf{1986},
  \emph{127}, 402--407\relax
\mciteBstWouldAddEndPuncttrue
\mciteSetBstMidEndSepPunct{\mcitedefaultmidpunct}
{\mcitedefaultendpunct}{\mcitedefaultseppunct}\relax
\EndOfBibitem
\bibitem[Grochowski and Trylska(2008)Grochowski, and
  Trylska]{grochowski2008continuum}
Grochowski,~P.; Trylska,~J. Continuum molecular electrostatics, salt effects,
  and counterion binding—a review of the Poisson--Boltzmann theory and its
  modifications. \emph{Biopolymers: Original Research on Biomolecules}
  \textbf{2008}, \emph{89}, 93--113\relax
\mciteBstWouldAddEndPuncttrue
\mciteSetBstMidEndSepPunct{\mcitedefaultmidpunct}
{\mcitedefaultendpunct}{\mcitedefaultseppunct}\relax
\EndOfBibitem
\bibitem[Netz(2001)]{netz2001electrostatistics}
Netz,~R.~R. Electrostatistics of counter-ions at and between planar charged
  walls: From Poisson-Boltzmann to the strong-coupling theory. \emph{The
  European Physical Journal E} \textbf{2001}, \emph{5}, 557--574\relax
\mciteBstWouldAddEndPuncttrue
\mciteSetBstMidEndSepPunct{\mcitedefaultmidpunct}
{\mcitedefaultendpunct}{\mcitedefaultseppunct}\relax
\EndOfBibitem
\bibitem[de~Souza \latin{et~al.}(2020)de~Souza, Goodwin, McEldrew, Kornyshev,
  and Bazant]{Pedro2020}
de~Souza,~J.~P.; Goodwin,~Z.~A.; McEldrew,~M.; Kornyshev,~A.~A.; Bazant,~M.~Z.
  Interfacial layering in the electrical double layer of ionic liquids.
  \emph{Phys. Rev. Lett.} \textbf{2020}, \emph{125}, 116001\relax
\mciteBstWouldAddEndPuncttrue
\mciteSetBstMidEndSepPunct{\mcitedefaultmidpunct}
{\mcitedefaultendpunct}{\mcitedefaultseppunct}\relax
\EndOfBibitem
\bibitem[Avni \latin{et~al.}(2020)Avni, Adar, and Andelman]{avni2020charge}
Avni,~Y.; Adar,~R.~M.; Andelman,~D. Charge oscillations in ionic liquids: A
  microscopic cluster model. \emph{Physical Review E} \textbf{2020},
  \emph{101}, 010601\relax
\mciteBstWouldAddEndPuncttrue
\mciteSetBstMidEndSepPunct{\mcitedefaultmidpunct}
{\mcitedefaultendpunct}{\mcitedefaultseppunct}\relax
\EndOfBibitem
\bibitem[Adar \latin{et~al.}(2019)Adar, Safran, Diamant, and
  Andelman]{adar2019screening}
Adar,~R.~M.; Safran,~S.~A.; Diamant,~H.; Andelman,~D. Screening length for
  finite-size ions in concentrated electrolytes. \emph{Physical Review E}
  \textbf{2019}, \emph{100}, 042615\relax
\mciteBstWouldAddEndPuncttrue
\mciteSetBstMidEndSepPunct{\mcitedefaultmidpunct}
{\mcitedefaultendpunct}{\mcitedefaultseppunct}\relax
\EndOfBibitem
\bibitem[Levy \latin{et~al.}(2019)Levy, McEldrew, and Bazant]{levy2019spin}
Levy,~A.; McEldrew,~M.; Bazant,~M.~Z. Spin-glass charge ordering in ionic
  liquids. \emph{Physical Review Materials} \textbf{2019}, \emph{3},
  055606\relax
\mciteBstWouldAddEndPuncttrue
\mciteSetBstMidEndSepPunct{\mcitedefaultmidpunct}
{\mcitedefaultendpunct}{\mcitedefaultseppunct}\relax
\EndOfBibitem
\bibitem[Hiemenz and Rajagopalan(2016)Hiemenz, and
  Rajagopalan]{hiemenz2016principles}
Hiemenz,~P.~C.; Rajagopalan,~R. \emph{Principles of Colloid and Surface
  Chemistry, revised and expanded}; CRC press, 2016\relax
\mciteBstWouldAddEndPuncttrue
\mciteSetBstMidEndSepPunct{\mcitedefaultmidpunct}
{\mcitedefaultendpunct}{\mcitedefaultseppunct}\relax
\EndOfBibitem
\bibitem[Chu and Bazant(2007)Chu, and Bazant]{chu2007surface}
Chu,~K.~T.; Bazant,~M.~Z. Surface conservation laws at microscopically diffuse
  interfaces. \emph{Journal of colloid and interface science} \textbf{2007},
  \emph{315}, 319--329\relax
\mciteBstWouldAddEndPuncttrue
\mciteSetBstMidEndSepPunct{\mcitedefaultmidpunct}
{\mcitedefaultendpunct}{\mcitedefaultseppunct}\relax
\EndOfBibitem
\bibitem[Mason \latin{et~al.}(2015)Mason, Ansell, Neilson, and
  Rempe]{mason2015neutron}
Mason,~P.; Ansell,~S.; Neilson,~G.; Rempe,~S. Neutron scattering studies of the
  hydration structure of Li+. \emph{The Journal of Physical Chemistry B}
  \textbf{2015}, \emph{119}, 2003--2009\relax
\mciteBstWouldAddEndPuncttrue
\mciteSetBstMidEndSepPunct{\mcitedefaultmidpunct}
{\mcitedefaultendpunct}{\mcitedefaultseppunct}\relax
\EndOfBibitem
\bibitem[Hanwell \latin{et~al.}(2012)Hanwell, Curtis, Lonie, Vandermeersch,
  Zurek, and Hutchison]{hanwell2012avogadro}
Hanwell,~M.~D.; Curtis,~D.~E.; Lonie,~D.~C.; Vandermeersch,~T.; Zurek,~E.;
  Hutchison,~G.~R. Avogadro: an advanced semantic chemical editor,
  visualization, and analysis platform. \emph{Journal of cheminformatics}
  \textbf{2012}, \emph{4}, 1--17\relax
\mciteBstWouldAddEndPuncttrue
\mciteSetBstMidEndSepPunct{\mcitedefaultmidpunct}
{\mcitedefaultendpunct}{\mcitedefaultseppunct}\relax
\EndOfBibitem
\bibitem[Kilic \latin{et~al.}(2007)Kilic, Bazant, and Ajdari]{kilic2007a}
Kilic,~M.~S.; Bazant,~M.~Z.; Ajdari,~A. Steric effects in the dynamics of
  electrolytes at large applied voltages. I. Double-layer charging.
  \emph{Physical review E} \textbf{2007}, \emph{75}, 021502\relax
\mciteBstWouldAddEndPuncttrue
\mciteSetBstMidEndSepPunct{\mcitedefaultmidpunct}
{\mcitedefaultendpunct}{\mcitedefaultseppunct}\relax
\EndOfBibitem
\bibitem[Budkov \latin{et~al.}(2018)Budkov, Kolesnikov, Goodwin, Kiselev, and
  Kornyshev]{BBKGK}
Budkov,~Y.~A.; Kolesnikov,~A.~L.; Goodwin,~Z. A.~H.; Kiselev,~M.;
  Kornyshev,~A.~A. Theory of Electrosorption of Water from Ionic Liquids.
  \emph{Electrochim. Acta} \textbf{2018}, \emph{284}, 346--354\relax
\mciteBstWouldAddEndPuncttrue
\mciteSetBstMidEndSepPunct{\mcitedefaultmidpunct}
{\mcitedefaultendpunct}{\mcitedefaultseppunct}\relax
\EndOfBibitem
\bibitem[Zheng \latin{et~al.}(2023)Zheng, Goodwin, Gopalakrishnan, Hoane, Han,
  Zhang, Hawthorne, Batteas, Gewirth, and Espinosa-Marzal]{Zheng2023water}
Zheng,~Q.; Goodwin,~Z.~A.; Gopalakrishnan,~V.; Hoane,~A.~G.; Han,~M.;
  Zhang,~R.; Hawthorne,~N.; Batteas,~J.~D.; Gewirth,~A.~A.;
  Espinosa-Marzal,~R.~M. Water in the Electrical Double Layer of Ionic Liquids
  on Graphene. \emph{ACS Nano} \textbf{2023}, \emph{17}, 9347--9360\relax
\mciteBstWouldAddEndPuncttrue
\mciteSetBstMidEndSepPunct{\mcitedefaultmidpunct}
{\mcitedefaultendpunct}{\mcitedefaultseppunct}\relax
\EndOfBibitem
\bibitem[Goodwin and Kornyshev(2017)Goodwin, and Kornyshev]{Goodwin2017}
Goodwin,~Z. A.~H.; Kornyshev,~A.~A. {Underscreening, overscreening and
  double-layer capacitance}. \emph{Electrochem. Commun.} \textbf{2017},
  \emph{82}, 129--133\relax
\mciteBstWouldAddEndPuncttrue
\mciteSetBstMidEndSepPunct{\mcitedefaultmidpunct}
{\mcitedefaultendpunct}{\mcitedefaultseppunct}\relax
\EndOfBibitem
\bibitem[Brug \latin{et~al.}(1984)Brug, van~den Eeden, Sluyters-Rehbach, and
  Sluyters]{brug1984analysis}
Brug,~G.; van~den Eeden,~A.~L.; Sluyters-Rehbach,~M.; Sluyters,~J.~H. The
  analysis of electrode impedances complicated by the presence of a constant
  phase element. \emph{Journal of electroanalytical chemistry and interfacial
  electrochemistry} \textbf{1984}, \emph{176}, 275--295\relax
\mciteBstWouldAddEndPuncttrue
\mciteSetBstMidEndSepPunct{\mcitedefaultmidpunct}
{\mcitedefaultendpunct}{\mcitedefaultseppunct}\relax
\EndOfBibitem
\bibitem[Andersson \latin{et~al.}(2005)Andersson, Krebs, and
  Morgner]{andersson2005activity}
Andersson,~G.; Krebs,~T.; Morgner,~H. Activity of surface active substances
  determined from their surface excess. \emph{Physical Chemistry Chemical
  Physics} \textbf{2005}, \emph{7}, 136--142\relax
\mciteBstWouldAddEndPuncttrue
\mciteSetBstMidEndSepPunct{\mcitedefaultmidpunct}
{\mcitedefaultendpunct}{\mcitedefaultseppunct}\relax
\EndOfBibitem
\bibitem[Chu and Bazant(2007)Chu, and Bazant]{chu2007}
Chu,~K.~T.; Bazant,~M.~Z. Surface conservation laws at microscopically diffuse
  interfaces. \emph{J. Colloid Interface Sci.} \textbf{2007}, \emph{315},
  319--329\relax
\mciteBstWouldAddEndPuncttrue
\mciteSetBstMidEndSepPunct{\mcitedefaultmidpunct}
{\mcitedefaultendpunct}{\mcitedefaultseppunct}\relax
\EndOfBibitem
\bibitem[Hoane \latin{et~al.}(2024)Hoane, Zheng, Maldonado, Espinosa-Marzal,
  and Gewirth]{hoane2024impact}
Hoane,~A.~G.; Zheng,~Q.; Maldonado,~N.~D.; Espinosa-Marzal,~R.~M.;
  Gewirth,~A.~A. Impact of Multivalent Cations on Interfacial Layering in
  Water-In-Salt Electrolytes. \emph{ACS Applied Energy Materials}
  \textbf{2024}, \relax
\mciteBstWouldAddEndPunctfalse
\mciteSetBstMidEndSepPunct{\mcitedefaultmidpunct}
{}{\mcitedefaultseppunct}\relax
\EndOfBibitem
\bibitem[Barducci \latin{et~al.}(2011)Barducci, Bonomi, and
  Parrinello]{barducci2011metadynamics}
Barducci,~A.; Bonomi,~M.; Parrinello,~M. Metadynamics. \emph{Wiley
  Interdisciplinary Reviews: Computational Molecular Science} \textbf{2011},
  \emph{1}, 826--843\relax
\mciteBstWouldAddEndPuncttrue
\mciteSetBstMidEndSepPunct{\mcitedefaultmidpunct}
{\mcitedefaultendpunct}{\mcitedefaultseppunct}\relax
\EndOfBibitem
\bibitem[Valsson \latin{et~al.}(2016)Valsson, Tiwary, and
  Parrinello]{valsson2016enhancing}
Valsson,~O.; Tiwary,~P.; Parrinello,~M. Enhancing important fluctuations: Rare
  events and metadynamics from a conceptual viewpoint. \emph{Annual review of
  physical chemistry} \textbf{2016}, \emph{67}, 159--184\relax
\mciteBstWouldAddEndPuncttrue
\mciteSetBstMidEndSepPunct{\mcitedefaultmidpunct}
{\mcitedefaultendpunct}{\mcitedefaultseppunct}\relax
\EndOfBibitem
\bibitem[H{\'e}nin \latin{et~al.}(2022)H{\'e}nin, Leli{\`e}vre, Shirts,
  Valsson, and Delemotte]{henin2022enhanced}
H{\'e}nin,~J.; Leli{\`e}vre,~T.; Shirts,~M.~R.; Valsson,~O.; Delemotte,~L.
  Enhanced sampling methods for molecular dynamics simulations. \emph{arXiv
  preprint arXiv:2202.04164} \textbf{2022}, \relax
\mciteBstWouldAddEndPunctfalse
\mciteSetBstMidEndSepPunct{\mcitedefaultmidpunct}
{}{\mcitedefaultseppunct}\relax
\EndOfBibitem
\bibitem[Hettige \latin{et~al.}(2014)Hettige, Araque, and
  Margulis]{hettige2014bicontinuity}
Hettige,~J.~J.; Araque,~J.~C.; Margulis,~C.~J. Bicontinuity and multiple length
  scale ordering in triphilic hydrogen-bonding ionic liquids. \emph{The Journal
  of Physical Chemistry B} \textbf{2014}, \emph{118}, 12706--12716\relax
\mciteBstWouldAddEndPuncttrue
\mciteSetBstMidEndSepPunct{\mcitedefaultmidpunct}
{\mcitedefaultendpunct}{\mcitedefaultseppunct}\relax
\EndOfBibitem
\bibitem[Zhang \latin{et~al.}(2024)Zhang, Goodwin, Hoane, Deptula, Markiewitz,
  Molinari, Zheng, Li, McEldrew, Kozinsky, Bazant, Leal, Atkin, Gewirth,
  Rutland, and Espinosa-Marzal]{Zhang2024}
Zhang,~X. \latin{et~al.}  Long-Range Interactions in Salt-in-Ionic Liquids.
  \emph{10.26434/chemrxiv-2024-hs7sf} \textbf{2024}, \relax
\mciteBstWouldAddEndPunctfalse
\mciteSetBstMidEndSepPunct{\mcitedefaultmidpunct}
{}{\mcitedefaultseppunct}\relax
\EndOfBibitem
\bibitem[Pajkossy and Kolb(2011)Pajkossy, and Kolb]{pajkossy2011interfacial}
Pajkossy,~T.; Kolb,~D.~M. The interfacial capacitance of Au (100) in an ionic
  liquid, 1-butyl-3-methyl-imidazolium hexafluorophosphate.
  \emph{Electrochemistry communications} \textbf{2011}, \emph{13},
  284--286\relax
\mciteBstWouldAddEndPuncttrue
\mciteSetBstMidEndSepPunct{\mcitedefaultmidpunct}
{\mcitedefaultendpunct}{\mcitedefaultseppunct}\relax
\EndOfBibitem
\bibitem[Dr{\"u}schler and Roling(2011)Dr{\"u}schler, and
  Roling]{druschler2011commentary}
Dr{\"u}schler,~M.; Roling,~B. Commentary on ‘The interface between Au (1 1 1)
  and an ionic liquid’. \emph{Electrochimica Acta} \textbf{2011}, \emph{56},
  7243--7245\relax
\mciteBstWouldAddEndPuncttrue
\mciteSetBstMidEndSepPunct{\mcitedefaultmidpunct}
{\mcitedefaultendpunct}{\mcitedefaultseppunct}\relax
\EndOfBibitem
\bibitem[Pajkossy(2011)]{pajkossy2011response}
Pajkossy,~T. Response to the Commentary of Marcel Dr{\"u}schler and Bernhard
  Roling on ‘The interface between Au (1 1 1) and an ionic liquid’.
  \emph{Electrochimica Acta} \textbf{2011}, \emph{56}, 7246--7247\relax
\mciteBstWouldAddEndPuncttrue
\mciteSetBstMidEndSepPunct{\mcitedefaultmidpunct}
{\mcitedefaultendpunct}{\mcitedefaultseppunct}\relax
\EndOfBibitem
\bibitem[Wang and Pilon(2012)Wang, and Pilon]{wang2012intrinsic}
Wang,~H.; Pilon,~L. Intrinsic limitations of impedance measurements in
  determining electric double layer capacitances. \emph{Electrochimica Acta}
  \textbf{2012}, \emph{63}, 55--63\relax
\mciteBstWouldAddEndPuncttrue
\mciteSetBstMidEndSepPunct{\mcitedefaultmidpunct}
{\mcitedefaultendpunct}{\mcitedefaultseppunct}\relax
\EndOfBibitem
\bibitem[Roling and Dr{\"u}schler(2012)Roling, and
  Dr{\"u}schler]{roling2012comments}
Roling,~B.; Dr{\"u}schler,~M. Comments on “Intrinsic limitations of impedance
  measurements in determining electric double layer capacitances” by H. Wang
  and L. Pilon [Electrochim. Acta 63 (2012) 55]. \emph{Electrochimica Acta}
  \textbf{2012}, \emph{76}, 526--528\relax
\mciteBstWouldAddEndPuncttrue
\mciteSetBstMidEndSepPunct{\mcitedefaultmidpunct}
{\mcitedefaultendpunct}{\mcitedefaultseppunct}\relax
\EndOfBibitem
\bibitem[Wang and Pilon(2012)Wang, and Pilon]{wang2012reply}
Wang,~H.; Pilon,~L. Reply to comments on “Intrinsic limitations of impedance
  measurements in determining electric double layer capacitances” by H. Wang,
  L. Pilon [Electrochimica Acta 63 (2012) 55]. \emph{Electrochimica Acta}
  \textbf{2012}, \emph{76}, 529--531\relax
\mciteBstWouldAddEndPuncttrue
\mciteSetBstMidEndSepPunct{\mcitedefaultmidpunct}
{\mcitedefaultendpunct}{\mcitedefaultseppunct}\relax
\EndOfBibitem
\bibitem[Fedorov and Kornyshev(2014)Fedorov, and Kornyshev]{Fedorov2014}
Fedorov,~M.~V.; Kornyshev,~A.~A. {Ionic Liquids at Electrified Interfaces}.
  \emph{Chem. Rev.} \textbf{2014}, \emph{114}, 2978--3036\relax
\mciteBstWouldAddEndPuncttrue
\mciteSetBstMidEndSepPunct{\mcitedefaultmidpunct}
{\mcitedefaultendpunct}{\mcitedefaultseppunct}\relax
\EndOfBibitem
\bibitem[Pajkossy(1994)]{pajkossy1994impedance}
Pajkossy,~T. Impedance of rough capacitive electrodes. \emph{Journal of
  Electroanalytical Chemistry} \textbf{1994}, \emph{364}, 111--125\relax
\mciteBstWouldAddEndPuncttrue
\mciteSetBstMidEndSepPunct{\mcitedefaultmidpunct}
{\mcitedefaultendpunct}{\mcitedefaultseppunct}\relax
\EndOfBibitem
\bibitem[Lockett \latin{et~al.}(2010)Lockett, Horne, Sedev, Rodopoulos, and
  Ralston]{lockett2010differential}
Lockett,~V.; Horne,~M.; Sedev,~R.; Rodopoulos,~T.; Ralston,~J. Differential
  capacitance of the double layer at the electrode/ionic liquids interface.
  \emph{Physical Chemistry Chemical Physics} \textbf{2010}, \emph{12},
  12499--12512\relax
\mciteBstWouldAddEndPuncttrue
\mciteSetBstMidEndSepPunct{\mcitedefaultmidpunct}
{\mcitedefaultendpunct}{\mcitedefaultseppunct}\relax
\EndOfBibitem
\bibitem[J{\"a}nsch \latin{et~al.}(2015)J{\"a}nsch, Wallauer, and
  Roling]{jansch2015influence}
J{\"a}nsch,~T.; Wallauer,~J.; Roling,~B. Influence of electrode roughness on
  double layer formation in ionic liquids. \emph{The Journal of Physical
  Chemistry C} \textbf{2015}, \emph{119}, 4620--4626\relax
\mciteBstWouldAddEndPuncttrue
\mciteSetBstMidEndSepPunct{\mcitedefaultmidpunct}
{\mcitedefaultendpunct}{\mcitedefaultseppunct}\relax
\EndOfBibitem
\bibitem[Oll \latin{et~al.}(2017)Oll, Romann, Siimenson, and
  Lust]{oll2017influence}
Oll,~O.; Romann,~T.; Siimenson,~C.; Lust,~E. Influence of chemical composition
  of electrode material on the differential capacitance characteristics of the
  ionic liquid| electrode interface. \emph{Electrochemistry Communications}
  \textbf{2017}, \emph{82}, 39--42\relax
\mciteBstWouldAddEndPuncttrue
\mciteSetBstMidEndSepPunct{\mcitedefaultmidpunct}
{\mcitedefaultendpunct}{\mcitedefaultseppunct}\relax
\EndOfBibitem
\bibitem[Torabi \latin{et~al.}(2017)Torabi, Cherry, Duijnstee, Le~Corre, Qiu,
  Hummelen, Palasantzas, and Koster]{torabi2017rough}
Torabi,~S.; Cherry,~M.; Duijnstee,~E.~A.; Le~Corre,~V.~M.; Qiu,~L.;
  Hummelen,~J.~C.; Palasantzas,~G.; Koster,~L. J.~A. Rough electrode creates
  excess capacitance in thin-film capacitors. \emph{ACS applied materials \&
  interfaces} \textbf{2017}, \emph{9}, 27290--27297\relax
\mciteBstWouldAddEndPuncttrue
\mciteSetBstMidEndSepPunct{\mcitedefaultmidpunct}
{\mcitedefaultendpunct}{\mcitedefaultseppunct}\relax
\EndOfBibitem
\bibitem[Aslyamov \latin{et~al.}(2021)Aslyamov, Sinkov, and
  Akhatov]{aslyamov2021electrolyte}
Aslyamov,~T.; Sinkov,~K.; Akhatov,~I. Electrolyte structure near electrodes
  with molecular-size roughness. \emph{Physical Review E} \textbf{2021},
  \emph{103}, L060102\relax
\mciteBstWouldAddEndPuncttrue
\mciteSetBstMidEndSepPunct{\mcitedefaultmidpunct}
{\mcitedefaultendpunct}{\mcitedefaultseppunct}\relax
\EndOfBibitem
\bibitem[Zhang \latin{et~al.}(2023)Zhang, Yu, Suo, Zhuang, He, Zhang, Hong, and
  Tan]{zhang2023evolution}
Zhang,~L.; Yu,~Y.; Suo,~L.; Zhuang,~W.; He,~L.; Zhang,~X.; Hong,~L.; Tan,~P.
  The evolution of anionic nanoclusters at the electrode interface in
  water-in-salt electrolytes. \emph{Physical Chemistry Chemical Physics}
  \textbf{2023}, \emph{25}, 10301--10312\relax
\mciteBstWouldAddEndPuncttrue
\mciteSetBstMidEndSepPunct{\mcitedefaultmidpunct}
{\mcitedefaultendpunct}{\mcitedefaultseppunct}\relax
\EndOfBibitem
\bibitem[Finney and Salvalaglio(2024)Finney, and
  Salvalaglio]{finney2024properties}
Finney,~A.~R.; Salvalaglio,~M. Properties of aqueous electrolyte solutions at
  carbon electrodes: effects of concentration and surface charge on solution
  structure, ion clustering and thermodynamics in the electric double layer.
  \emph{Faraday Discussions} \textbf{2024}, \emph{249}, 334--362\relax
\mciteBstWouldAddEndPuncttrue
\mciteSetBstMidEndSepPunct{\mcitedefaultmidpunct}
{\mcitedefaultendpunct}{\mcitedefaultseppunct}\relax
\EndOfBibitem
\bibitem[Aggarwal \latin{et~al.}(2024)Aggarwal, Gordiz, Baskin, Vivona,
  Stenlid, Lawson, Grossman, and Shao-Horn]{aggarwal2024revealing}
Aggarwal,~A.; Gordiz,~K.; Baskin,~A.; Vivona,~D.; Stenlid,~J.~H.;
  Lawson,~J.~W.; Grossman,~J.~C.; Shao-Horn,~Y. Revealing the Molecular Origin
  of Driving Forces and Thermodynamic Barriers for Li+ Ion Transport to
  Electrode-Electrolyte Interfaces. \textbf{2024}, \relax
\mciteBstWouldAddEndPunctfalse
\mciteSetBstMidEndSepPunct{\mcitedefaultmidpunct}
{}{\mcitedefaultseppunct}\relax
\EndOfBibitem
\bibitem[Yu \latin{et~al.}(2024)Yu, Chiang, Dhinojwala, Bonn, Hunger, and
  Nagata]{yu2024flipping}
Yu,~C.-C.; Chiang,~K.-Y.; Dhinojwala,~A.; Bonn,~M.; Hunger,~J.; Nagata,~Y.
  Flipping Water Orientation at the Surface of Water-in-Salt and Salt-in-Water
  Solutions. \emph{The Journal of Physical Chemistry Letters} \textbf{2024},
  \emph{15}, 10265--10271\relax
\mciteBstWouldAddEndPuncttrue
\mciteSetBstMidEndSepPunct{\mcitedefaultmidpunct}
{\mcitedefaultendpunct}{\mcitedefaultseppunct}\relax
\EndOfBibitem
\bibitem[Wu \latin{et~al.}(2023)Wu, McDowell, and Qi]{wu2023effect}
Wu,~Q.; McDowell,~M.~T.; Qi,~Y. Effect of the electric double layer (EDL) in
  multicomponent electrolyte reduction and solid electrolyte interphase (SEI)
  formation in lithium batteries. \emph{Journal of the American Chemical
  Society} \textbf{2023}, \emph{145}, 2473--2484\relax
\mciteBstWouldAddEndPuncttrue
\mciteSetBstMidEndSepPunct{\mcitedefaultmidpunct}
{\mcitedefaultendpunct}{\mcitedefaultseppunct}\relax
\EndOfBibitem
\bibitem[Xu \latin{et~al.}(2007)Xu, Lam, Zhang, Jow, and
  Curtis]{xu2007solvation}
Xu,~K.; Lam,~Y.; Zhang,~S.~S.; Jow,~T.~R.; Curtis,~T.~B. Solvation sheath of
  Li+ in nonaqueous electrolytes and its implication of graphite/electrolyte
  interface chemistry. \emph{The Journal of Physical Chemistry C}
  \textbf{2007}, \emph{111}, 7411--7421\relax
\mciteBstWouldAddEndPuncttrue
\mciteSetBstMidEndSepPunct{\mcitedefaultmidpunct}
{\mcitedefaultendpunct}{\mcitedefaultseppunct}\relax
\EndOfBibitem
\bibitem[von Wald~Cresce \latin{et~al.}(2012)von Wald~Cresce, Borodin, and
  Xu]{von2012correlating}
von Wald~Cresce,~A.; Borodin,~O.; Xu,~K. Correlating Li+ solvation sheath
  structure with interphasial chemistry on graphite. \emph{The Journal of
  Physical Chemistry C} \textbf{2012}, \emph{116}, 26111--26117\relax
\mciteBstWouldAddEndPuncttrue
\mciteSetBstMidEndSepPunct{\mcitedefaultmidpunct}
{\mcitedefaultendpunct}{\mcitedefaultseppunct}\relax
\EndOfBibitem
\bibitem[Pinson and Bazant(2012)Pinson, and Bazant]{pinson2012theory}
Pinson,~M.~B.; Bazant,~M.~Z. Theory of SEI formation in rechargeable batteries:
  capacity fade, accelerated aging and lifetime prediction. \emph{Journal of
  the Electrochemical Society} \textbf{2012}, \emph{160}, A243\relax
\mciteBstWouldAddEndPuncttrue
\mciteSetBstMidEndSepPunct{\mcitedefaultmidpunct}
{\mcitedefaultendpunct}{\mcitedefaultseppunct}\relax
\EndOfBibitem
\bibitem[Rakov \latin{et~al.}(2020)Rakov, Chen, Ferdousi, Li, Pathirana,
  Simonov, Howlett, Atkin, and Forsyth]{rakov2020engineering}
Rakov,~D.~A.; Chen,~F.; Ferdousi,~S.~A.; Li,~H.; Pathirana,~T.; Simonov,~A.~N.;
  Howlett,~P.~C.; Atkin,~R.; Forsyth,~M. Engineering high-energy-density sodium
  battery anodes for improved cycling with superconcentrated ionic-liquid
  electrolytes. \emph{Nature materials} \textbf{2020}, \emph{19},
  1096--1101\relax
\mciteBstWouldAddEndPuncttrue
\mciteSetBstMidEndSepPunct{\mcitedefaultmidpunct}
{\mcitedefaultendpunct}{\mcitedefaultseppunct}\relax
\EndOfBibitem
\bibitem[Gebbie \latin{et~al.}(2013)Gebbie, Valtiner, Banquy, Fox, Henderson,
  and Israelachvili]{Gebbie2013}
Gebbie,~M.~A.; Valtiner,~M.; Banquy,~X.; Fox,~E.~T.; Henderson,~W.~A.;
  Israelachvili,~J.~N. {Ionic liquids behave as dilute electrolyte solutions}.
  \emph{Proceedings of the National Academy of Sciences} \textbf{2013},
  \emph{110}, 9674--9679\relax
\mciteBstWouldAddEndPuncttrue
\mciteSetBstMidEndSepPunct{\mcitedefaultmidpunct}
{\mcitedefaultendpunct}{\mcitedefaultseppunct}\relax
\EndOfBibitem
\bibitem[Espinosa-Marzal \latin{et~al.}(2014)Espinosa-Marzal, Arcifa, Rossi,
  and Spencer]{Marzal2014}
Espinosa-Marzal,~R.~M.; Arcifa,~A.; Rossi,~A.; Spencer,~N.~D. Microslips to
  ``Avalanches'' in Confined, Molecular Layers of Ionic Liquids. \emph{J. Phys.
  Chem. Lett.} \textbf{2014}, \emph{5}, 179--184\relax
\mciteBstWouldAddEndPuncttrue
\mciteSetBstMidEndSepPunct{\mcitedefaultmidpunct}
{\mcitedefaultendpunct}{\mcitedefaultseppunct}\relax
\EndOfBibitem
\bibitem[Gebbie \latin{et~al.}(2015)Gebbie, Dobes, Valtiner, and
  Israelachvili]{Gebbie2015}
Gebbie,~M.~A.; Dobes,~H.~A.; Valtiner,~M.; Israelachvili,~J.~N. {Long-range
  electrostatic screening in ionic liquids}. \emph{Proceedings of the National
  Academy of Sciences} \textbf{2015}, \emph{112}, 7432–7437\relax
\mciteBstWouldAddEndPuncttrue
\mciteSetBstMidEndSepPunct{\mcitedefaultmidpunct}
{\mcitedefaultendpunct}{\mcitedefaultseppunct}\relax
\EndOfBibitem
\bibitem[Smith \latin{et~al.}(2016)Smith, Lee, and Perkin]{Smith2016}
Smith,~A.~M.; Lee,~A.~A.; Perkin,~S. {The Electrostatic Screening Length in
  Concentrated Electrolytes Increases with Concentration}. \emph{The Journal of
  Physical Chemistry Letters} \textbf{2016}, \emph{7}, 2157--2163\relax
\mciteBstWouldAddEndPuncttrue
\mciteSetBstMidEndSepPunct{\mcitedefaultmidpunct}
{\mcitedefaultendpunct}{\mcitedefaultseppunct}\relax
\EndOfBibitem
\bibitem[Han \latin{et~al.}(2020)Han, Kim, Leal, Negrito, Batteas, and
  Espinosa-Marzal]{Han2020}
Han,~M.; Kim,~H.; Leal,~C.; Negrito,~M.; Batteas,~J.~D.; Espinosa-Marzal,~R.~M.
  Insight into the Electrical Double Layer of Ionic Liquids Revealed through
  Its Temporal Evolution. \emph{Adv Mater Interfaces} \textbf{2020}, \emph{7},
  2001313\relax
\mciteBstWouldAddEndPuncttrue
\mciteSetBstMidEndSepPunct{\mcitedefaultmidpunct}
{\mcitedefaultendpunct}{\mcitedefaultseppunct}\relax
\EndOfBibitem
\bibitem[Fung and Perkin(2023)Fung, and Perkin]{fung2023structure}
Fung,~Y.~C.; Perkin,~S. Structure and anomalous underscreening in ethylammonium
  nitrate solutions confined between two mica surfaces. \emph{Faraday
  Discussions} \textbf{2023}, \emph{246}, 370--386\relax
\mciteBstWouldAddEndPuncttrue
\mciteSetBstMidEndSepPunct{\mcitedefaultmidpunct}
{\mcitedefaultendpunct}{\mcitedefaultseppunct}\relax
\EndOfBibitem
\bibitem[Feng \latin{et~al.}(2019)Feng, Chen, Bi, Goodwin, Postnikov,
  Brilliantov, Urbakh, and Kornyshev]{feng2019free}
Feng,~G.; Chen,~M.; Bi,~S.; Goodwin,~Z.~A.; Postnikov,~E.~B.; Brilliantov,~N.;
  Urbakh,~M.; Kornyshev,~A.~A. Free and Bound States of Ions in Ionic Liquids,
  Conductivity, and Underscreening Paradox. \emph{Phys. Rev. X} \textbf{2019},
  \emph{9}, 021024\relax
\mciteBstWouldAddEndPuncttrue
\mciteSetBstMidEndSepPunct{\mcitedefaultmidpunct}
{\mcitedefaultendpunct}{\mcitedefaultseppunct}\relax
\EndOfBibitem
\bibitem[Jones \latin{et~al.}(2021)Jones, Coupette, H{\"a}rtel, \latin{et~al.}
  others]{jones2021bayesian}
Jones,~P.; Coupette,~F.; H{\"a}rtel,~A., \latin{et~al.}  Bayesian unsupervised
  learning reveals hidden structure in concentrated electrolytes. \emph{The
  Journal of Chemical Physics} \textbf{2021}, \emph{154}\relax
\mciteBstWouldAddEndPuncttrue
\mciteSetBstMidEndSepPunct{\mcitedefaultmidpunct}
{\mcitedefaultendpunct}{\mcitedefaultseppunct}\relax
\EndOfBibitem
\bibitem[Krucker-Velasquez and Swan(2021)Krucker-Velasquez, and
  Swan]{krucker2021underscreening}
Krucker-Velasquez,~E.; Swan,~J.~W. Underscreening and hidden ion structures in
  large scale simulations of concentrated electrolytes. \emph{The Journal of
  Chemical Physics} \textbf{2021}, \emph{155}\relax
\mciteBstWouldAddEndPuncttrue
\mciteSetBstMidEndSepPunct{\mcitedefaultmidpunct}
{\mcitedefaultendpunct}{\mcitedefaultseppunct}\relax
\EndOfBibitem
\bibitem[Espinosa-Marzal \latin{et~al.}(2023)Espinosa-Marzal, Goodwin, Zhang,
  and Zheng]{Espinosa2023rev}
Espinosa-Marzal,~R.~M.; Goodwin,~Z.~A.; Zhang,~X.; Zheng,~Q. Colloidal
  Interactions in Ionic Liquids—The Electrical Double Layer Inferred from Ion
  Layering and Aggregation. One Hundred Years of Colloid Symposia: Looking Back
  and Looking Forward. 2023; pp 123--148\relax
\mciteBstWouldAddEndPuncttrue
\mciteSetBstMidEndSepPunct{\mcitedefaultmidpunct}
{\mcitedefaultendpunct}{\mcitedefaultseppunct}\relax
\EndOfBibitem
\bibitem[J{\"a}ger \latin{et~al.}(2023)J{\"a}ger, Schlaich, Yang, Lian,
  Kondrat, and Holm]{jager2023screening}
J{\"a}ger,~H.; Schlaich,~A.; Yang,~J.; Lian,~C.; Kondrat,~S.; Holm,~C. A
  screening of results on the decay length in concentrated electrolytes.
  \emph{Faraday Discussions} \textbf{2023}, \emph{246}, 520--539\relax
\mciteBstWouldAddEndPuncttrue
\mciteSetBstMidEndSepPunct{\mcitedefaultmidpunct}
{\mcitedefaultendpunct}{\mcitedefaultseppunct}\relax
\EndOfBibitem
\bibitem[H{\"a}rtel \latin{et~al.}(2023)H{\"a}rtel, B{\"u}ltmann, and
  Coupette]{hartel2023anomalous}
H{\"a}rtel,~A.; B{\"u}ltmann,~M.; Coupette,~F. Anomalous underscreening in the
  restricted primitive model. \emph{Physical Review Letters} \textbf{2023},
  \emph{130}, 108202\relax
\mciteBstWouldAddEndPuncttrue
\mciteSetBstMidEndSepPunct{\mcitedefaultmidpunct}
{\mcitedefaultendpunct}{\mcitedefaultseppunct}\relax
\EndOfBibitem
\bibitem[Kumar \latin{et~al.}(2022)Kumar, Cats, Alotaibi, Ayirala, Yousef, van
  Roij, Siretanu, and Mugele]{kumar2022absence}
Kumar,~S.; Cats,~P.; Alotaibi,~M.~B.; Ayirala,~S.~C.; Yousef,~A.~A.; van
  Roij,~R.; Siretanu,~I.; Mugele,~F. Absence of anomalous underscreening in
  highly concentrated aqueous electrolytes confined between smooth silica
  surfaces. \emph{Journal of colloid and interface science} \textbf{2022},
  \emph{622}, 819--827\relax
\mciteBstWouldAddEndPuncttrue
\mciteSetBstMidEndSepPunct{\mcitedefaultmidpunct}
{\mcitedefaultendpunct}{\mcitedefaultseppunct}\relax
\EndOfBibitem
\bibitem[Zeman \latin{et~al.}(2020)Zeman, Kondrat, and Holm]{zeman2020bulk}
Zeman,~J.; Kondrat,~S.; Holm,~C. Bulk ionic screening lengths from extremely
  large-scale molecular dynamics simulations. \emph{Chemical Communications}
  \textbf{2020}, \emph{56}, 15635--15638\relax
\mciteBstWouldAddEndPuncttrue
\mciteSetBstMidEndSepPunct{\mcitedefaultmidpunct}
{\mcitedefaultendpunct}{\mcitedefaultseppunct}\relax
\EndOfBibitem
\bibitem[Perez-Martinez \latin{et~al.}(2017)Perez-Martinez, Smith, Perkin,
  \latin{et~al.} others]{perez2017underscreening}
Perez-Martinez,~C.~S.; Smith,~A.~M.; Perkin,~S., \latin{et~al.}  Underscreening
  in concentrated electrolytes. \emph{Faraday discussions} \textbf{2017},
  \emph{199}, 239--259\relax
\mciteBstWouldAddEndPuncttrue
\mciteSetBstMidEndSepPunct{\mcitedefaultmidpunct}
{\mcitedefaultendpunct}{\mcitedefaultseppunct}\relax
\EndOfBibitem
\bibitem[Molinari \latin{et~al.}(2019)Molinari, Mailoa, and
  Kozinsky]{molinari2019general}
Molinari,~N.; Mailoa,~J.~P.; Kozinsky,~B. General Trend of a Negative Li
  Effective Charge in Ionic Liquid Electrolytes. \emph{J. Phys. Chem. Lett.}
  \textbf{2019}, \emph{10}, 2313--2319\relax
\mciteBstWouldAddEndPuncttrue
\mciteSetBstMidEndSepPunct{\mcitedefaultmidpunct}
{\mcitedefaultendpunct}{\mcitedefaultseppunct}\relax
\EndOfBibitem
\bibitem[Molinari \latin{et~al.}(2019)Molinari, Mailoa, Craig, Christensen, and
  Kozinsky]{molinari2019transport}
Molinari,~N.; Mailoa,~J.~P.; Craig,~N.; Christensen,~J.; Kozinsky,~B. Transport
  anomalies emerging from strong correlation in ionic liquid electrolytes.
  \emph{J. Power Sources} \textbf{2019}, \emph{428}, 27--36\relax
\mciteBstWouldAddEndPuncttrue
\mciteSetBstMidEndSepPunct{\mcitedefaultmidpunct}
{\mcitedefaultendpunct}{\mcitedefaultseppunct}\relax
\EndOfBibitem
\bibitem[Attard(1993)]{attard1993asymptotic}
Attard,~P. Asymptotic analysis of primitive model electrolytes and the
  electrical double layer. \emph{Physical Review E} \textbf{1993}, \emph{48},
  3604\relax
\mciteBstWouldAddEndPuncttrue
\mciteSetBstMidEndSepPunct{\mcitedefaultmidpunct}
{\mcitedefaultendpunct}{\mcitedefaultseppunct}\relax
\EndOfBibitem
\bibitem[Leote~de Carvalho and Evans(1994)Leote~de Carvalho, and
  Evans]{leote1994decay}
Leote~de Carvalho,~R.; Evans,~R. The decay of correlations in ionic fluids.
  \emph{Molecular Physics} \textbf{1994}, \emph{83}, 619--654\relax
\mciteBstWouldAddEndPuncttrue
\mciteSetBstMidEndSepPunct{\mcitedefaultmidpunct}
{\mcitedefaultendpunct}{\mcitedefaultseppunct}\relax
\EndOfBibitem
\bibitem[Goodwin \latin{et~al.}(2021)Goodwin, de~Souza, Bazant, and
  Kornyshev]{Goodwin2021Rev}
Goodwin,~Z. A.~H.; de~Souza,~J.~P.; Bazant,~M.~Z.; Kornyshev,~A.~A. Mean-Field
  Theory of the Electrical Double Layer in Ionic Liquids. \emph{Encyclopedia of
  Ionic Liquids} \textbf{2021}, 1--13\relax
\mciteBstWouldAddEndPuncttrue
\mciteSetBstMidEndSepPunct{\mcitedefaultmidpunct}
{\mcitedefaultendpunct}{\mcitedefaultseppunct}\relax
\EndOfBibitem
\bibitem[Fraggedakis \latin{et~al.}(2021)Fraggedakis, McEldrew, Smith,
  Krishnan, Zhang, Bai, Chueh, Shao-Horn, and Z.Bazant]{Fraggedakis2021}
Fraggedakis,~D.; McEldrew,~M.; Smith,~R.~B.; Krishnan,~Y.; Zhang,~Y.; Bai,~P.;
  Chueh,~W.~C.; Shao-Horn,~Y.; Z.Bazant,~M. Theory of coupled ion-electron
  transfer kinetics. \emph{Electrochim. Acta} \textbf{2021}, \emph{367},
  137432\relax
\mciteBstWouldAddEndPuncttrue
\mciteSetBstMidEndSepPunct{\mcitedefaultmidpunct}
{\mcitedefaultendpunct}{\mcitedefaultseppunct}\relax
\EndOfBibitem
\bibitem[Bazant(2023)]{bazant2023unified}
Bazant,~M.~Z. Unified quantum theory of electrochemical kinetics by coupled
  ion--electron transfer. \emph{Faraday Discussions} \textbf{2023}, \emph{246},
  60--124\relax
\mciteBstWouldAddEndPuncttrue
\mciteSetBstMidEndSepPunct{\mcitedefaultmidpunct}
{\mcitedefaultendpunct}{\mcitedefaultseppunct}\relax
\EndOfBibitem
\bibitem[Teixeira and Sciortino(2019)Teixeira, and Sciortino]{Teixeira2019}
Teixeira,~P. I.~C.; Sciortino,~F. Patchy particles at a hard wall:
  Orientation-dependent bonding. \emph{J. Phys. Chem.} \textbf{2019},
  \emph{151}, 174903\relax
\mciteBstWouldAddEndPuncttrue
\mciteSetBstMidEndSepPunct{\mcitedefaultmidpunct}
{\mcitedefaultendpunct}{\mcitedefaultseppunct}\relax
\EndOfBibitem
\bibitem[Zhang \latin{et~al.}(2020)Zhang, Lan, Qian, Zhao, and
  Wang]{zhang2020microscale}
Zhang,~G.; Lan,~H.; Qian,~L.; Zhao,~J.; Wang,~F. A microscale 3D printing based
  on the electric-field-driven jet. \emph{3D Printing and Additive
  Manufacturing} \textbf{2020}, \emph{7}, 37--44\relax
\mciteBstWouldAddEndPuncttrue
\mciteSetBstMidEndSepPunct{\mcitedefaultmidpunct}
{\mcitedefaultendpunct}{\mcitedefaultseppunct}\relax
\EndOfBibitem
\bibitem[Cohen and Benedek(1982)Cohen, and Benedek]{cohen1982equilibrium}
Cohen,~R.~J.; Benedek,~G.~B. Equilibrium and kinetic theory of polymerization
  and the sol-gel transition. \emph{The Journal of Physical Chemistry}
  \textbf{1982}, \emph{86}, 3696--3714\relax
\mciteBstWouldAddEndPuncttrue
\mciteSetBstMidEndSepPunct{\mcitedefaultmidpunct}
{\mcitedefaultendpunct}{\mcitedefaultseppunct}\relax
\EndOfBibitem
\bibitem[Danks \latin{et~al.}(2016)Danks, Hall, and
  Schnepp]{danks2016evolution}
Danks,~A.~E.; Hall,~S.~R.; Schnepp,~Z. The evolution of ‘sol--gel’chemistry
  as a technique for materials synthesis. \emph{Materials Horizons}
  \textbf{2016}, \emph{3}, 91--112\relax
\mciteBstWouldAddEndPuncttrue
\mciteSetBstMidEndSepPunct{\mcitedefaultmidpunct}
{\mcitedefaultendpunct}{\mcitedefaultseppunct}\relax
\EndOfBibitem
\bibitem[Descartes(1637)]{Descartes}
Descartes,~R. \emph{La g{\'e}om{\'e}trie}; 1637\relax
\mciteBstWouldAddEndPuncttrue
\mciteSetBstMidEndSepPunct{\mcitedefaultmidpunct}
{\mcitedefaultendpunct}{\mcitedefaultseppunct}\relax
\EndOfBibitem
\bibitem[Plimpton(1995)]{plimpton1995}
Plimpton,~S. Fast Parallel Algorithms for Short-Range Molecular Dynamics.
  \emph{J. Comput. Phys.} \textbf{1995}, \emph{117}, 1 -- 19\relax
\mciteBstWouldAddEndPuncttrue
\mciteSetBstMidEndSepPunct{\mcitedefaultmidpunct}
{\mcitedefaultendpunct}{\mcitedefaultseppunct}\relax
\EndOfBibitem
\bibitem[Mart{\'\i}nez \latin{et~al.}(2009)Mart{\'\i}nez, Andrade, Birgin, and
  Mart{\'\i}nez]{martinez2009packmol}
Mart{\'\i}nez,~L.; Andrade,~R.; Birgin,~E.~G.; Mart{\'\i}nez,~J.~M. PACKMOL: a
  package for building initial configurations for molecular dynamics
  simulations. \emph{J. Comput. Chem.} \textbf{2009}, \emph{30},
  2157--2164\relax
\mciteBstWouldAddEndPuncttrue
\mciteSetBstMidEndSepPunct{\mcitedefaultmidpunct}
{\mcitedefaultendpunct}{\mcitedefaultseppunct}\relax
\EndOfBibitem
\bibitem[Lopes and P{\'a}dua(2012)Lopes, and P{\'a}dua]{lopes2012}
Lopes,~J. N.~C.; P{\'a}dua,~A.~A. CL\&P: A generic and systematic force field
  for ionic liquids modeling. \emph{Theoretical Chemistry Accounts}
  \textbf{2012}, \emph{131}, 1129\relax
\mciteBstWouldAddEndPuncttrue
\mciteSetBstMidEndSepPunct{\mcitedefaultmidpunct}
{\mcitedefaultendpunct}{\mcitedefaultseppunct}\relax
\EndOfBibitem
\bibitem[Hockney and Eastwood(1988)Hockney, and Eastwood]{hockney1988}
Hockney,~R.~W.; Eastwood,~J.~W. \emph{Computer simulation using particles}; CRC
  Press, 1988\relax
\mciteBstWouldAddEndPuncttrue
\mciteSetBstMidEndSepPunct{\mcitedefaultmidpunct}
{\mcitedefaultendpunct}{\mcitedefaultseppunct}\relax
\EndOfBibitem
\bibitem[Wertheim(1984)]{wertheim1984fluids1}
Wertheim,~M.~S. Fluids with highly directional attractive forces. I.
  Statistical thermodynamics. \emph{Journal of statistical physics}
  \textbf{1984}, \emph{35}, 19--34\relax
\mciteBstWouldAddEndPuncttrue
\mciteSetBstMidEndSepPunct{\mcitedefaultmidpunct}
{\mcitedefaultendpunct}{\mcitedefaultseppunct}\relax
\EndOfBibitem
\bibitem[Wertheim(1984)]{wertheim1984fluids2}
Wertheim,~M.~S. Fluids with highly directional attractive forces. II.
  Thermodynamic perturbation theory and integral equations. \emph{Journal of
  statistical physics} \textbf{1984}, \emph{35}, 35--47\relax
\mciteBstWouldAddEndPuncttrue
\mciteSetBstMidEndSepPunct{\mcitedefaultmidpunct}
{\mcitedefaultendpunct}{\mcitedefaultseppunct}\relax
\EndOfBibitem
\bibitem[Wertheim(1986)]{wertheim1986fluids1}
Wertheim,~M. Fluids with highly directional attractive forces. III. Multiple
  attraction sites. \emph{Journal of statistical physics} \textbf{1986},
  \emph{42}, 459--476\relax
\mciteBstWouldAddEndPuncttrue
\mciteSetBstMidEndSepPunct{\mcitedefaultmidpunct}
{\mcitedefaultendpunct}{\mcitedefaultseppunct}\relax
\EndOfBibitem
\bibitem[Wertheim(1986)]{wertheim1986fluids2}
Wertheim,~M. Fluids with highly directional attractive forces. IV. Equilibrium
  polymerization. \emph{Journal of statistical physics} \textbf{1986},
  \emph{42}, 477--492\relax
\mciteBstWouldAddEndPuncttrue
\mciteSetBstMidEndSepPunct{\mcitedefaultmidpunct}
{\mcitedefaultendpunct}{\mcitedefaultseppunct}\relax
\EndOfBibitem
\bibitem[Lar{\'\i}a \latin{et~al.}(1990)Lar{\'\i}a, Corti, and
  Fern{\'a}ndez-Prini]{laria1990cluster}
Lar{\'\i}a,~D.; Corti,~H.~R.; Fern{\'a}ndez-Prini,~R. The cluster theory for
  electrolyte solutions. Its extension and its limitations. \emph{Journal of
  the Chemical Society, Faraday Transactions} \textbf{1990}, \emph{86},
  1051--1056\relax
\mciteBstWouldAddEndPuncttrue
\mciteSetBstMidEndSepPunct{\mcitedefaultmidpunct}
{\mcitedefaultendpunct}{\mcitedefaultseppunct}\relax
\EndOfBibitem
\bibitem[Blum and Bernard(1995)Blum, and Bernard]{blum1995general}
Blum,~L.; Bernard,~O. The general solution of the binding mean spherical
  approximation for pairing ions. \emph{Journal of statistical physics}
  \textbf{1995}, \emph{79}, 569--583\relax
\mciteBstWouldAddEndPuncttrue
\mciteSetBstMidEndSepPunct{\mcitedefaultmidpunct}
{\mcitedefaultendpunct}{\mcitedefaultseppunct}\relax
\EndOfBibitem
\bibitem[Simonin \latin{et~al.}(1999)Simonin, Bernard, and
  Blum]{simonin1999ionic}
Simonin,~J.-P.; Bernard,~O.; Blum,~L. Ionic solutions in the binding mean
  spherical approximation: thermodynamic properties of mixtures of associating
  electrolytes. \emph{The Journal of Physical Chemistry B} \textbf{1999},
  \emph{103}, 699--704\relax
\mciteBstWouldAddEndPuncttrue
\mciteSetBstMidEndSepPunct{\mcitedefaultmidpunct}
{\mcitedefaultendpunct}{\mcitedefaultseppunct}\relax
\EndOfBibitem
\bibitem[Sciortino \latin{et~al.}(2007)Sciortino, Bianchi, Douglas, and
  Tartaglia]{sciortino2007self}
Sciortino,~F.; Bianchi,~E.; Douglas,~J.~F.; Tartaglia,~P. Self-assembly of
  patchy particles into polymer chains: A parameter-free comparison between
  Wertheim theory and Monte Carlo simulation. \emph{The Journal of chemical
  physics} \textbf{2007}, \emph{126}\relax
\mciteBstWouldAddEndPuncttrue
\mciteSetBstMidEndSepPunct{\mcitedefaultmidpunct}
{\mcitedefaultendpunct}{\mcitedefaultseppunct}\relax
\EndOfBibitem
\bibitem[Dr{\"u}schler \latin{et~al.}(2011)Dr{\"u}schler, Huber, and
  Roling]{druschler2011capacitive}
Dr{\"u}schler,~M.; Huber,~B.; Roling,~B. On Capacitive Processes at the
  Interface between 1-Ethyl-3-methylimidazolium tris (pentafluoroethyl)
  trifluorophosphate and Au (111). \emph{The Journal of Physical Chemistry C}
  \textbf{2011}, \emph{115}, 6802--6808\relax
\mciteBstWouldAddEndPuncttrue
\mciteSetBstMidEndSepPunct{\mcitedefaultmidpunct}
{\mcitedefaultendpunct}{\mcitedefaultseppunct}\relax
\EndOfBibitem
\bibitem[Klein \latin{et~al.}(2019)Klein, Panichi, and
  Gurkan]{klein2019potential}
Klein,~J.~M.; Panichi,~E.; Gurkan,~B. Potential dependent capacitance of
  [EMIM][TFSI],[N 1114][TFSI] and [PYR 13][TFSI] ionic liquids on glassy
  carbon. \emph{Physical Chemistry Chemical Physics} \textbf{2019}, \emph{21},
  3712--3720\relax
\mciteBstWouldAddEndPuncttrue
\mciteSetBstMidEndSepPunct{\mcitedefaultmidpunct}
{\mcitedefaultendpunct}{\mcitedefaultseppunct}\relax
\EndOfBibitem
\bibitem[Roling \latin{et~al.}(2012)Roling, Dr{\"u}schler, and
  Huber]{roling2012slow}
Roling,~B.; Dr{\"u}schler,~M.; Huber,~B. Slow and fast capacitive process
  taking place at the ionic liquid/electrode interface. \emph{Faraday
  discussions} \textbf{2012}, \emph{154}, 303--311\relax
\mciteBstWouldAddEndPuncttrue
\mciteSetBstMidEndSepPunct{\mcitedefaultmidpunct}
{\mcitedefaultendpunct}{\mcitedefaultseppunct}\relax
\EndOfBibitem
\bibitem[Reichert \latin{et~al.}(2018)Reichert, Kj{\ae}r, van Driel, Mars,
  Ochsmann, Pontoni, Deutsch, Nielsen, and Mezger]{reichert2018molecular}
Reichert,~P.; Kj{\ae}r,~K.~S.; van Driel,~T.~B.; Mars,~J.; Ochsmann,~J.~W.;
  Pontoni,~D.; Deutsch,~M.; Nielsen,~M.~M.; Mezger,~M. Molecular scale
  structure and dynamics at an ionic liquid/electrode interface. \emph{Faraday
  discussions} \textbf{2018}, \emph{206}, 141--157\relax
\mciteBstWouldAddEndPuncttrue
\mciteSetBstMidEndSepPunct{\mcitedefaultmidpunct}
{\mcitedefaultendpunct}{\mcitedefaultseppunct}\relax
\EndOfBibitem
\bibitem[Lockett \latin{et~al.}(2008)Lockett, Sedev, Ralston, Horne, and
  Rodopoulos]{lockett2008differential}
Lockett,~V.; Sedev,~R.; Ralston,~J.; Horne,~M.; Rodopoulos,~T. Differential
  capacitance of the electrical double layer in imidazolium-based ionic
  liquids: influence of potential, cation size, and temperature. \emph{The
  Journal of Physical Chemistry C} \textbf{2008}, \emph{112}, 7486--7495\relax
\mciteBstWouldAddEndPuncttrue
\mciteSetBstMidEndSepPunct{\mcitedefaultmidpunct}
{\mcitedefaultendpunct}{\mcitedefaultseppunct}\relax
\EndOfBibitem
\bibitem[Hirschorn \latin{et~al.}(2010)Hirschorn, Orazem, Tribollet, Vivier,
  Frateur, and Musiani]{hirschorn2010determination}
Hirschorn,~B.; Orazem,~M.~E.; Tribollet,~B.; Vivier,~V.; Frateur,~I.;
  Musiani,~M. Determination of effective capacitance and film thickness from
  constant-phase-element parameters. \emph{Electrochimica acta} \textbf{2010},
  \emph{55}, 6218--6227\relax
\mciteBstWouldAddEndPuncttrue
\mciteSetBstMidEndSepPunct{\mcitedefaultmidpunct}
{\mcitedefaultendpunct}{\mcitedefaultseppunct}\relax
\EndOfBibitem
\bibitem[Small and Wheeler(2014)Small, and Wheeler]{small2014influence}
Small,~L.~J.; Wheeler,~D.~R. Influence of analysis method on the experimentally
  observed capacitance at the gold-ionic liquid interface. \emph{Journal of The
  Electrochemical Society} \textbf{2014}, \emph{161}, H260\relax
\mciteBstWouldAddEndPuncttrue
\mciteSetBstMidEndSepPunct{\mcitedefaultmidpunct}
{\mcitedefaultendpunct}{\mcitedefaultseppunct}\relax
\EndOfBibitem
\end{mcitethebibliography}

\end{document}